\documentclass[letterpaper,11pt]{article}
\usepackage[margin=1in]{geometry}
\usepackage{my_style}
\usepackage{subfiles}
\usepackage{float}
\usepackage{wrapfig}
\usepackage{microtype} %if unwanted, comment out or use option "draft"
\usepackage{thmtools}
\usepackage{thm-restate}

\usepackage{floatflt}
\usepackage{graphics}

\begin{document}
\pagenumbering{gobble}
\begin{titlepage}

\title{Sensitivity Oracles for All-Pairs Mincuts}
%%%%%%%%%%%%%%%%%%%%%%Uncomment to show the authors
\author{
  Surender Baswana\thanks{Department of Computer Science \& Engineering, IIT Kanpur, Kanpur -- 208016, India, sbaswana@cse.iitk.ac.in}
  \and
  Abhyuday Pandey\thanks{Department of Computer Science \& Engineering, IIT Kanpur, Kanpur -- 208016, India, pandey.abhyuday07@gmail.com}
}
%%%%%%%%%%%%%%%%%%%%%%Uncomment to show the authors
\maketitle

\begin{abstract}
% Let $G=(V,E)$ be an undirected unweighted graph on $n$ vertices and $m$ edges. We address the problem of fault-tolerant data structure for all-pairs mincuts in $G$ defined as follows.

% Build a compact data structure that, on receiving a pair of vertices $s,t\in V$ and any edge $(x,y)$ as query, can efficiently report the value of the mincut between $s$ and $t$ upon failure of the edge $(x,y)$.

% To the best of our knowledge, there exists no data structure for this problem which takes $o(mn)$ space and a non-trivial query time. We present two compact data structures for this problem.
% \begin{enumerate}
% \item Our first data structure guarantees ${\cal O}(1)$ query time. The space occupied by this data structure is ${\cal O}(n^2)$ which matches the worst-case size of a graph on $n$ vertices.
% \item
% Our second data structure takes ${\cal O}(m)$ space which 
% % is optimal
% matches the size of the graph. The query time is ${\cal O}(\min(m,n c_{s,t}))$ where $c_{s,t}$ is the value of the mincut between $s$ and $t$ in $G$. The query time guaranteed by our data structure is faster by a factor of $\Omega(\min(m^{1/3},\sqrt{n}))$ compared to the best known deterministic algorithm \cite{DBLP:conf/focs/GoldbergR97a,DBLP:conf/stoc/KargerL98,DBLP:journals/corr/abs-2003-08929} to compute a $(s,t)$-mincut.
% \end{enumerate}

% Both these data structures can also report the resulting $(s,t)$-mincut incorporating the failure in ${\cal O}(\min(m,n c_{s,t}))$ time.

{
Let $G=(V,E)$ be an undirected unweighted graph on $n$ vertices and $m$ edges. We address the problem of sensitivity oracle for all-pairs mincuts in $G$ defined as follows.

Build a compact data structure that, on receiving any pair of vertices $s,t\in V$ and failure (or insertion) of any edge as query, can efficiently report the mincut between $s$ and $t$ after the failure (or the insertion).

To the best of our knowledge, there exists no data structure for this problem which takes $o(mn)$ space and a non-trivial query time.
% Recently, Baswana, Gupta, and Knollman \cite{DBLP:conf/esa/BaswanaGK20} gave a data structure that can handle single edge insertion in ${\cal O}(n^2)$ space and ${\cal O}(1)$ query time. 
We present the following results.

\begin{enumerate}
    \item Our first data structure occupies ${\cal O}(n^2)$ space and guarantees ${\cal O}(1)$ query time to report the value of resulting $(s,t)$-mincut upon failure (or insertion) of any edge. Moreover, the set of vertices defining a resulting $(s,t)$-mincut after the update can be reported in ${\cal O}(n)$ time which is worst-case optimal.
    % Our data structure also subsumes the previous results of Baswana, Gupta, and Knollman \cite{DBLP:conf/esa/BaswanaGK20} which only handles an edge insertion.
\item
Our second data structure optimizes space at the expense of increased query time. It takes ${\cal O}(m)$ space -- which is also the space taken by $G$. The query time is ${\cal O}(\min(m,n c_{s,t}))$ where $c_{s,t}$ is the value of the mincut between $s$ and $t$ in $G$. This query time is faster by a factor of $\Omega(\min(m^{1/3},\sqrt{n}))$ compared to the best known deterministic algorithm \cite{DBLP:conf/focs/GoldbergR97a,DBLP:conf/stoc/KargerL98,DBLP:journals/corr/abs-2003-08929} to compute a $(s,t)$-mincut from scratch.

\item 
If we are only interested in knowing if failure (or insertion) of an edge changes the value of $(s,t)$-mincut, we can distribute our ${\cal O}(n^2)$ space data structure evenly among $n$ vertices. For any failed (or inserted) edge we only require the data structures stored at its endpoints to determine if the value of $(s,t)$-mincut has changed for any $s,t \in V$. 
Moreover, using these data structures we can also output efficiently a compact encoding of all pairs of vertices whose mincut value is changed after the failure (or insertion) of the edge.

\end{enumerate}
}
\end{abstract}
\end{titlepage}
\pagebreak
\pagenumbering{arabic}
\section{Introduction}
Graph mincut is a fundamental structure in graph theory with numerous applications. Let $G=(V,E)$ be an undirected unweighted connected graph on $n=|V|$ vertices and $m=|E|$ edges.
Two most common types of mincuts are global mincuts and pairwise mincuts. A set of edges with the least cardinality whose removal disconnects the graph is called a global mincut. For any pair of vertices $s,t\in V$, 
a set of edges with the least cardinality whose removal disconnects $t$ from $s$ is called a pairwise mincut for $s,t$ or simply a $(s,t)$-mincut. A more general notion is that of Steiner mincuts. For any given set $S\subseteq V$ of vertices, a set of edges with the least cardinality whose removal disconnects $S$ is called a Steiner mincut for $S$. It is easy to observe that the Steiner mincuts for $S=V$ are the global mincuts and for $S=\{s,t\}$ are $(s,t)$-mincuts.

% {\color{red} (-)
% While designing an algorithm for a graph problem, one usually assumes that the underlying graph is static. But, this assumption is unrealistic for most of the real-world graphs where vertices and/or edges do fail, though occasionally. While it is impractical to assume that a real-world graph is immune to such failures, it is also a fact that these failures are transient - an edge (or a vertex) once failed, becomes active after some time due to the simultaneous repair mechanism that is undertaken in most of the real-world graphs. So the set of failed edges (or vertices), though small in size, keeps changing with time. Therefore, a more realistic way to model a real-world graph is to assume that, at any time, there will be at most $k$ failed vertices or edges for some $k$ defined suitably. Solving a graph problem in this model requires building a compact data structure such that given any set of at most $k$ failed vertices or edges, the solution of the problem can be reported efficiently.
% }

 Typically a graph algorithmic problem is solved assuming that the underlying graph is static. However, changes are inevitable in most real world graphs. For example, real world graphs are prone to failures of edges. These failures are transient in nature. As a result, the set of failed edges at any time, though small in size, may keep changing as time goes by. Therefore, a data structure for a problem on a real world graph has to consider a given set of failed edges as well to answer a query correctly. It is also important to efficiently know the {\em change} in the solution of the problem for a given set of failed edges or a given set of new edges. This knowledge can help to efficiently determine the {\em impact} of the failure/insertion of an edge on the solution of the problem. Solving a graph problem in this model requires building a compact data structure such that given any set of at most $k$ changes, i.e. failures (or insertions) of edges (or vertices), the solution of the problem can be reported efficiently. It is important to note that this model subsumes the fault-tolerant model, in which we are only concerned about failure of edges (or vertices).

In the past, many elegant fault-tolerant data structures have been designed for various classical problems, namely,  connectivity \cite{DBLP:journals/algorithmica/FrigioniI00,DBLP:journals/siamcomp/ChanPR11,DBLP:conf/soda/DuanP17}, shortest-paths \cite{DBLP:conf/stoc/BernsteinK09,DBLP:journals/siamcomp/DemetrescuTCR08,DBLP:conf/stoc/ChechikC20}, graph spanners \cite{DBLP:journals/siamcomp/ChechikLPR10,DBLP:journals/tcs/BraunschvigCPS15}, SCC \cite{DBLP:journals/algorithmica/BaswanaCR19} , DFS tree \cite{DBLP:journals/siamcomp/BaswanaCC019}, and BFS structure \cite{DBLP:journals/talg/ParterP16,DBLP:journals/talg/ParterP18}. Recently, a data structure was designed which could report the mincut between a pair of vertices amidst insertion of an edge \cite{DBLP:conf/esa/BaswanaGK20}. However, little is known about fault-tolerant data structures for various types of mincuts.

The problem of sensitivity oracle for all-pairs mincuts aims at preprocessing a given graph to build a compact data structure so that the following queries can be answered efficiently for any $s,t\in V$ and $(x,y) \in V \times V$.

\begin{itemize}
    \item {\textsc{ft-Mincut}}$(s,t,x,y)$: Report a $(s,t)$-mincut in $G$ after the failure of edge $(x,y)\in E$.
    \item {\textsc{in-Mincut}}$(s,t,x,y)$: Report a $(s,t)$-mincut in $G$ upon insertion of edge $(x,y)\in V \times V$.
\end{itemize}
%}

% \noindent
% FT-mincut$(u,v,x,y)$: Report the value of $(u,v)$-mincut in $G$ after the failure/removal of edge $(x,y)$, if exists, from $E$.\\
% \noindent
% {\textsc{ft-mincut}}$(s,t,x,y)$: Report a $(s,t)$-mincut in $G$ after the failure of edge $(x,y)$, if exists, from $E$.\\

% Upon failure of edge $(x,y)$, the value of $(s,t)$-mincut for any $s,t\in V$ either decreases by unity or remains unchanged. We now state the necessary and sufficient condition for the value of $(s,t)$-mincut to decrease upon failure of edge $(x,y)$.

% It follows from Fact \ref{fact:(x,y)-lies-in-(s,t)-mincut} that any {\textsc{ft-mincut}}($s,t,x,y$) query can be answered by determining whether $(x,y)$ belongs to any $(s,t)$-mincut. Unfortunately, there can be exponential number of $(s,t)$-mincuts in a given graph \cite{DBLP:journals/mp/PicardQ82}. Thus, designing a compact data structure to answer this query efficiently turns out to be a challenging task.

\paragraph{Previous Results:} 
\label{sec:previous-results}

There exists a classical ${\cal O}(n)$ size data structure that stores all-pairs mincuts \cite{GH61} known as Gomory-Hu tree. It is a tree on the vertex set $V$ that compactly stores a mincut between each pair of vertices. However, we cannot determine using a Gomory-Hu tree whether the failure of an edge will affect the $(s,t)$-mincut unless this edge belongs to the $(s,t)$-mincut present in the tree. We can get a fault-tolerant data structure by storing $m$ Gomory-Hu trees, one for each edge failure. The overall data structure occupies ${\cal O}(mn)$ space and takes ${\cal O}(1)$ time to report the value of $(s,t)$-mincut for any $s,t\in V$ upon failure of a given edge. 
% However, the ${\cal O}(mn)$ space occupied by this data structure is far from the size of the graph. 
% To the best of our knowledge, even after $60$ years of existence of Gomory-Hu tree, there exists no data structure for this problem which takes $o(mn)$ space and a non-trivial query time.
%{ \color{blue}
Recently, for the case of single edge insertion only, Baswana, Gupta, and Knollman \cite{DBLP:conf/esa/BaswanaGK20} designed an ${\cal O}(n^2)$ data structure that can report the value of a $(s,t)$-mincut upon insertion of an edge $(x,y)$ in ${\cal O}(1)$ time for any given $s,t,x,y \in V$.  To the best of our knowledge, there exists no sensitivity data structure for all-pairs mincuts problem which takes $o(mn)$ space and has a non-trivial query time.
%}

\paragraph{Our Contribution:} We present the following data structures for the all-pairs mincuts sensitivity problem in an undirected unweighted graph.

\begin{enumerate}
    \item {\bf Query Optimal:}~~ 
%{\color{blue}
   Our first data structure occupies ${\cal O}(n^2)$ space and guarantees ${\cal O}(1)$ query time to report the value of a $(s,t)$-mincut for any $s,t \in V$ upon failure/insertion of any edge. Moreover, the set of vertices defining a resulting $(s,t)$-mincut after the update can be reported in ${\cal O}(n)$ time which is worst-case optimal. Therefore, our data structure also subsumes the all-pairs sensitivity result of Baswana, Gupta, and Knollman \cite{DBLP:conf/esa/BaswanaGK20} which only handles an edge insertion.

  An interesting byproduct of our data structure, which is of independent interest, is that it can output a compact representation \cite{DBLP:journals/mp/PicardQ82} storing all $(s,t)$-mincuts for any $s,t \in V$ in ${\cal O}(m)$ time which is optimal (see Appendix \ref{appendix:xxx}). 
%   { \color{red}
%   For any given edge failure (or insertion) our data structure can also report all pairs of vertices whose mincut decreases (or increases) in time ${\cal O}()$. Our data structure also subsumes the previous results of Baswana, Gupta, and Knollman \cite{DBLP:conf/esa/BaswanaGK20} which only handles an edge insertion.
%   }
 
%}

    \item {\bf Compact:}~~
Our second data structure occupies ${\cal O}(m)$ space.
% and is indeed optimal (Lemma \ref{lem:lower-bound})
The query time guaranteed by this data structure to report a $(s,t)$-mincut for any $s,t\in V$ upon failure/insertion of any edge is ${\cal O}(\min(m,nc_{s,t}))$ where $c_{s,t}$ is the value of $(s,t)$-mincut in graph $G$. This query time 
is faster by a factor of $\Omega(\min(m^{1/3},\sqrt{n}))$ (details in Appendix \ref{appendix:non-trivial-query-time}) compared to the best known deterministic algorithm \cite{DBLP:conf/focs/GoldbergR97a,DBLP:conf/stoc/KargerL98,DBLP:journals/corr/abs-2003-08929} to compute a $(s,t)$-mincut from scratch.

\begin{remark}
There exists a Las-Vegas algorithm \cite{DBLP:conf/stoc/KargerL98} to compute a $(s,t)$-mincut in ${\cal O}(\max(m,nc_{s,t}))$ time. 
% However, it remains an open problem if there exists a deterministic algorithm for static $(s,t)$-mincut with ${\cal O}(m^{1+o(1)})$ time.
However, no deterministic algorithm till date matches this bound even within the factor of $m^{o(1)}$ \cite{DBLP:conf/focs/GoldbergR97a,DBLP:conf/stoc/KargerL98,DBLP:journals/corr/abs-2003-08929}.
\end{remark}

    \item {\bf Distributed:}~~
If we are only interested in just knowing if failure/insertion of any edge \textit{changes} the value of $(s,t)$-mincut for any $s,t\in V$, we can have the following distributed data structure. 
%we can distribute our ${\cal O}(n^2)$ space data structure among $n$ vertices. 
Each vertex $v \in V$ stores an ${\cal O}(n)$ space data structure ${\cal L}_v$. For any failed/inserted edge $(x,y)$ we require only ${\cal L}_x$ and ${\cal L}_y$ to determine if the value of $(s,t)$-mincut has changed for any $s,t \in V$. Using ${\cal L}_x$ and ${\cal L}_y$, we can also output a compact encoding of all pairs of vertices whose mincut value is changed upon the failure/insertion of edge $(x,y)$. This encoding takes ${\cal O}(\min(n,k))$ space where $k$ is the number of affected pairs.

\end{enumerate}

Our first data structure achieves optimal query time but uses ${\cal O}(n^2)$ space. The second data structure, on the other hand, takes linear space but has ${\cal O}(\min(m,nc_{s,t}))$ query time.
%{\color{blue} It is natural to ask if we can improve the query time and space simultaneously. In other words, the following problem is still open.}
Therefore, the following problem still remains open. 

\textit{For a sparse undirected graph, does there exist an $o(n^2)$ size data structure that can answer all-pairs mincut sensitivity queries in $o(m)$ time?} 

\begin{remark}
The above problem appears very hard even for static all-pairs reachability data structure for a directed graph \cite{DBLP:journals/siamcomp/Patrascu11,DBLP:conf/wads/GoldsteinKLP17} which is a more fundamental problem.
\end{remark}

In order to design our data structures, we present an efficient solution for a related problem of independent interest, called edge-containment query on a mincut defined as follows.\\
%}

\noindent
% {\textsc{Edge-Contained}}$(s,t,(x,y))$: Check if an edge $(x,y)\in E$ belong to some $(s,t)$-mincut.\\
{\textsc{Edge-Contained}}$(s,t,E_y)$: Check if a given set of edges $E_y\subset E$ sharing a common endpoint $y$ belong to some $(s,t)$-mincut.\\

Using Fact \ref{fact:(x,y)-lies-in-(s,t)-mincut}, any fault-tolerant query can be answered using old value of $(s,t)$-mincut and edge-containment query on $(s,t)$-mincut for $E_y = \{(x,y)\}$. Fact \ref{fact:(x,y)-insertion} helps us in answering the query of $(s,t)$-mincut upon the insertion of any edge.

%  Upon failure of edge $(x,y)$, the value of $(s,t)$-mincut for any $s,t\in V$ either decreases by unity or remains unchanged. We now state the necessary and sufficient condition for the value of $(s,t)$-mincut to decrease upon failure of edge $(x,y)$.
 
 \begin{fact}
\label{fact:(x,y)-lies-in-(s,t)-mincut}
The value of $(s,t)$-mincut decreases (by unity) on failure of an edge $(x,y)$ if and only if $(x,y)$ lies in \textit{at least one} $(s,t)$-mincut.
\end{fact}

 \begin{fact}
\label{fact:(x,y)-insertion}
The value of $(s,t)$-mincut increases (by unity) on insertion of an edge $(x,y)$ if and only if $x$ and $y$ are separated in \textit{all} $(s,t)$-mincuts.
\end{fact}
% Using a data structure for this problem, we can answer a fault-tolerant query upon failure of edge $(x,y)$ as follows. We perform the corresponding edge-containment query for $E_y = \{(x,y)\}$. The value of $(s,t)$-mincut will reduce by unity if the edge-containment query evaluates to true. We can keep a Gomory-Hu tree of ${\cal O}(n)$ size as an auxiliary data structure to lookup the old value of $c_{s,t}$. Our first data structure can answer edge-containment queries for $|E_y|=1$ in  ${\cal O}(1)$ time. On the other hand, our second data structure works for any given set $E_y$ but takes ${\cal O}(\min(m,nc_{s,t}))$ time. 

% \begin{remark}
% It is worthwhile to note that edge-containment queries cannot be extended to handle more than one edge failure. This is because Fact \ref{fact:(x,y)-lies-in-(s,t)-mincut} doesn't hold for multiple edges failures.
% \end{remark}

\paragraph{Related Work:} A related problem is that of maintaining mincuts in a dynamic environment. Until recently, most of the work on this problem has been limited to global mincuts. Thorup \cite{DBLP:journals/combinatorica/Thorup07} gave a Monte-Carlo algorithm for maintaining a global mincut of polylogarithmic size with ${\tilde {\cal O}}(\sqrt{n})$ update time.  He also showed how to maintain a global mincut of arbitrary size with $1+o(1)$-approximation within the same time-bound. Goranci, Henzinger and Thorup \cite{DBLP:journals/talg/GoranciHT18} gave a deterministic incremental algorithm for maintaining a global mincut with amortized ${\tilde{\cal O}}(1)$ update time and ${\cal O}(1)$ query time. Hartmann and Wagner \cite{DBLP:conf/isaac/HartmannW12} designed a fully dynamic algorithm for maintaining all-pairs mincuts  which provided significant speedup in many real-world graphs, however, its worst-case asymptotic time complexity is not better than the best static algorithm for an all-pairs mincut tree. Recently, there is a fully-dynamic algorithm \cite{DBLP:journals/corr/abs-2005-02368} that approximates all-pairs mincuts up to a nearly logarithmic factor in ${\tilde{\cal O}}(n^{2/3} )$ amortized time against an oblivious adversary, and ${\tilde{\cal O}}(m^{3/4} )$ time against an adaptive adversary. To the best of our knowledge, there exists no non-trivial dynamic algorithm for all-pairs \textit{exact} mincut. We feel that the insights developed in our paper may be helpful for this problem.

\noindent

% {
%     \color{blue}
%     \begin{itemize}
%         \item First para remains intact.
%         \item Only two components can be effectively used to answer ft queries for Steiner set.
%         \item Draw analogy between flesh and dag. How topological ordering can be used to efficiently report ft-mincut in a dag. Similar idea for flesh.
%         \item Additional augmentation to projection mapping can help report mincut efficiently.
%     \end{itemize}

% }

\paragraph{Overview of our results:} Dinitz and Vainshtein \cite{DBLP:conf/stoc/DinitzV94,DBLP:journals/siamcomp/DinitzV00}
presented a novel data structure called {\em connectivity carcass} that stores all Steiner mincuts for a given Steiner set $S\subseteq V$ in ${\cal O}(\min(m,nc_S))$ space, where $c_S$ is the value of the Steiner mincut. %We observe that this data structure can be used easily to design a fault-tolerant data structure for all-pairs mincuts that occupies $O(mn)$-space. 
Katz, Katz, Korman and Peleg \cite{DBLP:journals/siamcomp/KatzKKP04} presented a data structure of ${\cal O}(n)$ size for labeling scheme of all-pairs mincuts. This structure hierarchically partitions the vertices based on their connectivity in the form of a rooted tree. In this tree, each leaf node is a vertex in set $V$ and each internal node $\nu$ stores the Steiner mincut value of the set $S(\nu)$ of leaf nodes in the subtree rooted at $\nu$. %stores all-pairs mincut value.
% Apparently, both these data structures seem to be designed for a static graph. 
We observe that if each internal node $\nu$ of the hierarchy tree ${\cal T}_G$
is augmented with the connectivity carcass of $S(\nu)$, we get a data structure for the edge-containment query. This data structure occupies ${\cal O}(mn)$ space. An edge-containment query can be answered using the connectivity carcass at the Lowest Common Ancestor (LCA) of the given pair of vertices. However, this data structure will require ${\cal O}(m)$ time to report the mincut between the given pair of vertices upon failure of an edge. As mentioned in previous results, storing $m$ copies of Gomory-Hu tree trivially is a better data structure for this problem.

We focus our attention on a fixed $s,t$ pair. There exists a dag structure \cite{DBLP:journals/mp/PicardQ82,DBLP:journals/siamcomp/DinitzV00} that stores all $(s,t)$-mincuts. Exploiting the acyclic structure of the dag and the transversality of each $(s,t)$-mincut in this DAG, we design a data structure that occupies just ${\cal O}(n)$ space. Not only it can report the value of $(s,t)$-mincut after failure of an edge in ${\cal O}(1)$ time but also a resulting $(s,t)$-mincut in ${\cal O}(n)$ time. We can trivially store this data structure for all possible $(s,t)$ pairs to get a sensitivity oracle for all-pairs mincuts. However, the ${\cal O}(n^3)$ space taken by this data structure is worse than the trivial structure mentioned in previous results.

% {\color{red}
% Can we concisely convey what is topological ordering in this multi-terminal dag-like structure? Even vaguely. We are not storing multi-terminal dag-like structure.
% }

In their seminal work, Dinitz and Vainshtein \cite{DBLP:conf/stoc/DinitzV94,DBLP:journals/siamcomp/DinitzV00} also showed that there exists a multi-terminal dag-like structure that implicitly stores all Steiner mincuts. This structure is called \textit{flesh} and takes the majority of the space in the connectivity carcass. 
{ Now, the question arises whether we can exploit the ``acyclicity" of flesh and transversality of Steiner mincuts just like the way we did it for $(s,t)$-mincuts, to get a space efficient data structure for the problem. Indeed, we show that the concepts analogous to topological ordering can be extended to the flesh as well. Furthermore, we can bypass the requirement of storing the flesh by cleverly augmenting the remaining connectivity carcass. This data structure takes only ${\cal O}(n)$ space and can report a $(s,t)$-mincut upon failure (or insertion) of an edge if $s$ and $t$ are Steiner vertices separated by some Steiner mincut. The time taken to report the value of $(s,t)$-mincut and a resulting cut is ${\cal O}(1)$ and ${\cal O}(n)$ respectively.} 
% { Now, the question arises whether we can exploit the ``acyclicity" of flesh and transversality of Steiner mincuts just like the way we did it for $(s,t)$-mincuts, to get a space efficient data structure for the problem. Indeed, we show that the concepts analogous to topological ordering can be extended to the flesh as well. 
% We can bypass the requirement of storing the flesh by cleverly augmenting the remaining data structure to materialize these concepts.
% This data structure takes only ${\cal O}(n)$ space and can report the value of $(s,t)$-mincut  and a resulting such cut in ${\cal O}(1)$ and ${\cal O}(n)$ time respectively.
% Moreover, we are able to bypass the requirement of storing the flesh to materialize these concepts -- we cleverly augment the data structure at a node using ${\cal O}(n)$ space only. This data structure can report the value of $(s,t)$-mincut and a resulting such cut in ${\cal O}(1)$ and ${\cal O}(n)$ time respectively. }
By augmenting each internal node of the hierarchy tree ${\cal T}_G$ with this data structure, we get our ${\cal O}(n^2)$ space sensitivity oracle.

As we move down the hierarchy tree, the size of Steiner set associated with an internal node reduces. So, to make the data structure more compact, a possible approach is to associate a smaller graph $G_\nu$ for each internal node $\nu$ that is {\em small enough} to improve the overall space-bound, yet {\em large enough} to retain the internal connectivity of set $S(\nu)$. 
However, such a compact graph cannot directly answer the edge-containment query as it does not even contain the information about all edges in $G$. A possible way to overcome this challenge is to transform any edge-containment query in graph $G$ to an equivalent query in graph $G_{\nu}$. We show that not only such a transformation exists, but it can also be computed efficiently. We model the query transformation as a multi-step procedure. 
% The following result captures a single step of this procedure. 
We now provide the crucial insight that captures a single step of this procedure.

% The key observation that allows us to do query transformation is as follows. 
Given an undirected graph $G=(V,E)$ and a Steiner set $S\subseteq V$, let $S'\subset S$ be any maximal set with connectivity strictly greater than that of $S$. We can build a quotient graph $G_{S'}=(V_{S'},E_{S'})$ such that $S' \subset V_{S'}$ with the following property.

{\em
For any two vertices $s,t \in S'$ and any set of edges $E_y$ incident on vertex $y$ in $G$, there exists a set of edges $E_{y'}$ incident on a vertex $y'$ in $G_{S'}$ such that $E_y$ lies in a $(s,t)$-mincut in $G$ if and only if $E_{y'}$ lies in a $(s,t)$-mincut in $G_{S'}$. 
}

We build graph $G_\mu$ associated with each internal node $\mu$ of hierarchy tree ${\cal T}_G$ as follows. For the root node $r$, $G_r = G$. For any other internal node $\mu$, we use the above property to build the graph $G_{\mu}$ from $G_{\mu'}$, where $\mu'$ is the parent of $\mu$. 
Our ${\cal O}(m)$ space data-structure is the hierarchy tree ${\cal T}_G$ where each internal node $\mu$ is augmented with the connectivity carcass for $G_\mu$ and the Steiner set $S(\mu)$.
% A high-level description of our query algorithm is as follows. We traverse the path from the root node to the LCA of $s$ and $t$. We keep transforming the edge-containment query for each edge in this path. At the LCA of $s$ and $t$, we stop and perform the query using the connectivity carcass stored at this node. Following a rigorous analysis, we show that this data structure takes only ${\cal O}(m)$ space and can answer any edge-containment query in ${\cal O}(\min(m,nc_{s,t}))$ time.

\paragraph{Organization of the paper:}

In addition to the basic preliminaries, Section \ref{sec:prelimiaries} presents compact representation for various mincuts. Section \ref{sec:n^2-space-sensitivity-oracle} presents the ${\cal O}(n^2)$ space sensitivity oracle. Section \ref{sec:query-transformation} gives insights into $3$-vertex mincuts that form the foundation for transforming an edge-containment query in original graph to a compact graph. 
% Section \ref{sec:insights-nearest-mincuts} builds tools for handling edge-insertion in linear space data structure. 
We give the construction of the compact graph for query transformation in Section \ref{sec:compact-graph-section}. Using this compact graph as a building block, we present the linear space sensitivity data structure in Section \ref{sec:final-ds}. Section \ref{sec:distributed-sensitivity-oracle} describes the distributed data structure for all-pairs mincut sensitivity problem.

\section{Preliminaries} \label{sec:prelimiaries}

%\subsection{Notations and lemmas on mincuts}
Let $G=(V,E)$ be an undirected unweighted multigraph without self-loops. To contract (or compress) a set of vertices $U\subseteq V$ means to replace all vertices in $U$ by a single vertex $u$, delete all edges with both endpoints in $u$ and for every edge which has one endpoint in $U$, replace this endpoint by $u$. A graph obtained by performing a sequence of vertex contractions is called a {\em quotient} graph of $G$.

For any given $A,B\subset V$ such that $A\cap B=\emptyset$, we use $c(A,B)$ to denote the number of edges with one endpoint in $A$
and another in $B$. Overloading the notation, we shall use $c(A)$ for $c(A,\bar{A})$.

\begin{definition}[$(s,t)$-cut]
A subset of edges whose removal disconnects $t$ from $s$ is called a $(s,t)$-cut. An $(s,t)$-mincut is a $(s,t)$-cut of minimum cardinality and its value is denoted by $c_{s,t}$. 
\label{def:(u,v)-cut}
\end{definition}

\begin{definition}[set of vertices defining a cut]
A subset $A\subset V$ is said to define a ($s,t$)-cut if $s\in A$ and $t\notin A$. The corresponding cut is denoted by cut$(A,\bar{A})$ or more compactly cut$(A)$.  
\label{def:set-definiting-a-cut}
\end{definition}

Detailed Preliminaries for Section \ref{sec:query-transformation} and beyond can be found in Appendix \ref{appendix:extended-preliminaries}.

\subsection{Compact representation for all \texorpdfstring{$(s,t)$}{(s,t)}-mincuts}
Dinitz and Vainshtein \cite{DBLP:journals/siamcomp/DinitzV00} showed that there exists a quotient graph of $G$ that compactly stores all $(s,t)$-mincuts, called strip ${\cal D}_{s,t}$. The 2 node to which $s$ and $t$ are mapped in ${\cal D}_{s,t}$ are called the terminal nodes, denoted by ${\bf s}$ and ${\bf t}$ respectively. Every other node is called a non-terminal node. We now elaborate some interesting properties of the strip ${\cal D}_{s,t}$.
% by Dinitz and Vainshtein \cite{DBLP:journals/siamcomp/DinitzV00}

 Consider any non-terminal node $v$, and let $E_v$ be the set of edges incident on it in ${\cal D}_{s,t}$. There exists a unique partition, called {\em inherent partition}, of $E_v$ into 2 subsets of equal sizes. These subsets are called the 2 sides of the inherent partition of $E_v$. 
 %Dinitz and Vainshtein established the following very interesting property of this inherent partition.
If we traverse ${\cal D}_{s,t}$ such that upon visiting any non-terminal node using an edge from one side of its inherent partition, the edge that we traverse while leaving it belong to the other side of the inherent partition, then no node will be visited again. Such a path is called a {\em coherent} path in ${\cal D}_{s,t}$. Furthermore, if we begin traversal from a non-terminal node $u$ along one side of its inherent partition and keep following a coherent path we are bound to reach the terminal ${\bf s}$ or terminal ${\bf t}$. So the two sides of the inherent partitions can be called side-${\bf s}$ and side-${\bf t}$ respectively.
It is because of these properties
that the strip ${\cal D}_{s,t}$ can be viewed as an undirected analogue of a directed acyclic graph with a single source and a single sink. 

A cut in the strip ${\cal D}_{s,t}$ is said to be a \textit{transversal} if each coherent path in ${\cal D}_{s,t}$ intersects it at most once. The following lemma provides the key insight for representing all $(s,t)$-mincuts through the strip ${\cal D}_{s,t}$.
\begin{lemma}[\cite{DBLP:journals/siamcomp/DinitzV00}]
    $A\subset V$ defines a $(s,t)$-mincut if and only if $A$ is a transversal in ${\cal D}_{s,t}$.
    \label{lem:mincut-transversal}
\end{lemma}

The following lemma can be viewed as a corollary of Lemma \ref{lem:mincut-transversal}.

\begin{lemma}
A $(s,t)$-mincut contains an edge $(x,y)$ if and only it appears in strip ${\cal D}_{s,t}$.
\label{lem:E_y-edges-same-side}
\end{lemma}

% \begin{lemma} 
% If $A\subset V$ defines a $(s,t)$-mincut with $s\in A$, then $A$ can be merged with the terminal  node ${\mathbf s}$ in ${\cal D}_{s,t}$ to get the strip ${\cal D}_{A,t}$ that stores all those $(s,t)$-mincuts that enclose $A$.
% \label{lem:strip-A}
% \end{lemma}

% Another simple observation helps us describe the nearest mincuts in the strip.

% \begin{lemma}
% The mincuts defined by $\mathbf{s}$ and $\mathbf{t}$ are the nearest mincut from $s$ to $t$ and $t$ to $s$ respectively.
% \label{lem:nearest-mincut-strip}
% \end{lemma}

Consider any non-terminal node $x$. Let ${\cal R}_s(x)$ be the set of all the nodes $y$ in ${\cal D}_{s,t}$ that are reachable from $x$ through coherent paths that originate from the side-${\bf s}$ of the inherent partition of $x$ -- notice that all these paths will terminate at ${\bf s}$. 
It follows from the construction that ${\cal R}_s(x)$ defines a transversal in
${\cal D}_{s,t}$. We call ${\cal R}_s(x)$ the \textit{reachability cone} of $x$ towards $s$. It follows from the definition that each transversal in the strip ${\cal D}_{s,t}$ is defined by the union of reachability cones in the direction of ${\mathbf s}$ for a set of non-terminals.

%{\color{brown} JOURNAL:\\ Exploiting the acyclic structure of the strip ${\cal D}_{s,t}$ and the transversal structure of any $(s,t)$-mincut, we present an ${\cal O}(n)$ size fault tolerant data structure for $(s,t)$-mincut in Section \ref{subsec:fixed-s-t}. This data structure takes ${\cal O}(1)$ time to determine if the failure of any given edge in the graph reduces the value of the $(s,t)$-mincut. Moreover, the mincut after the failure of the edge can also be reported in ${\cal O}(n)$ time.}

% The $(s,t)$-mincut defined by ${\cal R}_s(x)$ is the nearest mincut from $\{s,x\}$ to $t$. 
% Interestingly, each transversal in ${\cal D}_{s,t}$, and hence each $(s,t)$-mincut, is a union of the reachability cones of a subset of nodes of ${\cal D}_{s,t}$ in the direction of $s$. We now state the following Lemma that we shall crucially use.

% \begin{lemma}[\cite{DBLP:journals/siamcomp/DinitzV00}]
% If $x_1,\ldots, x_k$ are any non-terminal nodes in strip ${\cal D}_{s,t}$,  the union of the reachability cones of $x_i$'s in the direction of ${\mathbf s}$ defines the nearest mincut between $\{s, x_1,\ldots, x_k\}$ and $t$.
% \label{lem:reachability-cones}
% \end{lemma} 
\subsection{Compact representation for all Global mincuts}

% Dinitz, Karzanov, and Lomonosov \cite{DL76} showed that there exists a cactus graph ${\cal H}_V$ of size $O(n)$ that compactly stores all global mincuts of $G$. A \textit{cactus graph} is a tree-like graph such that two simple cycles intersect at not more than one node. For details, see Appendix \ref{appendix:cactus}.

% Let $c_V$ denote the value of the global mincut of the graph $G$.
% Dinitz, Karzanov, and Lomonosov \cite{DL76} showed that there exists a graph ${\cal H}_V$ of size ${\cal O}(n)$ that compactly stores all global mincuts of $G$. 
% %In order to maintain the distinction between the two graphs,
% Henceforth, we shall use nodes and structural edges for vertices and edges of ${\cal H}_V$ respectively. There exists a projection mapping $\pi:V(G)\rightarrow V({\cal H}_V)$ assigning a vertex of graph $G$ to a node in graph ${\cal H}_V$. In this way, any cut $(A,{\bar A})$ in cactus ${\cal H}_V$ is associated to a cut $(\pi^{-1}(A),\pi^{-1}(\bar A))$ in the original graph $G$.
% The graph ${\cal H}_V$ has a nice tree-like structure with the following properties.

% {\color{blue}
Dinitz, Karzanov, and Lomonosov \cite{DL76} showed that there exists a  cactus graph ${\cal H}_V$ of size ${\cal O}(n)$ that compactly stores all global mincuts of $G$. 
In order to maintain the distinction between $G$ and ${\cal H}_V$,
henceforth, we shall use nodes and structural edges for vertices and edges of ${\cal H}_V$ respectively.  Each vertex in $G$ is mapped to a unique node in ${\cal H}_V$.
%There exists a projection mapping $\pi:V(G)\rightarrow V({\cal H}_V)$ assigning a vertex of graph $G$ to a node in graph ${\cal H}_V$. In this way, any cut $(A,{\bar A})$ in cactus ${\cal H}_V$ is associated to a cut $(\pi^{-1}(A),\pi^{-1}(\bar A))$ in the original graph $G$.
The graph ${\cal H}_V$ has a nice tree-like structure with the following properties.
% }
\begin{enumerate}
    \item Any two distinct simple cycles of ${\cal H}_V$ have at most one node in common. So each structural edge of ${\cal H}_V$ belongs to at most one simple cycle. As a result, each cut in ${\cal H}_V$ either corresponds to a tree edge or a pair of cycle edges in the same cycle.
    \item Let $c_V$ denote the value of the global mincut of the graph $G$. If a structural edge belongs to a simple cycle, it is called a \textit{cycle edge} and its weight is $\frac{c_V}{2}$. Otherwise, the structural edge is called a \textit{tree edge} and its weight is $c_V$.
    \item For any cut in the cactus ${\cal H}_V$, the associated cut in graph $G$ is a global mincut. Moreover, any global mincut in $G$ must have at least one cut in ${\cal H}_V$ associated with it.
\end{enumerate}

% Let $\nu$ and $\mu$ be any two nodes in the cactus ${\cal H}_V$. If they belong to the same cycle, say $c$, there are two paths between them on the cycle $c$ itself - their union forms the cycle itself. Using the fact that any two cycles in  ${\cal H}_V$ can have at most one common node, it can be seen that these are the only paths between $\nu$ and $\mu$. Using the same fact, if $\nu$ and $\mu$ are two arbitrary nodes in the cactus, there exists a unique path of cycles and tree edges between these two nodes. Any global mincut that separates $\nu$ from $\mu$ must correspond to a cut in this path.

% {\color{blue} 
Property $1$ implies that there exists a unique path of cycles and tree edges between any two arbitrary nodes $\nu$ and $\mu$. Any global mincut separating $\nu$ from $\mu$ corresponds to a cut in this path.

The cactus ${\cal H}_V$ can be stored in a tree-structure \cite{DBLP:journals/algorithmica/DinitzW98}, denoted by $T({\cal H}_V)$. The vertex set of $T({\cal H}_V)$ consists of all the cycles and the nodes of the cactus. For any node $\nu$ of the cactus ${\cal H}_V$, let $v(\nu)$ denote the corresponding vertex in $T({\cal H}_V)$. Likewise, for any cycle $\pi$ in the cactus, let $v(\pi)$ denote the corresponding vertex in $T({\cal H}_V)$. We now describe the edges of  $T({\cal H}_V)$. For a tree edge $(\nu_1,\nu_2) \in {\cal H}_V$ we add an edge between $v(\nu_1)$ and $v(\nu_2)$ as well.
Let $\nu$ be any node of ${\cal H}_V$ and let there be $j$ cycles - $\pi_1,\ldots,\pi_j$ that pass through it. We add an edge between $v(\nu)$ and $v(\pi_i)$ for each $1\le i\le j$. Moreover, for each vertex $\nu(\pi)$ in $T{({\cal H}_V)}$ we store all its neighbours in the order in which they appear in the cycle $\pi$ in ${\cal H}_V$ to ensure that cycle information is retained. This completes the description of $T({\cal H}_V)$. 

The fact that the graph structure $T({\cal H}_V)$ is a tree follows from the property that any two cycles in a cactus may have at most one vertex in common. It is a simple exercise to show that the size of $T({\cal H}_V)$ is of the order of the number of nodes of ${\cal H}_V$, which is 
${\cal O}(n)$. 

We root the tree $T({\cal H}_V)$ at any arbitrary vertex and augment it suitably so that it can answer any LCA query in $\mathcal O(1)$ time (using \cite{DBLP:journals/jal/BenderFPSS05}). Henceforth, we use \textit{skeleton tree} to refer to this structure.

Suppose $\nu_1$ and $\nu_2$ are two non-cycle nodes in the skeleton tree $T({\cal H}_V)$ and $P$ is the path between them. We can suitably compress $T({\cal H}_V)$ to get another tree structure such that all global mincuts separating $\nu_1$ and $\nu_2$ are retained. This tree structure is called the link associated to path $P$ and is denoted by $L(P)$ (see Appendix \ref{appendix:link-associated-to-a-path} for details) .

\paragraph{Extendability of Proper Paths}
A \textit{proper path} in a cactus refers to a path which contains at most one structural edge from a cycle it passes through. It is easy to observe that there is at most one proper path between a pair of nodes in the cactus. We describe a transitive relation between proper paths in a cactus graph called \textit{extendable in a direction}.

% {\color{red} extendable in direction $\nu_2$ looks a bit. Try to redefine only in terms of projection mapping paths.}
% \subsubsection{Extendable in a direction}
\begin{definition}[Extendable in a direction]
\label{def:extendable-in-a-direction}
% Suppose $u$ and $v$ are two stretched units projected to proper paths $P(\nu_1,\nu_2)$ and $P(\nu_3,\nu_4)$ respectively. $v$ is said to be extendable in direction $\nu_2$ of $u$ if proper paths $P(\nu_1,\nu_2)$ and $P(\nu_3,\nu_4)$ are extendable to a proper path $P(\nu,\nu')$ with $P(\nu_1,\nu_2)$ as the initial part and $P(\nu_3,\nu_4)$ as the final part.
Consider two proper paths $P_1 = P(\nu_1,\nu_2)$ and $P_2 = P(\nu_3,\nu_4)$ in a cactus graph. $P_2$ is said to be extendable from $P_1$ in direction $\nu_2$ if proper paths $P_1$ and $P_2$ are extendable to a proper path $P(\nu,\nu')$ with $P_1$ as its prefix and $P_2$ as its suffix.
\label{def:extendable}
\end{definition}

% It follows from Theorem \ref{lem:path-extendable} that if the stretched unit $v$ is reachable from $u$ in the direction $\nu_2$ through a coherent path, then $v$ is extendable in direction $\nu_2$ from $u$.

\begin{lemma}
\label{lem:skeleton-tree-queries}
Given two paths $P_1 = P(\nu_1,\nu_2)$ and $P_2 = P(\nu_3,\nu_4)$ in cactus ${\cal H}_V$, the following queries can be answered using skeleton tree ${T}({\cal H}_V)$ using ${\cal O}(1)$ LCA queries.\\
1. Determine if $P_1$ intersects $P_2$ at a tree edge or cycle, and report the intersection (if exists).\\
2. Determine if $P_2$ is extendable from $P_1$ in direction $\nu_2$ (given that $P_1,P_2$ are proper paths), and report the extended path (if exists).
\end{lemma}
% It is interesting to note that verifying if $P(\nu_3,\nu_4)$ is extendable from $P(\nu_1,\nu_2)$ in direction $\nu_2$ can be done in ${\cal O}(1)$ LCA queries on the skeleton tree.

\subsection{Compact representation for all Steiner mincuts} \label{subsec:connectivity-carcass}

% Let $G=(V,E)$ be an undirected unweighted graph and $S\subseteq V$ be a subset (Steiner set) of its vertices. 
Dinitz and Vainshtein \cite{DBLP:conf/stoc/DinitzV94} designed a data structure $\mathfrak{C}_S = ({\cal F}_S,{\cal H}_S, \pi_S)$ that stores all the Steiner mincuts (or $S$-mincuts) for a Steiner set $S\subseteq V$ in the graph. 
% We present a summary of this data structure.
% This data structure can be seen as a generalization of two already discussed data structures,
This data structure generalizes --
~(i) strip ${\cal D}_{s,t}$ storing all $(s,t)$-mincuts, and
~(ii) cactus graph ${\cal H}_V$ storing all global mincuts.

Two $S$-mincuts are said to be equivalent if they divide the Steiner set $S$ in the same way. The equivalence classes thus formed are known as the \textit{bunches}. Similarly, two vertices are said to be equivalent if they are not separated by any Steiner mincut. The equivalence classes thus formed are known as \textit{units}. A unit is called a \textit{Steiner unit} if it contains at least a Steiner vertex.

Let $(S_B,{\bar S_B})$ be the $2-$partition of Steiner set induced by a bunch $\cal B$. If we compress all vertices in $S_B$ to $s$ and all vertices in ${\bar S_B}$ to $t$, ${\cal D}_{s,t}$ will store all cuts in ${\cal B}$. We shall denote this strip by ${\cal D}_{\cal B}$. Any such strip has the following property -- if two non-terminals of two strips intersect at even one vertex then these nodes along with the inherent partitions coincide.

The first component of ${\mathfrak C}_S$, \textit{flesh graph} ${\cal F}_S$, is a generalization of the strip. It is a quotient graph of $G$. The vertices of ${\cal F}_S$, denoted by {\em units}, can be obtained by contracting each unit of $G$ to a single vertex. In addition to it, each unit of ${\cal F}_S$ is assigned a $2-$partition known as the \textit{inherent partition} on the set of edges incident on it. Any unit that appears as a non-terminal in the strip corresponding to some bunch is called a \textit{stretched unit}. Otherwise, it is called a \textit{terminal unit}. 
% Another distinction between these two units follows from the two observations made on strip corresponding to a bunch mentioned above. 
The inherent partition assigned to a stretched unit consists of two sets of equal cardinality. On the other hand, inherent partition assigned to a terminal unit is a trivial partition (one of the set is empty). Note that all Steiner units are terminal units but the reverse is not true.
% The concept of reachability is slightly modified in ${\cal F}_S$. Whenever we say that a unit $u$ is reachable from unit $u'$, it means that there exists a coherent path between $u$ and $u'$. A \textit{coherent path} refers to a sequence of units and edges in flesh $(u_1,e_1,u_2,e_2,\ldots,u_k)$ such that any $e_i$ is incident on $u_{i-1}$ and $u_i$ and for any $u_i$ $e_{i-1}$ and $e_i$ lie in different side of the inherent partition. The structure of the flesh graph implies that it is not possible for a coherent path to start and finish at a single unit and hence, ${\cal F}_S$ is in a sense acyclic. A \textit{transversal} refers to a $2-$partition of units such that any coherent path intersects it at most once. It can be shown that each transversal in the flesh ${\cal F}_S$ corresponds to a Steiner mincut.
The concept of reachability in ${\cal F}_S$ is similar to the strip. A unit $u$ is said to be reachable from unit $u'$ if there exists a coherent path between $u$ and $u'$. The structure of ${\cal F}_S$ is such that a coherent path cannot start and finish at same unit and hence, ${\cal F}_S$ is in a sense acyclic. There is a one-to-one correspondence between transversals in ${\cal F}_S$ and $S$-mincuts in $G$.

The second component of ${\mathfrak C}_S$ is a cactus graph called {\em skeleton} ${\cal H}_S$. 
% To avoid confusion with the original graph, the vertices and edges of the skeleton will be referred to as nodes and structural edges respectively. 
% A structural edge in the skeleton is a tree-edge if it is not part of a cycle, otherwise, it is a cycle-edge. 
% If $c_S$ is the value of the Steiner mincut, then each tree-edge is assigned weight $c_S$ and each cycle-edge is assigned weight $\frac{c_S}{2}$. 
Each terminal unit of ${\cal F}_S$ is mapped to a node in the ${\cal H}_S$ by projection mapping ${\pi}_S$. A stretched unit on the other hand is mapped to a set of edges corresponding to a proper path in ${\cal H}_S$ by ${\pi}_S$. 
% A \textit{proper path} in the skeleton refers to an alternating sequence of nodes and structural edges $(\nu_1,\epsilon_1,\nu_2,...,\nu_k)$ such that $\epsilon_i$ is incident on $\nu_{i-1}$ and $\nu_i$ and it intersects each cycle of the skeleton at at most one structural edge. 
% A \textit{subbunch} is a subset of a bunch that can be represented by a strip. 
All the bunches can be stored in ${\cal H}_S$ in the form of subbunches (disjoint subsets of a bunch). Each cut in ${\cal H}_S$ corresponds to a subbunch. The strip ${\cal D}_{\cal B}$ corresponding to this subbunch $\cal B$ can be obtained as follows. Let the cut in the skeleton separates it into two subcactuses ${\cal H}_S(\cal B)$ and ${\bar {\cal H}_S(\cal B)}$. If $P(\nu_1,\nu_2)$ is the path in the skeleton to which a unit $u$ is mapped, it will be placed in ${\cal D}_{\cal B}$ as follows.\\

\noindent
1. If both $\nu_1$ and $\nu_2$ lie in ${\cal H}_S(\cal B)$ (or ${\bar {\cal H}_S(\cal B)}$) $u$ is contracted in source (or sink).\\
2. Otherwise, $u$ is kept as a non-terminal unit.\\

Following lemma conveys the relation between reachability of a stretched unit $u$ and $\pi_S(u)$.

\begin{lemma}[\cite{DBLP:conf/soda/DinitzV95}]
Let $u$ be a stretched unit and $u'$ be any arbitrary unit in the flesh ${\cal F}_S$ and $\pi_S(u) = P(\nu_1,\nu_2)$, $\pi_S(u') = P(\nu_3,\nu_4)$. If $u'$ is reachable from $u$ in direction $\nu_2$, then $P(\nu_3,\nu_4)$ is extendable from $P(\nu_1,\nu_2)$ in direction $\nu_2$. (see Definition \ref{def:extendable-in-a-direction})
\label{lem:path-extendable}
\end{lemma}

\begin{lemma}[\cite{DBLP:conf/stoc/DinitzV94}]
\label{lem:strip-from-carcass}
Let $s,t \in S$ such that $c_{s,t}=c_S$. Given the connectivity carcass ${\mathfrak C}_S$ storing all Steiner mincuts, the strip ${\cal D}_{s,t}$ can be constructed in time linear in the size of flesh graph.
\end{lemma}

% It is important to note that nearest $s$ to $t$ and $t$ to $s$ mincuts are easier to identify in the connectivity carcass. The following lemma conveys the fact.

% \begin{lemma}[\cite{DBLP:conf/stoc/DinitzV94}]
% \label{lem:u-nearest-s-t-mincut}
% Let $s,t \in S$ such that $c_{s,t}=c_S$. Determining if a unit $u$ lies in nearest $s$ to $t$ mincut (or vice-versa) can be done using skeleton ${\cal H}_S$ and projection mapping $\pi_S$ using ${\cal O}(1)$ LCA queries on skeleton tree.
% \end{lemma}

% The size of flesh ${\cal F}_S$ is ${\cal O}(\min(m,\tilde{n}c_S))$ where $\tilde{n}$ is the number of units in ${\cal F}_S$. The size taken by skeleton is linear in the number of Steiner units. Thus, overall space taken by the connectivity carcass is ${\cal O}(\min(m,\tilde{n}c_S))$.
The notion of projection mapping can be extended for an edge $(x,y) \in E$ as follows. If $x$ and $y$ belong to the same unit, then $P(x,y) = \varnothing$. If $x$ and $y$ belong to distinct terminal units mapped to nodes, say $\nu_1$ and $\nu_2$, in the skeleton ${\cal H}_S$, then $P(x,y) = P(\nu_1,\nu_2)$. If at least one of them belongs to a stretched unit, $P(x,y)$ is the extended path defined in Lemma \ref{lem:path-extendable}. Projection mapping of $(x,y)$ can be computed in ${\cal O}(1)$ time (using Lemma \ref{lem:skeleton-tree-queries}). The following lemma establishes a relation between projection mapping of an edge and the subbunches in which it appears.

\begin{lemma}[\cite{DBLP:conf/stoc/DinitzV94}]
Edge $(x,y)\in E$ appears in the strip corresponding to a subbunch if and only if one of the structural edge in the cut of ${\cal H}_S$ corresponding to this subbunch lies in $P(x,y)$.
\label{lem:edge-path-intersect-subbunch}
\end{lemma}

\subsection{Compact representation of all-pairs mincuts values} \label{subsec:all-pairs-mincuts-values}

We describe a hierarchical tree structure of Katz, Katz, Korman and Peleg \cite{DBLP:journals/siamcomp/KatzKKP04}
that was used for compact labeling scheme for all-pairs mincuts, denoted by ${\cal T}_G$. The key insight on which this tree is built is an equivalence relation defined for a Steiner set $S\subseteq V$ as follows.

\begin{definition}[Relation ${\cal R}_S$]
Any two vertices $u,v\in S$ are said to be related by ${\cal R}_S$, that is $(u,v)\in {\cal R}_S$, if
$c_{u,v}>c_S$, where $c_S$ is the value of a Steiner mincut of $S$.
\end{definition}

% \noindent
% The fact that ${\cal R}_S$ is an equivalence relation defined over $S$ can be easily derived using Lemma \ref{lem:triangle-inequality}.
% It can be observed that for any vertex $x\in S$, the equivalence class $[x]$ defined by ${\cal R}_S$ consists of all those vertices $y\in S$ such that the value of $(x,y)$-mincut is strictly greater than $c_S$. 

By using ${\cal R}_S$ for various carefully chosen instances of $S$, we can build the tree structure ${\cal T}_G$ in a top-down manner as follows. Each node $\nu$ of the tree will be associated with a Steiner set, denoted by $S(\nu)$, and the equivalence relation ${\cal R}_{S(\nu)}$. To begin with, for the root node $r$, we associate $S(r)=V$.
Using ${\cal  R}_{S(\nu)}$, we partition $S(\nu)$ into equivalence classes. For each equivalence class, we create a unique child node of $\nu$; the Steiner set associated with this child will be the corresponding equivalence class. We process the children of $\nu$ recursively along the same lines. We stop when the corresponding Steiner set is a single vertex. 

It follows from the construction described above that the tree ${\cal T}_G$ will have $n$ leaves - each corresponding to a vertex of $G$. The size of ${\cal T}_G$ will be ${\cal O}(n)$ since each internal node has at least 2 children. Notice that $S(\nu)$ is the set of vertices present at the leaf nodes of the subtree of ${\cal T}_G$ rooted at $\nu$. The following observation captures the relationship between a parent and child node in 
${\cal T}_G$.

\begin{observation}
\label{obs:maximal-subset-subtree}
Suppose $\nu \in {\cal T}_G$ and $\mu$ is its parent. $S(\nu)$ comprises of a maximal subset of vertices in $S(\mu)$ with connectivity strictly greater than that of $S(\mu)$.
\end{observation}

The following observation allows us to use ${\cal T}_G$ for looking up $(s,t)$-mincut values for any $s,t\in V$.

\begin{observation}
\label{obs:(s,t)-mincut-lca}
Suppose $s,t \in V$ are two vertices and $\mu$ is their LCA in ${\cal T}_G$ then $c_{s,t}=c_{S(\mu)}$.
\end{observation}

\section{\texorpdfstring{${\cal O}(n^2)$}{Quadratic} space sensitivity oracle for all-pairs mincuts} \label{sec:n^2-space-sensitivity-oracle}

% {
% \color{blue}

% \begin{itemize}
% \item Edge-containment query for fixed s,t.
% \item Mincut containing $(x,y)$ using $(s,t)$-mincut for fixed $s,t$.
% \begin{itemize}
%     \item $O(m)$ space $O(m)$ time.
%     \item Augmented topological numbers.
%     \item $O(n)$ space $O(n)$ time.
% \end{itemize}

% \item Edge-containment query for $s,t \in S$ and $c_{s,t}=c_S$.
% \item Generalize $O(n)$ space $O(1)$ time.
% \item Mincut containing $(x,y)$ using $(s,t)$-mincut for $s,t \in S$ and $c_{s,t} = c_S$.
% \begin{itemize}
%     \item Idea of augmentation.
%     \item Given a stretched unit u and a bunch. Report the set of all stretched units that precede them in some topological ordering.
%     \item $O(n)$ space $O(n)$ time.
% \end{itemize}
% \end{itemize}
% }

% {\color{red} Describe 3.1 and highlight how it shall be useful in all-pairs case.}

% \boxed{
% $\tau$, $P$, $\pi$
% }
In this section, we shall present data structures that can handle edge-containment query for -- $(i)$ fixed source and destination pair $s,t\in V$ (in Section \ref{subsec:fixed-s-t}) and $(ii)$ for $s,t\in S$ and $c_{s,t}=c_S$ (in Section \ref{subsec:edge-containment-ds-steiner}). In Section \ref{subsec:all-pairs-mincuts}, we augment the tree structure of Katz, Katz, Korman and Peleg \cite{DBLP:journals/siamcomp/KatzKKP04} to get a sensitivity oracle for all-pairs mincuts. 

\subsection{Edge-containment query for fixed \texorpdfstring{$s,t \in V$}{s,t in V}} \label{subsec:fixed-s-t}

Consider  any edge $(x,y)\in E$. It follows from the construction of strip  ${\cal D}_{s,t}$ that edge $(x,y)$ lies in a $(s,t)$-mincut if and only if $x$ and $y$ are mapped to different nodes in ${\cal D}_{s,t}$. This query can be answered in ${\cal O}(1)$ time if we store the mapping from $V$ to nodes of ${\cal D}_{s,t}$. This requires only ${\cal O}(n)$ space.

Reporting a $(s,t)$-mincut that contains edge $(x,y)$ requires more insights. Without loss of generality, assume that $(x,y)$ lies in side-$\mathbf t$ of $\mathbf x$. If $\mathbf x$ is the same as $\mathbf s$, the set of vertices mapped to node $\mathbf s$ define a $(s,t)$-mincut that contains $(x,y)$. Thus, assume that $\mathbf x$ is a non-terminal in the strip ${\cal D}_{s,t}$. Observe that the set of vertices mapped to the nodes in the reachability cone of $\mathbf x$ towards $\mathbf s$, ${\cal R}_s({\mathbf x})$, defines a $(s,t)$-mincut that contains edge $(x,y)$. Unfortunately, reporting this mincut requires ${\cal O}(m)$ time and ${\cal O}(m)$ space. However, exploiting the acyclic structure of the strip and the transversality of each $(s,t)$-mincut, we get an important insight stated in the following lemma. This lemma immediately leads to an ${\cal O}(n)$ size data structure that can report a $(s,t)$-mincut containing edge $(x,y)$ in ${\cal O}(n)$ time. (Proof in Appendix \ref{appendix:s-t-mincut-containing-x-y-topological-fixed}).
% that leads to an ${\cal O}(n)$ size data structure that can also report the $(s,t)$-mincut in ${\cal O}To achieve better space and query time, we report another $(s,t)$-mincut that contains $(x,y)$ and has a simpler structure. Suppose $\tau$ is a topological ordering on the node set of strip ${\cal D}_{s,t}$ with $\tau(\mathbf s) = 0$. We show that storing the node mapping of each vertex and topological number of each node $\tau$ of the strip ${\cal D}_{s,t}$ can be used to report a $(s,t)$-mincut efficiently. 
% Consider the set of nodes, $X = \{u \;|\; \tau(u) \leq \tau(\mathbf x)\}$, with topological numbers less than or equal to that of $\mathbf x$. The set of vertices mapped to nodes in $X$ defines one such $(s,t)$-mincut. 

\begin{lemma}
\label{lem:s-t-mincut-containing-x-y-topological-fixed}
Consider the strip ${\cal D}_{s,t}$ with ${\mathbf x}$ as a non-terminal and edge $(x,y)$ lying on side-${\mathbf t}$ of ${\mathbf x}$. Suppose $\tau$ is a topological order on the nodes in the strip. The set of vertices mapped to the nodes in set $X = \{u \;|\; \tau(u) \leq \tau(\mathbf x)\}$ defines a $(s,t)$-mincut that contains edge $(x,y)$.
\end{lemma}
% \begin{proof}
% Consider $u \in X$ to be a non-terminal in strip ${\cal D}_{s,t}$. We shall show that ${\cal R}_s(u) \setminus \mathbf{s} \subseteq X$, i.e. reachability cone of $u$ towards source ${\mathbf s}$ in the strip ${\cal D}_{s,t}$ avoiding $\mathbf s$ is a subset of $X$. Consider any non-terminal $v \in {\cal R}_s(u) \setminus \mathbf{s}$. Since $u$ is reachable from $v$ in direction $\mathbf t$, therefore $\tau(v) < \tau(u)$. Therefore, $v \in X$. Therefore, ${\mathbf s} \cup X$ defines a transversal in the strip ${\cal D}_{s,t}$ (from Lemma \ref{lem:mincut-transversal}) and thus defines a $(s,t)$-mincut. The fact that $(x,y)$ lies in this $(s,t)$-mincut follows from the fact that $\tau(y) > \tau(x)$ and thus, $y \not \in X$.
% \end{proof}

\subsection{Edge-containment query for Steiner set \texorpdfstring{$S$}{S}} \label{subsec:edge-containment-ds-steiner}

Suppose $S$ is a designated Steiner set and $s,t\in S$ are Steiner vertices separated by some Steiner mincut that is $c_{s,t}=c_S$. It follows from Lemma \ref{lem:s-t-mincut-containing-x-y-topological-fixed} that we can determine if an edge $(x,y)\in E$ belongs to some $(s,t)$-mincut using the strip ${\cal D}_{s,t}$. Though this strip can be built from the connectivity carcass
$\mathfrak{C}_S$, the time to construction it will be ${\cal O}(\min(m,nc_S))$. Interestingly, we show that only the skeleton and the projection mapping of $\mathfrak{C}_S$ are sufficient for answering this query in constant time. Moreover, the skeleton and the projection mapping occupy only ${\cal O}(n)$ space compared to the ${\cal O}(\min(m,nc_S))$ space occupied by $\mathfrak{C}_S$.

The data structure $\mathfrak{D}(S)$ for edge containment query consists of the following components.
\begin{enumerate}
\item The skeleton tree $T({\cal H}_S)$. 
\item The projection mapping ${\pi}_S$ of all units in flesh.
\end{enumerate}

% In the following lemma, we show that this data structure can be used to efficiently report the edge-containment query for a Steiner Set $S$. 

% We state the necessary and sufficient condition for an edge $(x,y)$ to lie in an $(s,t)$-mincut. Note that two paths are said to intersect in the skeleton if the unique path of cycle and tree edges in both the paths intersect at some cycle or tree edge.\\

% {\color{red} 
% 1. There is some ambiguity here : skeleton versus skeleton tree.\\ 2. Moreover our contribution which is the following lemma is not in black and white.\\
% 3. Is the proof needed to be placed here ? Nontriviality of the proof ?}

% The skeleton ${\cal H}_{S(\nu)}$ and the corresponding projection mapping $\pi_{S(\nu)}$ have the sufficient information to infer whether any edge $(x,y)\in E$ belongs to some $(s,t)$-mincut as stated by the following lemma. We say that two paths intersect if the unique path of cycle and tree edges in both the paths intersect at some cycle or tree edge.
\begin{lemma}
\label{lem:path-intersects-tree}
% Given an undirected unweighted multigraph $G=(V,E)$ on $n=|V|$ vertices and a designated steiner set $S\subseteq V$. There exists an ${\cal O}(n)$ size data structure that can report if an edge $(x,y)$ lies in some $(s,t)$-mincut in ${\cal O}(1)$ time for given $s,t \in S$, $c_{s,t}=c_S$ and $(x,y)\in E$.
Given an undirected unweighted multigraph $G=(V,E)$ on $n=|V|$ vertices and a designated steiner set $S\subseteq V$, $\mathfrak{D}(S)$ takes ${\cal O}(n)$ space and can report if an edge $(x,y)$ lies in some $(s,t)$-mincut in ${\cal O}(1)$ time for any given $s,t \in S$ separated by some Steiner mincut.
\end{lemma}
\begin{proof}
We shall first show that an edge $(x,y)\in E$ belongs to a $(s,t)$-mincut if and only if the proper path $P(x,y)$ intersects a path between the nodes containing $s$ and $t$ in $\mathcal H_{S}$. We say that two paths {\em intersect} if they intersect at a tree edge or cycle.
% {\color{blue} make it short.}

% Observe that an edge $(x,y)$ lies in a $(s,t)$-mincut if and only if it appears in the strip ${\cal D}_{s,t}$ (follows from Lemma \ref{lem:E_y-edges-same-side}). Infact, we can extend this notion for subbunch as well. The edge $(x,y)$ lies in some $(s,t)$-mincut if and only if it appears in the strip corresponding to some subbunch that separates $s$ from $t$.

% Consider each subbunch that separates $s$ from $t$.
Let $\nu_1$ and $\nu_2$ be the nodes in ${\cal H}_S$ containing $s$ and $t$ respectively. A cut in ${\cal H}_S$ corresponding to any tree-edge (or a pair of cycle edges in same cycle) in the path from $\nu_1$ to $\nu_2$ defines a subbunch separating $s$ from $t$. Moreover, it follows from the structure of the skeleton that no other cut in the skeleton corresponds to a subbunch separating $s$ from $t$. Suppose $(x,y)$ lies in some $(s,t)$-mincut. Thus, it must be in some subbunch ${\cal B}$ separating $s$ from $t$. ${\cal B}$ must correspond to a cut $\cal{C}$ in the path from $\nu_1$ to $\nu_2$ in skeleton ${\cal H}_S$. Also, $P(x,y)$ must contain (one of) the structural edge(s) defining $\cal{C}$ (from Lemma \ref{lem:edge-path-intersect-subbunch}). Thus, $P(x,y)$ intersects the path from $\nu_1$ to $\nu_2$ in skeleton ${\cal H}_S$.

Now, consider the other direction of this proof. Suppose $P(x,y)$ and the path from $\nu_1$ to $\nu_2$ intersect at some cycle (or tree edge) $c$. Let $e_1$ and $e_2$ be structural edges belonging to the cycle $c$ that are part of $P(x,y)$ and the path from $\nu_1$ to $\nu_2$ respectively (in the case of tree edge $e_1=e_2=c$). Consider the cut in the skeleton corresponding to structural edges $e_1$ and $e_2$. It follows from Lemma \ref{lem:edge-path-intersect-subbunch} that $(x,y)$ lies in the strip corresponding to this subbunch. Since this cut separates $\nu_1$ from $\nu_2$ in ${\cal H}_S$, therefore the subbunch separates $s$ from $t$.

We can check if paths $P(\nu_1,\nu_2)$ and $P(s,t)$ in the skeleton ${\cal H}_S$ intersect with ${\cal O}(1)$ LCA queries on skeleton tree ${\cal T}({\cal H}_S)$ (from Lemma \ref{lem:skeleton-tree-queries}). Thus, we can determine if an edge $(x,y)$ lies in an $(s,t)$-mincut in ${\cal O}(1)$ time. The data structure takes only ${\cal O}(n)$ space.
\end{proof}

% {\color{red} Why $D_{B}$ instead of $D_{s,t}$ helps?}

Reporting a $(s,t)$-mincut that contains edge $(x,y)$ again requires more insights. Assume that $P(s,t)$ and $P(\nu_1,\nu_2)$ intersect at some tree edge or cycle. Let $e$ be a tree or cycle-edge in proper path $P(\nu_1,\nu_2)$ that lies in intersection of these two paths. Suppose ${\cal B}$ is a subbunch corresponding to a cut in the skeleton ${\cal H}_S$ that contains $e$ and separates $s$ from $t$ and ${\cal D}_{\cal B}$ be the strip corresponding to this subbunch. Without loss in generality, assume that $\nu_1$ lies in the side of source $s$ in this strip (denoted by ${\mathbf s}$). Using Lemma \ref{lem:edge-path-intersect-subbunch}, it is evident that edge $(x,y)$ lies in strip ${\cal D}_{\cal B}$. Assume $\mathbf x$ is a stretched unit in this strip, otherwise the source $\mathbf s$ is the required $(s,t)$-mincut. Suppose $(x,y)$ lies in side-$\mathbf t$ of $\mathbf x$. The set of vertices mapped to ${\cal R}_s(\mathbf{x})$, i.e. reachability cone of $\mathbf x$ towards $\mathbf s$ in this strip, defines a $(s,t)$-mincut that contains edge $(x,y)$. However, reporting this mincut is a tedious task. We must have the flesh ${\cal F}_S$ to construct the strip ${\cal D}_{\cal B}$ and then report the set ${\cal R}_s(\mathbf{x})$. This would require ${\cal O}(m)$ space and ${\cal O}(m)$ time.

% {\color{red} TODO: Go through this and the following paragraph again.}

% It is important to observe that the difficulty we face here looks more challenging compared to the one highlighted in Section \ref{subsec:fixed-s-t}. To achieve better space and query time, we use similar ideas as used in Section \ref{subsec:fixed-s-t}. 

Using insights developed in Section \ref{subsec:fixed-s-t}, we strive to report another $(s,t)$-mincut that contains $(x,y)$ and has a simpler structure. In particular, we aim to report a set of units $Y = \{ u \;|\; \tau_{\cal B}(u)\leq \tau_{\cal B}(\mathbf{x})\}$ for some topological ordering $\tau_{\cal B}$ of nodes in strip ${\cal D}_{\cal B}$. Using Lemma \ref{lem:s-t-mincut-containing-x-y-topological-fixed}, we know that set of vertices mapped to nodes in set $Y$ defines a $(s,t)$-mincut that contains edge $(x,y)$. Unfortunately, we cannot store topological order of each stretched unit for each possible bunch in which it appear as a non-terminal. This is because doing so will require ${\cal O}(n.|S|^2)$ space. We show that we can augment ${\mathfrak D}(S)$ with an additional mapping $\tau$ that takes only ${\cal O}(n)$ space and can still efficiently report the set $Y$. $\tau$ maps each stretched unit in ${\mathfrak D}(S)$ to a number as follows.
For all stretched units mapped to path $P(\nu_1,\nu_2)$, $\tau$ assigns a topological order on these stretched units as they appear in the $(\nu_1,\nu_2)$-strip. This additional augmentation will help us determine all those units of $Y$ which are mapped to same proper path as $\mathbf{x}$. The challenge now is to identify the units of $Y$ that are not mapped to the same path as $\mathbf{x}$. We use the notion of extendability of proper paths (Definition \ref{def:extendable-in-a-direction}) to find such units.

% In order to make the ideas more simple, we describe a transitive relation between proper paths on skeleton called \textit{extendable in a direction}.

% {\color{red} extendable in direction $\nu_2$ looks a bit. Try to redefine only in terms of projection mapping paths.}
% \subsubsection{Extendable in a direction}
% \begin{definition}[Extendable in a direction]
% % Suppose $u$ and $v$ are two stretched units projected to proper paths $P(\nu_1,\nu_2)$ and $P(\nu_3,\nu_4)$ respectively. $v$ is said to be extendable in direction $\nu_2$ of $u$ if proper paths $P(\nu_1,\nu_2)$ and $P(\nu_3,\nu_4)$ are extendable to a proper path $P(\nu,\nu')$ with $P(\nu_1,\nu_2)$ as the initial part and $P(\nu_3,\nu_4)$ as the final part.
% Consider two proper paths $P_1 = P(\nu_1,\nu_2)$ and $P_2 = P(\nu_3,\nu_4)$. $P_2$ is said to be extendable from $P_1$ in direction $\nu_2$ if proper paths $P_1$ and $P_2$ are extendable to a proper path $P(\nu,\nu')$ with $P_1$ as the initial part and $P_2$ as the final part.
% \label{def:extendable}
% \end{definition}

% It follows from Theorem \ref{lem:path-extendable} that if the stretched unit $v$ is reachable from $u$ in the direction $\nu_2$ through a coherent path, then $v$ is extendable in direction $\nu_2$ from $u$.
% Moreover, verifying if $P(\nu_3,\nu_4)$ is extendable from $P(\nu_1,\nu_2)$ in direction $\nu_2$ can be done in ${\cal O}(1)$ LCA queries on the skeleton tree. {\color{red} Add reference.}

Suppose stretched unit $\mathbf x$ is mapped to path $P(\nu,\nu')$ ($\nu$ lies in source $\mathbf{s}$). Let $X$ be the set of stretched units appearing as non-terminals in strip ${\cal D}_{\cal B}$ for which one of the following holds true -- (i) the stretched unit (say $v$) is mapped to $P(\nu,\nu')$ and $\tau(v) \leq \tau(\mathbf x)$, or (ii) the stretched unit is not mapped to $P(\nu,\nu')$ but $\pi_S(v)$ is extendable from $P(\nu,\nu')$ in direction $\nu$. The following lemma shows that $\mathbf{s}\cup X$ defines a desired $(s,t)$-mincut.

\begin{lemma}
The vertices mapped to units in ${\mathbf s}\cup X$ define a $(s,t)$-mincut and contains $(x,y)$.
\end{lemma}
\begin{proof}
% {\color{red} Fix this proof.}
Consider $u \in X$ to be a non-terminal unit in ${\cal D}_{\cal B}$. We shall show that ${\cal R}_{\mathbf s}(u)\setminus \mathbf{s} \subseteq X$, i.e. reachability cone of $u$ towards source $\mathbf{s}$ in the strip ${\cal D}_{\cal B}$ avoiding $\mathbf s$ is a subset of $X$. Suppose $\pi_S(u)=P(\mu,\mu')$ where $\mu$ is in source $\mathbf{s}$. It follows from the construction that either $P(\mu,\mu') = P(\nu,\nu')$ or $P(\mu,\mu')$ is extendable in direction $\nu$ from $P(\nu,\nu')$. Consider any unit $v$ in ${\cal R}_{\mathbf s}(u)\setminus \mathbf{s}$.
Suppose $v$ is projected to $P(\nu,\nu')$. In this case, clearly $P(\mu,\mu') = P(\nu,\nu')$ (using Lemma \ref{lem:path-extendable}). Since, $v$ is reachable from $u$ in direction $\nu$, it follows that $\tau(v) < \tau(u) < \tau(\mathbf{x})$. Thus, $v \in X$. Now, suppose $v$ is not projected to $P(\nu,\nu')$. In this case, $\pi_S(v)$ is extendable from $P(\mu,\mu')$ in direction $\mu$ (from Lemma \ref{lem:path-extendable}). It follows from the transitivity of Definition \ref{def:extendable} that $\pi_S(v)$ is extendable from $P(\nu,\nu')$ in direction $\nu$. Thus, $v \in X$. Therefore, ${\mathbf s}\cup X$ defines a $(s,t)$-mincut (from Lemma \ref{lem:mincut-transversal}).

%{\color{blue} 
%Let $u \in X$ be any non-terminal unit in ${\cal D}_{\cal B}$. We shall show that ${\cal R}_{\mathbf s}(u)\setminus \mathbf{s} \subseteq X$, i.e. reachability cone of $u$ towards source $\mathbf{s}$ in the strip ${\cal D}_{\cal B}$ avoiding $\mathbf s$ is a subset of $X$. This would imply using Lemma \ref{lem:mincut-transversal} that ${\mathbf s}\cup X$ defines a $(s,t)$-mincut. Notice that ${\mathbf s}$ belongs to the same direction as $\nu$ from $\nu'$.
%
%
%It follows from the construction of $X$ that $u$ is either projected to $P(\nu,\nu')$ or $\pi_S(u)$ is extendable in direction $\nu$ from $P(\nu,\nu')$. 
%
%Let us consider the case when $u$ is projected to $\pi(\nu,\nu')$. 
%Noti
%Consider any unit $v$ in ${\cal R}_{\mathbf s}(u)\setminus \mathbf{s}$. Since, $v$ is reachable from $u$ in direction ${\mathbf s}$, it follows that $\tau(v) < \tau(u) < \tau(\mathbf{x})$. Thus, $v \in X$. 
%
%Let us consider the case when suppose $v$ is not projected to $P(\nu,\nu')$. Since $v$ is reachable from $u$ in the direction of $s$, $\pi_S(v)$ is extendable from $\pi_S(u)$ in the direction of $s$. 
%It follows from the construction of $X$ this case, $\pi_S(v)$ is extendable from $P(\nu,\nu')$ in direction $\nu$ (from Theorem \ref{lem:path-extendable}). It follows from the transitivity of Definition \ref{def:extendable} that $\pi_S(v)$ is extendable from $\pi_S(u)$ in direction $\nu$. Thus, $v \in X$.
%}

Consider edge $(x,y)$. If $y$ is in $\mathbf t$ then $y \not \in X$ from the construction. Thus, assume $y$ is a non-terminal unit in ${\cal D}_{\cal B}$. If $y$ is projected to path $P(\nu,\nu')$ then $\tau(y) > \tau(u)$. Thus, $y \not \in X$. Otherwise $\pi_S(y)$ is extendable from $P(\nu,\nu')$ in direction $\nu'$. It follows from Definition \ref{def:extendable} that $y \not \in X$. Thus, the cut defined by ${\mathbf s} \cup X$ contains edge $(x,y)$.

\end{proof}

This data structure $\mathfrak{D}(S)$ now occupies ${\cal O}(n)$ space and can report a $(s,t)$-mincut containing edge $(x,y)$ in ${\cal O}(n)$ time. Using this data structure, we build a sensitivity oracle for all-pairs mincuts.

\subsection{Edge-containment Query for all-pairs Mincuts}
\label{subsec:all-pairs-mincuts}

The hierarchical tree structure ${\cal T}_G$ \cite{DBLP:journals/siamcomp/KatzKKP04} can be suitably augmented to design a sensitivity oracle for all-pairs mincuts. We augment each internal node $\nu$ of the hierarchy tree ${\cal T}_G$ with ${\mathfrak D}(S(\nu))$. Determining if given edge $(x,y)$ lies in some $(s,t)$-mincut for given pair of vertices $s,t\in V$ can be done using Algorithm \ref{algo:quadratic-space-query} in constant time. 

\begin{algorithm}%[H]
    \caption{Single edge-containment queries in ${\cal O}(n^2)$ data structure}
    \label{algo:quadratic-space-query}
    \begin{algorithmic}[1] % The number tells where the line numbering should start
        \Procedure{Edge-Contained}{$s,t,x,y$}
            \State{${\mu}\gets$ LCA($\mathcal T_G,s,t$)}
            \State $\mathcal P_1 \gets P(\pi_{S(\mu)}(s),\pi_{S(\mu)}(t))$
            \State $\mathcal P_2 \gets P(x,y)$
            \If{${\cal P}_1$ and ${\cal P}_2$ intersect at a tree edge or cycle} \Comment{ Using Lemma \ref{lem:skeleton-tree-queries}} 
            \State \textbf{return} True
            \Else 
            \State \textbf{ return} False
            \EndIf
        \EndProcedure
    \end{algorithmic}
\end{algorithm}

We state the following theorem (for edge-insertion see Appendix \ref{appendix: edge-insertion-n^2-space-ds}).

\begin{theorem}
Given an undirected unweighted multigraph $G=(V,E)$ on $n=|V|$ vertices, there exists an
${\cal O}(n^2)$ size sensitivity oracle that can report the value of $(s,t)$-mincut for any $s,t \in V$ upon failure (or insertion) of an edge in ${\cal O}(1)$ time. Moreover, a $(s,t)$-mincut 
%incorporating 
after the failure (or insertion) 
% of an edge 
can be reported in ${\cal O}(n)$ time.
\label{thm:O(n^2)-size-data-structure}
\end{theorem}

\section{Insights into \texorpdfstring{$3$}{3}-vertex mincuts} \label{sec:query-transformation}
%\subsection{Query Transformation}
% \label{sec:query-transformation}

The main result of this section is the following theorem.
% The main result of this section is an important insight into the mincuts of three vertices captured in the following theorem. This insight will form the foundation for transforming an  edge-containment query in original graph to a more compact graph.

\begin{theorem}
Let $s,r,t\in V$ be any 3 vertices such that $c_{s,r}\ge c_{s,t}$. Let $A\subset V$ define a $(s,t)$-mincut with  $s,r\in A$ and $t\in \bar{A}$. For any subset $E_y$ of edges incident on any vertex $y\in \bar{A}$, there exists a subset $E_A$ of edges from the mincut defined by $A$ such that the following assertion holds.

There is a $(r,s)$-mincut containing $E_y$ if and only if there is a $(r,s)$-mincut containing $E_A$. 
\label{thm:query-transform}
\end{theorem}

\noindent
In order to prove the theorem stated above, we first prove the following lemma of independent interest.

\begin{lemma}[$3$-vertex Lemma]
Let $s,r,t \in V$ be any three vertices and $c_{r,s} \ge c_{s,t}$. Let $A\subset V$ define an $(s,t)$-mincut as well as
an $(r,t)$-mincut
with $r,s\in A$ and $t\notin A$. Let $B\subset V$ define a $(r,s)$-mincut with $r\in B$. Without loss of generality, assume $t \in {\bar A}\cap {\bar B}$, then the following assertions hold:
\begin{enumerate}
\item
$c(\bar{A}\cap B, A\cap \bar{B})=0$
\item  $A\cap B$ defines a $(r,s)$-mincut.
\item  ${\bar A}\cap {\bar B}$ defines a $(s,t)$-mincut as well as a $(r,t)$-mincut.
\end{enumerate}

For a better illustration refer to Figure \ref{fig:non-S-crossing} ($ii$).
% (Proof in Appendix \ref{appendix:3-vertex-lemma})
\label{lem:3-vertex-lemma}
\end{lemma}

% \section{Proof of Lemma \ref{lem:3-vertex-lemma} (3-vertex lemma)} \label{appendix:3-vertex-lemma}

\begin{proof}
Let $\alpha = c(\bar{A}\cap B, A\cap {B}),
~\beta = c(\bar{A} \cap B, A\cap \bar{B}),~\gamma=c(\bar{A} \cap B, \bar{A}\cap \bar{B})$.
Refer to Figure \ref{fig:non-S-crossing} ($i$).
% that illustrates these edges and the respective cuts.
\begin{figure}[H]
\centering
\includegraphics[width=0.75\textwidth]{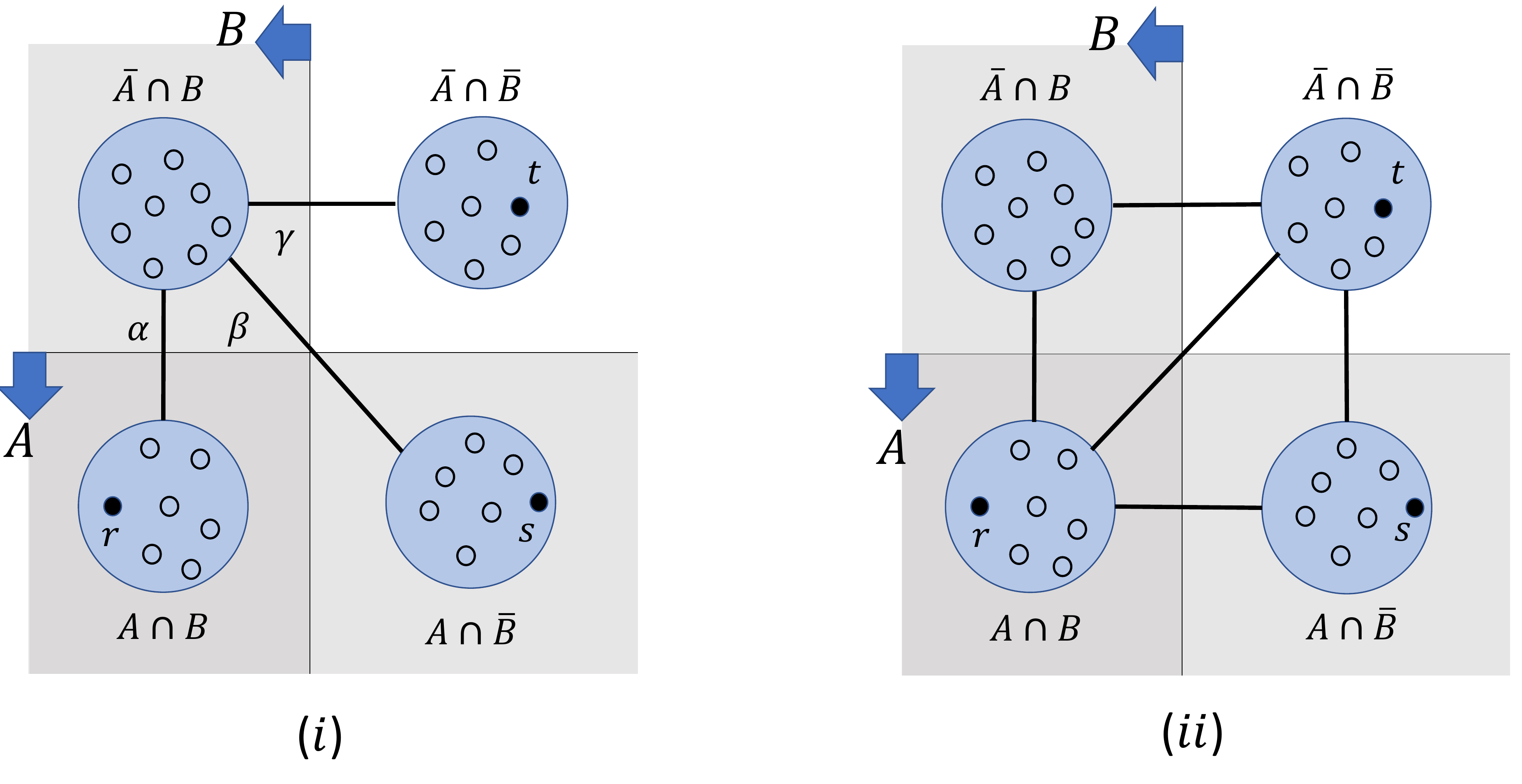}
    \caption{(i) $\alpha$, $\beta$ and $\gamma$ denote the capacities of edges incident on $\bar{A}\cap B$ from $A\cap B$, $A\cap \bar{B}$, and $\bar{A} \cap \bar{B}$ respectively. (ii) There are no edges along the diagonal between $\bar{A}\cap {B}$ and $A\cap \bar{B}$.}
\label{fig:non-S-crossing}
\end{figure}

Applying Lemma \ref{lem:subset-property-of-min-cut}
on $(s,t)$-mincut with $S=A$ and $S'= \bar{A}\cap B$, we get
\begin{equation}
    \alpha + \beta \le \gamma 
\label{eq:alpha+beta-le-gamma}
\end{equation}
Applying Lemma \ref{lem:subset-property-of-min-cut} on $(r,s)$-mincut with $S=\bar B$ and  $S'= \bar{A} \cap B$, we get $\gamma + \beta \le \alpha$. This inequality combined with Inequality \ref{eq:alpha+beta-le-gamma} implies that $\beta=0$. That is, $c(\bar{A}\cap B, A\cap \bar{B})=0$. This completes the proof of Assertion (1). Refer to Figure \ref{fig:non-S-crossing} (ii) for an illustration. 
It follows from (1) that $\alpha=\gamma$, i.e.,
$\bar{A} \cap B$ has equal number edges incident from $\bar{A}\cap \bar{B}$ as from $A\cap B$.  This fact can be easily used to infer Assertions (2) and (3) by 
removing $\bar{A}\cap B$ from $B$ and
$\bar{A}$ respectively.
\end{proof}

Let $s,r,t \in V$ be any three vertices such that $c_{s,r}\geq c_{s,t}$. Also suppose $A \subset V$ define a $(s,t)$-mincut with $s,r \in A$ and $t \in {\bar A}$. Let $E_y$ be a set of edges incident on vertex $y\in {\bar A}$.
Consider the strip ${\cal D}_{A,t}$ with source $A$ and sink $t$ (see Lemma \ref{lem:strip-A}). The source and sink nodes of ${\cal D}_{A,t}$ are denoted by ${\bf s}$ and ${\bf t}$ respectively. Let $B$ be any $(r,s)$-mincut with $r\in B$.
%containing edges $E_y$. Without loss of generality, assume that $r\in B$ and $s,t\notin B$. 
% The following lemma states a crucial property of the cut defined by $A\cup B$ in ${\cal D}_{A,t}$.
% % (Proof in Appendix \ref{appendix:AUB-contains-E_y}).
%
% \begin{lemma}
% $A\cup B$ will be a transversal in strip ${\cal D}_{A,t}$, and all edges in $E_y$ are present in this cut.
%
% \label{lem:AUB-contains-E_y}
% \end{lemma}
% \begin{proof}
% It follows from Lemma \ref{lem:3-vertex-lemma}(3) that $A\cup B$ will be a $(s,t)$-mincut. Hence $A\cup B$ will be a transversal in strip ${\cal D}_{A,t}$ that stores all
% $(s,t)$-mincuts.
% From definition, $y$ belongs to $\bar{A}$. Refer to Figure \ref{fig:non-S-crossing}($ii$).  If $y\in \bar{A}\cap B$, then it follows from Lemma \ref{lem:3-vertex-lemma}(1) that all neighbors of $y$ corresponding to $E_y$ will belong to $\bar{A}\cap \bar{B}$. So $E_y$ belongs to the cut defined by $A\cup B$. The same holds for the case $y\in \bar{A}\cap\bar{B}$ as well since $B\subset A\cup B$.
% \end{proof}
%
% {\color{red} Sir, there is some mismatch in $s,t$ and $r,s$ at some places}
%
It follows from Lemma \ref{lem:3-vertex-lemma}(3) that $A\cup B$ will be a $(s,t)$-mincut.
${\cal D}_{A,t}$ stores all $(s,t)$-mincuts that enclose the set $A$ (see Lemma \ref{lem:strip-A}), and thus the mincut defined by $A\cup B$ as well. So all nodes of the strip ${\cal D}_{A,t}$, excluding the terminal node ${\mathbf{s}}$ remain intact in the cut defined by $B$.
Therefore, if the subgraph of $G$ induced by $\bar{A}$ is replaced by the strip ${\cal D}_{A,t}$ excluding ${\mathbf{s}}$, the resulting graph, denoted by $G_A$, will preserve all $(r,s)$-mincuts of graph $G$. So, henceforth, we may focus on $G_A$ instead of $G$.

We now proceed towards proving Theorem \ref{thm:query-transform}.
Suppose the $(r,s)$-mincut defined by $B$ cuts edges $E_y$. Since $A\cup B$ is a $(s,t)$-mincut, it follows from the construction of $G_A$ and Lemma \ref{lem:E_y-edges-same-side} that all edges in $E_y$ must belong to the same side of the inherent partition of the node containing $y$ in strip ${\cal D}_{A,t}$. Otherwise, there is no $(r,s)$-mincut which contains set of edges $E_y$. In the latter case, we can choose $E_A = E(A,{\bar A})$ and Theorem \ref{thm:query-transform} trivially holds. So henceforth, we assume that $E_y$ belongs to
side-${\mathbf s}$ or side-${\mathbf t}$ of the node containing $y$ in the strip ${\cal D}_{A,t}$. Exploiting the strip structure of one side of the cut $(A,\bar{A})$ in the graph $G_A$, we now state and prove the following lemmas. These lemmas states an important relationship between a $(s,r)$-mincut and the reachability cone of 
any vertex $y\in \bar{A}$ in strip ${\cal D}_{A,t}$.

\begin{figure}[h]
    \centering
    \includegraphics[width=\textwidth]{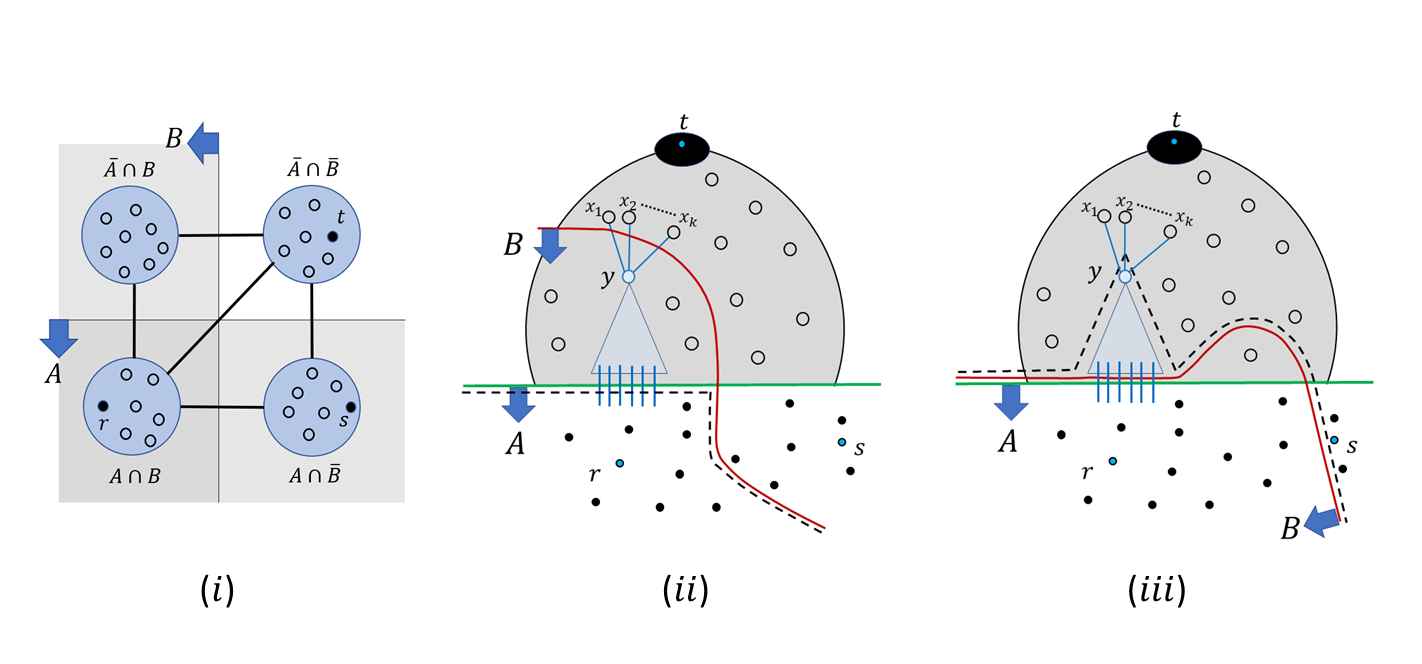}{}
    \caption{($i$) There are no edges along the diagonal between $\bar{A}\cap {B}$ and $A\cap \bar{B}$.~($ii$) $B$ cuts edges $\{(y,x_1),\ldots,(y,x_k)\}$.~$(iii)$ $B$ cuts all outgoing edges of $R = {\cal R}_s(y)\setminus{\bf s}$.}
    \label{fig:B-crosses-A}
\end{figure}

\begin{lemma}
Let $y$ be any vertex in $\bar{A}$ and let $B$ define a $(s,r)$-mincut with $s,t \in \bar{B}$. Moreover, $R = {\cal R}_s(y)\setminus \mathbf{s}$. The following holds true.
\begin{enumerate}
    \item[$(i)$] If $y\in \bar{A}\cap B$, there exists a $(s,r)$-mincut that contains all edges in $E(R,A)$
    \item[$(ii)$] If cut defined by $B$ contains all edges in $E(R,A)$ then $B\cup R$ also defines a $(s,r)$-mincut. 
\end{enumerate}
\label{lem:y-cone-in-sr-mincut}
\end{lemma}
%
%\begin{lemma}
%Let $y$ be any vertex in $\bar{A}\cap B$ and let $(B,\bar{B})$ define a $(s,r)$-mincut with $r,t\in \bar{B}$. There exists a $(s,r)$-mincut that cuts all edges between $R$ and $A$ where $R = {\cal R}_s(y)\backslash \{{\mathbf s}\}$.
%\label{lem:y-cone-in-sr-mincut}
%\end{lemma}
\begin{proof}
Refer to Figure \ref{fig:B-crosses-A}($ii$) that demonstrates $A$ and $B$ in the graph $G_A$.
Since $A\cup B$ is $(s,t)$-mincut, $A\cup B$ is a transversal in ${\cal D}_{A,t}$. So $R\subset A\cup B$. It follows from the construction that $R$ lies outside $A$, therefore, $R$ must lie inside $\bar{A}\cap B$ as well. Using this fact and Lemma \ref{lem:3-vertex-lemma} (1), all edges in $E(R,A)$ are incident only on the vertices from set $A\cap B$ only. But $A\cap B$ defines a $(r,s)$-mincut as follows from Lemma \ref{lem:3-vertex-lemma} (2). This completes the proof of part $(i)$.

%\end{proof}
%
%\begin{lemma}
%Let $y\in \bar{A}$ and let $(B,\bar{B})$ with $s\in \bar{B}$ define a $(s,r)$-mincut that contains all edges between $R$ and $A$, where $R={\cal R}_s(y)\backslash \{\bf s\}$. Then $B\cup R$ is also a $(s,r)$-mincut.
%\label{lem:BUR-is-a-(s,r)mincut}
%\end{lemma}
%\begin{proof}
Refer to Figure \ref{fig:B-crosses-A}($iii$). 
It follows from construction that $R\subset \bar{A}$. So
all edges from $E(R,A)$ are present in the cut defined by $A\cup B$. Since $A\cup B$ is a transversal in ${\cal D}_{A,t}$, it follows that $(A\cup B) \cap R = \emptyset$; otherwise it would imply a coherent path in strip ${\cal D}_{A,t}$ that intersects $A\cup B$ twice -- a contradiction. So $B \cap R = \emptyset$ too. That is, $R$ lies entirely on the side of $t$ in the cut defined by $B$. Treating $R$ as a single vertex, observe that the number of edges incident on $R$ from side-$A$ is same as that from side-$\mathbf{t}$ in ${\cal D}_{A,t}$. So it follows that $B\cup R$ is a $(s,r)$-mincut. This completes the proof for $(ii)$.
\end{proof}

We shall now prove Theorem \ref{thm:query-transform}. We first define $E_A$.
%Let $y\in \bar{A}$ be any vertex and $E_y$ be any subset of edges incident on $y$. 
Let $x_1,\ldots,x_k$ be the neighbors of $y$ defining $E_y$; that is, $E_y = \{(y,x_i)|\; 1 \leq i\leq k\}$. 
If all edges in $E_y$ lie in side-$\mathbf{s}$,
$R=\bigcup_{i=1}^{k}{\cal R}_s(x_i) \setminus \{\mathbf{s}\}$, that is, the union of the reachability cones of $x_i$'s in the strip ${\cal D}_{A,t}$ towards $\mathbf{s}$ excluding the terminal ${\mathbf{s}}$. If all edges in $E_y$ lie in side-$\mathbf{t}$,
$R={\cal R}_s(y) \setminus \{\mathbf{s}\}$.
We define $E_A$ to be the set of edges which are incident from $R$ to terminal $\mathbf{s}$ as well as those edges in $E_y$ having one endpoint in set $A$. 

% {\color{red} TODO: Add second side of proof.}

\begin{lemma}
\label{lem:query-transformation}
\noindent
There is a $(r,s)$-mincut containing edges $E_y$ if and only if there is a $(r,s)$-mincut containing all edges in set $E_A$. 
\end{lemma}

\begin{proof}
Let $B$ define a $(r,s)$-mincut containing edges $\{(y,x_1),\ldots,(y,x_k)\}$. 
Let us consider the case when $E_y$ is in side-$\mathbf{t}$ of the unit containing $y$ in ${\cal D}_{A,t}$. Observe that $y \in {\bar A}\cap B$ (see Fig \ref{fig:B-crosses-A} $(ii)$). So it follows from Lemma \ref{lem:y-cone-in-sr-mincut} ($i$) that the cut defined by $A\cap B$ is the desired $(r,s)$-cut that contains all edges in $E_A$.
Let us now consider the case when $E_y$ is in side-$\mathbf{s}$ of the unit containing $y$, that is, $y\in \bar{B}$. Consider an $x_i$ for $1\le i \le k$. Observe that $x_i \in B$. If $x_i \in {A \cap B}$, then $A \cap B$ cuts the edge $(y,x_i)$. If $x_i \in {\bar A}\cap B$, then clearly $A\cap B$ is a $(r,s)$-cut that cuts all edges from ${\cal R}_s(x_i)\setminus \{{\mathbf s}\}$ to $\mathbf s$ (using Lemma \ref{lem:y-cone-in-sr-mincut} $(i)$). Thus, $A\cap B$ is the desired $(r,s)$-cut that contains all edges in $E_A$.

% {\color{red} This proofs looks concise and complete proof for other side is very repetitive.

% Similarly, consider the case when $E_y$ is on side-$\mathbf{t}$, that is $y \in B$. Thus, it follows from Lemma \ref{lem:y-cone-in-sr-mincut} ($i$) that cut defined by $A\cap B$ is the desired $(r,s)$-mincut that contains all edges in $E_A$.
% }

Let $B$ define a $(r,s)$-mincut that contains all edges in set $E_A$. If $E_y$ lie in side-$\mathbf{s}$, it follows from Lemma \ref{lem:y-cone-in-sr-mincut} ($ii$) that $B\cup {\cal R}_s(x_i)$ define a $(r,s)$-mincut for each $x_i$. Taking union over all $x_i,1\le i\le k$, we get a 
 $(r,s)$-mincut. It can be observed that this cut contains edges $(y,x_1),\ldots,(y,x_k)$. Likewise, if $E_y$ lie in side-$\mathbf{t}$, $B\cup {\cal R}_s(y)$ define a $(r,s)$-mincut which contains all edges in $E_y$.
 
\end{proof}

Theorem \ref{thm:query-transform} follows from Lemma \ref{lem:query-transformation}. The following corollary follows from the construction in Proof of Lemma \ref{lem:query-transformation}.

\begin{corollary}
\label{cor:query-transformation}
Given a $(r,s)$-mincut that contains all edges in $E_A$ another $(r,s)$-mincut can be constructed that contains all edges in $E_y$ in time linear in the size of strip ${\cal D}_{A,t}$.
\end{corollary}

\subsection{Insights into Nearest Mincuts}
\label{sec:insights-nearest-mincuts}
Suppose $G'$ is the graph obtained by compressing the set ${\bar A}$ to a single vertex. To keep the ideas simple, we shall denote this compressed vertex by ${\bar A}$. Let 
%Using the notations from Section \ref{appendix:extended-preliminaries}, sets
$s^N_r$ define the $(s,r)$-mincut which is nearest from $s$. The following lemma states a necessary condition for $y \in s^N_r$ in $G$. (Proof in Appendix \ref{appendix:y-sNr-G})

% {\color{red} Gomory-Hu lemma move to extended preliminaries.}
\begin{lemma}
\label{lem:y-in-nearest-(r,s)-G_A}
If $y$ belongs to $s^N_r$ in $G$, then ${\bar A}$ lies in $s^N_r$ in $G'$.
\end{lemma}
% Another important observation can be derived from Lemma \ref{lem:3-vertex-lemma}. We can state the following lemma.

% \begin{lemma}
% ${\bar A}$ lies in nearest $r$ to $s$ mincut in $G_A$ if and only if $t \in r_s^N$.
% \end{lemma}

% Suppose $\mathbf s$ and $\mathbf t$ correspond to source $A$ and sink $t$ respectively of strip ${\cal D}_{A,t}$. Suppose $R={\cal R}_s(y)\setminus\{\mathbf{s}\}$ denote the set of vertices that form the reachability cone of $y$ towards source $\mathbf s$ in this strip (excluding $\mathbf s$). We define $E_A$ to be the set of edges which are incident from $R$ to terminal $\mathbf s$. We state the following lemma which concisely captures the condition for $y$ to lie in nearest $s$ to $r$ mincut.

% {\color{red} TODO: Define $R$ from focs submission}
% {\color{blue} I have added it but it still looks a lot of repetitive. Is there a better and compact way ? May be by defining $E_A$ in the statement of Lemma itself ? I am not sure...}
% Suppose $\mathbf s$ and $\mathbf t$ correspond to source $A$ and sink $t$ respectively of strip ${\cal D}_{A,t}$.
Suppose $R={\cal R}_s(y)\setminus\{\mathbf{s}\}$ denote the set of vertices that form the reachability cone of $y$ towards source $\mathbf s$ in this strip (excluding $\mathbf s$). We define $E_A$ to be the set of edges which are incident from $R$ to $A$, that is $E_A = E(R,A)$. We state the following lemma which concisely captures the condition for $y$ to lie in nearest $s$ to $r$ mincut.

\begin{lemma}

\label{lem:y-in-nearest-r-s-mincut}
Suppose $y\in \bar{A}$ such that ${\bar A}$ belongs to $s^N_r$ in $G'$. $y$ lies in the nearest $s$ to $r$ mincut in $G$ if and only there is no $(s,r)$-mincut that contains all edges in $E_A$.
\end{lemma}
\begin{proof}
 Suppose $s^N_r = {\bar B}$ in $G$. Since $\bar{A}$ belongs to $s^N_r$ in $G'$, $t$ must belong to $\bar{B}$. Otherwise, as follows from Lemma \ref{lem:3-vertex-lemma} ($2$), $A\cap {\bar B}$ will define a $(s,r)$-mincut which does not contain ${\bar A}$ -- a contradiction. 

Suppose $y$ does not belong to $s^N_r$, that is, $y \in {\bar A} \cap B$. In this case, it follows from Lemma \ref{lem:y-cone-in-sr-mincut}($i$) that $A\cap {\bar B}$ defines a $(s,r)$-mincut that contains all edges in $E_A$. %Thus, if $E_A$ does not lies in some $(s,r)$-mincut, $y$ lies in nearest $s$ to $r$ mincut.
%{\color{red} TODO: refer 4.4 in AUB transversal. refer fig 2 in this proof.}

Suppose $y$ belongs to $s^N_r$. If $y \in \mathbf{t}$,  there can not be any $(s,r)$-mincut that contains all edges in $E_A$ (note that $R = {\bar A}$ in this case). Thus, assume $y$ is a non-terminal in strip ${\cal D}_{A,t}$. Suppose there exists a $(s,r)$-mincut defined by set of vertices $B'$ such that it cuts all edges in $E_A$ and $s,t \in {\bar {B'}}$. 
It follows from Lemma \ref{lem:y-cone-in-sr-mincut} ($ii$) that $(\overline{B'\cup R}, B'\cup R)$ defines a $(s,r)$-mincut in which $y$ is not on the side of $s$ -- a contradiction. Therefore, if $y$ belongs to $s^N_r$, no $(s,r)$-mincut may contain all edges in $E_A$.
\end{proof}

The following lemma can be seen as a corollary of Lemma \ref{lem:query-transformation} and \ref{lem:y-in-nearest-r-s-mincut}.

\begin{corollary}
\label{cor:y-lies-in-nearest-r-s-mincut}
Let $A$ define a $(s,t)$-mincut in $G$ with $s,r \in A$ and $c_{s,r}\geq c_{s,t}$. Let $G'$ be the graph obtained by compressing ${\bar A}$ to a single vertex.
A vertex $y \in A$ lies in the nearest $s$ to $r$ mincut in $G$ if and only if the following conditions hold:
\begin{enumerate}
    \item ${\bar A}$ lies in nearest $s$ to $r$ mincut in $G'$.
    \item There is no $(s,r)$-mincut that contains all edges in side-$\mathbf{s}$ of $y$.
\end{enumerate}
\end{corollary}

\vspace{-6mm}
\section{A Compact Graph for Query Transformation}
\vspace{-2mm}
\label{sec:compact-graph-section}
Let $S\subseteq V$ be the Steiner set of vertices. Suppose $S'\subset S$ be any maximal subset of vertices with connectivity greater than that of $S$.  Observe that the entire set $S'$ will be mapped to a single node, say $\nu$, in the skeleton ${\cal H}_S$. In this section, we present the construction of a compact graph $G_{S'}$ such that any query \textsc{Edge-Contained}$(s,r,E_y)$ in graph $G$ can be efficiently transformed to a query \textsc{Edge-Contained}$(s,r,E_{y'})$ in graph $G_{S'}$ for any $s,r\in S'$.
% on which can answer \textsc{Edge-Contained}$(s,r,E_y)$ efficiently for any $s,r\in S'$ and any subset $E_y$ of edges emanating from $y\in V$.
\vspace{-2mm}
\subsection{Construction of Compact Graph \texorpdfstring{$G_{S'}$}{}}\label{sec:compact-graph}

The construction of $G_{S'}$ from the graph $G$, flesh ${\cal F}_S$ and skeleton ${\cal H}_S$ is a $2-$step process. In the first step, we contract the subcactuses neighbouring to node $\nu$ using the following procedure.
%We start with the original graph and then contract the subcactuses using the following procedure,

{\textsc{Contract-Subcactuses}}: For each tree-edge incident on $\nu$ (or cycle $c$ passing through $\nu$) in skeleton ${\cal H}_S$, remove the tree-edge (or the pair of edges from $c$ incident on $\nu$) to get 2 subcactuses. Compress all the terminal units of ${\cal F}_S$ that belong to the subcactus not containing $\nu$ into a single vertex. Moreover, compress all the stretched units with both endpoints within this subcactus into the same vertex.

In the quotient graph obtained after first step, each contracted subcactus defines a Steiner mincut. 
However, this graph may not necessarily be compact since there may be many stretched units that are not yet compressed. 
%Consider one such stretched unit $u$ and suppose
Let $u$ be any such unit and suppose
the path to which it is mapped in the skeleton is $P(\nu_1,\nu_2)$. If one of $\nu_1$ or $\nu_2$ is $\nu$, we can compress the stretched unit to the contracted vertex corresponding to the other endpoint. To handle the case when the subcactuses containing $\nu_1$ and $\nu_2$ are compressed to different vertices,
%The challenge is to tackle the case when subcactuses containing $\nu_1$ and $\nu_2$ are compressed to different vertices. 
% At first sight, it may seem appropriate to arbitrarily compress such stretched unit to one of these vertices. However, this approach won't guarantee that the contracted subcactus corresponds to a Steiner mincut. It can be explained as follows. Suppose $u_1,u_2$ and $u_3$ are three stretched units and there is a coherent path in ${\cal F}_S$ that passes through them in the order $\langle u_1,u_2,u_3 \rangle$. If we compress $u_2$ to one vertex and $u_1,u_3$ to another, the cut defined by either of these vertices will intersect the coherent path twice, and hence it will not be a Steiner mincut.
%To tackle this problem,
we define a total ordering on the set containing all tree-edges and cycles in the skeleton. The second step uses this ordering to compress the stretched units as follows.
%Thenceforth, we contract each of the remaining stretched units to one of the contracted subcactus using the following procedure to get $G_{S'}$,

{\textsc{Contract-Stretched-Units}}: A stretched unit mapped to path $P(\nu_1,\nu_2)$, where $\nu_1\neq \nu \neq \nu_2$, is compressed to the contracted subcactus corresponding to lesser ordered cycle or tree-edge in which endpoints lie. If one of $\nu_1$ or $\nu_2$ is $\nu$, we compress it to the contracted subcactus corresponding to the cycle or tree-edge where the other endpoint lies.

The following lemma (Proof in Appendix \ref{appendix:contracted-subcactus-mincut}) states the property of the resulting graph.
% $G_{S'}$ obtained after the $2-$step contraction procedure. 
% that will be required for answering the queries.

\begin{lemma}
Let $S'\subset S$ be a maximal subset of vertices such that $c_{S'}>c_S$ and $\nu$ be the node in ${\cal H}_S$ corresponding to $S'$. Let $G_{S'}$ be the graph obtained after %applying the above procedure.
$2$-step contraction procedure. 
\begin{enumerate}
    \item The set of vertices compressed to a contracted vertex defines a Steiner mincut for set $S$.
    \item The number of contracted vertices equals the number of cycles and tree edges incident on node $\nu$ in ${\cal H}_S$.
\end{enumerate}
\label{lem:contracted-subcactus-mincut}
\end{lemma}

\vspace{-8mm}
\subsection{Query transformation in \texorpdfstring{$G_{S'}$}{compact graph}}
% {\color{red} Motivation for this subsection is missing in the following start line.}

% The graph $G_{S'}$ does not even contain all edges of graph $G$ because a large portion of vertices is contracted. However, we can use the findings of Section \ref{sec:query-transformation} to effectively handle this problem. 
We begin with a lemma that was used by Gomory and Hu to build a tree storing all-pairs mincuts. 

\begin{lemma}[Gomory and Hu \cite{GH61}]
Let $A$ defines a $(s,t)$-mincut with $s\in A$. Let $r\in A$ be any vertex.
For any $(s,r)$-mincut, say defined by $B$, there exists a $(s,r)$-mincut that keeps $\bar A$ intact and still contains all edges in cut defined by $B$ that don't have both endpoints in $\bar A$.
\label{lem:GH}
\end{lemma}

Consider any two vertices $s,r\in S'$. Recall that $S'$ is mapped to $\nu$ in the skeleton ${\cal H}_S$. 
% Consider any cycle $c$ that passes through $\nu$ in the cactus. 
Let $A$ be the subset of vertices compressed to a contracted vertex in $G_{S'}$. Notice that all those $(s,r)$-mincuts in $G$ that keep $\bar{A}$ intact remain preserved in $G_{S'}$. Moreover, it follows from Lemma \ref{lem:contracted-subcactus-mincut} and Lemma \ref{lem:GH} that there is at least one such $(s,r)$-mincut. So it suffices to work with graph $G_{S'}$ if one wishes to calculate the value of $(s,r)$-mincut in $G$ or simply report a $(s,r)$-mincut in $G$
for any $s,r\in S$. Moreover, we can answer a query \textsc{Edge-Contained}$(s,r,E_y)$ using $G_{S'}$ if all edges in $E_y$ remain intact in graph $G_{S'}$. 
However, answering a query \textsc{Edge-Contained}$(s,r,E_{y})$ for any arbitrary $E_y$ using $G_{S'}$ is still challenging. This is because 
% we need to find at least one $(s,r)$-mincut in $G$ that contains all edges in $E_y$ but 
$G_{S'}$ may not even preserve all $(s,r)$-mincuts. In particular, all those $(s,r)$-mincuts that cut the set associated with a contracted vertex in $G_{S'}$ get lost during the transformation from $G$ to $G_{S'}$. We shall now establish a mapping from the set of all such lost $(s,r)$-mincuts to the set of $(s,r)$-mincuts that are present in $G_{S'}$. 

Let $y$ belong to $\bar{A}$. It follows from Lemma \ref{lem:contracted-subcactus-mincut} that the cut $(A,\bar{A})$ is a $(s,t)$-mincut for any $t\in S\cap \bar{A}$. Hence, $A,s,t,r$ satisfy all conditions of Theorem  \ref{thm:query-transform}. Now notice that entire $\bar{A}$
is compressed to a single vertex, say $y'$, in $G_{S'}$. Hence we can state the following Theorem.

\begin{theorem} \label{thm:edge-correspondence}
Given an undirected graph $G=(V,E)$, a subset $S\subseteq V$, let $S'\subset S$ be a maximal subset of vertices such that $c_{S'}>c_S$.
There exists a quotient graph $G_{S'}$ with the following property.~
For any two vertices $r,s \in S'$ and a set of edges $E_y$ incident on vertex $y$ in $G$, there exists a set of edges $E_{y'}$ incident on a vertex $y'$ in $G_{S'}$ such that $E_y$ lies in a $(r,s)$-mincut in $G$ if and only if $E_{y'}$ lies in a $(r,s)$-mincut in $G_{S'}$. 
\end{theorem}

We have already seen the construction of $G_{S'}$. In order to transform  \textsc{Edge-Contained}$(s,r,E_y)$ to \textsc{Edge-Contained}$(s,r,E_{y'})$  using Theorem \ref{thm:edge-correspondence}, we give an efficient algorithm for computing $E_{y'}$. Moreover, once we find a $(r,s)$-mincut in $G_{S'}$ that contains $E_{y'}$ we can efficiently compute a $(r,s)$-mincut in $G$ that contains all edges in $E_{y}$. Interestingly, we have algorithms that run in time linear in the size of flesh for both these tasks, stated in the following Lemma (Proof in Appendix \ref{appendix:linear-time-qt}). 

\begin{lemma}
\label{lem:linear-time-qt}
Set of edges $E_{y'}$ in Theorem \ref{thm:edge-correspondence} can be obtained from $E_y$ given flesh ${\cal F}_S$ and skeleton $\mathcal H_S$ in time linear in the size of flesh.
\end{lemma}

\begin{lemma}
\label{lem:mincut-qt}
Given a $(r,s)$-mincut in $G_{S'}$ that contains the all the edges in set $E_{y'}$, we can construct a $(r,s)$-mincut in $G$ that contains all edges in set $E_y$ in time linear in the size of flesh ${\cal F}_S$.
% (Proof follows from Corollary \ref{cor:query-transformation})
\end{lemma}

\subsection{Nearest mincut queries using \texorpdfstring{$G_{S'}$}{compact graph}}

Let us see how to answer the following query -- Given $y \in V$ check if $y$ lies in nearest $s$ to $t$ mincut where $s,t \in S$. If $s$ and $t$ are mapped to different node in the skeleton ${\cal H}_S$, the procedure is fairly straightforward -- construct the strip ${\cal D}_{s,t}$ and check if $y$ lies in source vertex $\mathbf{s}$. So, we shall restrict our attention only to the case when $s$ and $t$ are mapped to same node $\nu$ of the skeleton ${\cal H}_S$. In Algorithm \ref{algo:y-lies-in-nearest-r-s-mincut} we show how this query can be efficiently computed using a recursive call to this query in $G_{S'}$ and an \textsc{Edge-Contained} query. The correctness of algorithm follows from Lemma \ref{cor:y-lies-in-nearest-r-s-mincut}.

% Algorithm \ref{algo:y-lies-in-nearest-r-s-mincut} highlights an algorithm for same. 

\begin{algorithm}%[H]
    \caption{Check if $y$ lies in nearest $s$ to $t$ mincut in $G$}
    \label{algo:y-lies-in-nearest-r-s-mincut}
    \begin{algorithmic}[1] % The number tells where the line numbering should start
        \Procedure{Check-Nearest-Mincut}{$G,s,t,y$}
            \If{$s$ and $t$ are mapped to different nodes in ${\cal H}_S$}
            \IfThenElse{$y \in s_t^N$}{\textbf{return} True}{\textbf{return} False} \Comment{ ref Appendix \ref{appendix: edge-insertion-n^2-space-ds}}
            \EndIf
            \State $\nu \gets$  node to which $s$ and $t$ are mapped in ${\cal H}_S$
            \IIf{$y$ is mapped to $\nu$} \textbf{return} \textsc{Check-Nearest-Mincut}$(G_{S'},s,t,y)$
            \State $y' \gets$ vertex to which $y$ is compressed \Comment{ $y'$ corresponds to set ${\bar A}$}
            \If{\textsc{Check-Nearest-Mincut}$(G_{S'},s,t,y')$ is \textbf{False}}
            \State \textbf{return} \textbf{False}
            \EndIf
            \State Pick a vertex $x \in S \cap {\bar A}$
            \State Construct the strip ${\cal D}_{A,x}$
            \State $E_y \gets$ Edges in side-$\mathbf{s}$ of $y$ in the strip ${\cal D}_{A,x}$
            \State \textbf{return} $\neg$ \textsc{Edge-Contained}$(s,t,E_y)$
        \EndProcedure
    \end{algorithmic}
\end{algorithm}
\vspace{-3mm}
\section{Compact Data Structure for Sensitivity Query} \label{sec:final-ds}
\vspace{-2mm}

In the following section we present our ${\cal O}(m)$ space sensitivity data structure.

\subsection{An \texorpdfstring{${\cal O}(m)$}{linear} size data structure}

Consider any node $\mu$ in the hierarchy tree \cite{DBLP:journals/siamcomp/KatzKKP04} ${\cal T}_G$. We associate a compact graph with node $\mu$, say $G_\mu=(V_\mu,E_\mu)$ with the following properties. 

\begin{enumerate}
    \item $G_\mu$ is a quotient graph of $G$ with $S(\mu)\subseteq V_\mu$
    \item  For each $s,t\in S(\mu)$ and a set of edges $E_y$ incident on vertex $y\in V$, there exists a set of edges $E_{y'}$ incident on vertex $y'\in V_\mu$ such that $E_y$ lies in a $(s,t)$-mincut in $G$ if and only if $E_{y'}$ lies in a $(s,t)$-mincut in $G_\mu$.
\end{enumerate}

For the root node $r$, $G_r=G$ and the two properties hold trivially.
We traverse ${\cal T}_G$ in a top down fashion to construct $G_\mu$ for each node $\mu \in {\cal T}_G$ as follows. Let $\mu$ be the parent of $\mu'$ in ${\cal T}_G$. Assume we have graph $G_{\mu}$ already built with the properties mentioned above. Thus, $S(\mu)\subseteq V_\mu$. Using Observation \ref{obs:maximal-subset-subtree} we know that $S(\mu')$ is a maximal subset of $S(\mu)$ with connectivity strictly greater than that of $S(\mu)$. Using Theorem \ref{thm:edge-correspondence} with $S=S(\mu)$ and $S'=S(\mu')$, it can be shown that there exists a graph $G_{S(\mu')}$ 
% such that -- for each $s,t \in S(\nu)$ and set of edges $E_y$ incident on vertex $y\in V_{\mu}$ there exists a set of edges $E_{y'}$ incident on vertex $y'\in V_\nu$ such that $E_y$ lies in a $(s,t)$-mincut in $G_\mu$ if and only if $E_{y'}$ lies in a $(s,t)$-mincut in $G_\nu$. This also implies that the above $2$ properties will hold for $G_\nu$.
that satisfies both the properties above. We define $G_{\mu'}$ to be $G_{S(\mu')}$.
The graph $G_{\mu'}$ itself
can be obtained using the $2$-step contraction procedure described in Section \ref{sec:compact-graph}.

% We use Property 1 and Observation \ref{obs:maximal-subset-subtree} to build the graph $G_\nu$ for the subset $S(\nu)$ from $G_\mu$ following the procedure described in Section \ref{sec:compact-graph}. Using Theorem \ref{thm:edge-correspondence}, it follows that $G_\nu$ also satisfies Property 2.

Our compact data structure will be ${\cal T}_G$ where each node $\mu$ will be augmented with the connectivity carcass of $S(\mu)$ in graph $G_\mu$. 

\subsubsection*{Query algorithm for edge-containment query}

A query \textsc{Edge-Contained}$(s,t,E_y)$ can be answered by the data structure as follows. We start from the root node of ${\cal T}_G$ and traverse the path to the node $\nu$ which is the LCA of $s$ and $t$. Consider any edge $(\mu,\mu')$ on this path, where $\mu$ is parent of $\mu'$. We modify the query \textsc{Edge-Contained}$(s,t,E_y)$ in $G_{\mu}$ to an equivalent query \textsc{Edge-Contained}$(s,t,E_{y'})$ in $G_{\mu'}$ as we move to $\mu'$ (see Theorem \ref{thm:edge-correspondence}). This computation can be carried out in time linear in size of flesh at node $\mu$ using Lemma \ref{lem:linear-time-qt}. We stop when $\mu = \nu$. Observe that $c_{s,t} = c_{S(\nu)}$ (using Observation \ref{obs:(s,t)-mincut-lca}) and thus must be separated by some Steiner mincut for $S(\nu)$.
Thus we compute the strip ${\cal D}_{s,t}$ at node $\nu$ and answer the edge-containment query using Lemma \ref{lem:E_y-edges-same-side}. If the query evaluates to true, we compute a $(s,t)$-mincut in $G_\nu$ using ${\cal D}_{s,t}$ that contains the required set of edges at this level. We retrace the path from $\nu$ to the root of ${\cal T}_G$.
% A $(s,t)$-mincut containing the edges in $E_y$ can be obtained as follows. We start from the LCA of $s$ and $t$ in ${\cal T}_G$, say $\nu$. We construct the strip ${\cal D}_{s,t}$ and obtain a $(s,t)$-mincut that contains the required set of edges. We traverse the path from $\nu$ to the root of ${\cal T}_G$.
Consider any edge $(\mu,\mu')$ on this path where $\mu$ is parent of $\mu'$. We find a corresponding $(s,t)$-mincut in graph $G_\mu$ from the $(s,t)$-mincut of $G_{\mu'}$ using Lemma \ref{lem:mincut-qt}. We stop at the root node and report the set of vertices defining the $(s,t)$-mincut in $G$.

A \textsc{Ft-Mincut}$(s,t,(x,y))$ can be accomplished using the query \textsc{Edge-Contained}$(s,t,\{(x,y)\})$ and the old value of $(s,t)$-mincut (using Fact \ref{fact:(x,y)-lies-in-(s,t)-mincut}).

\subsubsection*{Query algorithm for nearest mincut queries}

% {\color{red} For the sake of continuity, won't it be better to bring Section 5.3 and Algorithm 2 here itself ?}

A query \textsc{Check-Nearest-Mincut}$(s,t,y)$ can be answered using Algorithm \ref{algo:y-lies-in-nearest-r-s-mincut} on the graph $G$ with $S=V$. A naive implementation of the algorithm will however be inefficient. This is because there can be ${\cal O}(c_{s,t})$ internal nodes from the root node to $LCA(s,t)$ in hierarchy tree ${\cal T}_G$. Thus, the algorithm may invoke ${\cal O}(n)$ instances of edge-containment query which will give an ${\cal O}(\min(mc_{s,t},nc_{s,t}^2))$ time algorithm for accomplishing this query. This query time is worse than the best known deterministic static algorithm to compute a $(s,t)$-mincut. 

We crucially exploit the following fact from Algorithm \ref{algo:y-lies-in-nearest-r-s-mincut}. If $\textsc{Edge-Contained}(s,t,E_y)$ (Line $14$, Algorithm \ref{algo:y-lies-in-nearest-r-s-mincut}) in $G_\nu$ evaluates to True at some internal node, then we can conclude that $y \not \in s_t^N$ in $G$. Thus, the query can possibly be true only if $\textsc{Edge-Contained}(s,t,E_y)$ in $G_\nu$ is False for all possible levels. Algorithm \ref{algo:check-nearest-mincut-efficient} gives an efficient implementation of the \textsc{Check-Nearest-Mincut} query. The following lemma gives proof of correctness for the algorithm.

\begin{lemma}
Algorithm \ref{algo:check-nearest-mincut-efficient} determines if $y$ lies in nearest $s$ to $t$ mincut in ${\cal O}(\min(m,nc_{s,t}))$ time.
\end{lemma}
\begin{proof}
The fact that the run-time of algorithm is ${\cal O}(\min(m,nc_S))$ follows from analysis for edge-containment query (see Appendix \ref{appendix:size-time-analysis-compact-ds}). Thus, we shall focus on the correctness of algorithm.

Consider the two nodes $\mu$ and $\nu$ in the path from root node to $\ell = LCA(s,t)$ such that $\mu$ is the parent of $\nu$. Consider the execution of \textsc{Check-Nearest-Mincut}$(\mu,s,t,y,E_y)$ at node $\mu$. 
% Observe that $E_y$ lies in a $(s,t)$-mincut in $G_\mu$ if and only if $E_{y'}$ lies in a $(s,t)$-mincut in $G_\nu$ (ref Line $14$, Lemma \ref{lem:query-transformation}).
Suppose $E_\mu'$ denotes one side of the partition of unit containing $y$ in strip ${\cal D}_{A,z}$ such that $E_y \cap E_\mu' \neq \varnothing$. Observe that if $E_y$ lies in a $(s,t)$-mincut then it must be entirely in one-side of the partition of unit containing $y$ in this strip. Note that both sides of the partition of unit containing $y$ can be determined just be constructing ${\cal D}_{A,z}$ strip -- independent of the query procedure.

% Consider the nodes $\mu$ and its parent $\lambda = parent(\mu)$ on Line $8$. Observe that $E_y$ lies in a $(s,t)$-mincut in $G_\lambda$ if and only if $E_{y'}$ lies in a $(s,t)$-mincut in $G_\mu$ (from Lemma \ref{lem:query-transformation}). Suppose $E_\mu'$ denotes one-side of the inherent partition of $y'$ in $G_\mu$ such that $E_{y'}\cap E_\mu' \neq \varnothing$. Observe that if $E_{y'}$ lies in a $(s,t)$-mincut then $E_{y'} \subseteq E_\mu'$ (from Lemma \ref{lem:AUB-contains-E_y} and Lemma \ref{lem:E_y-edges-same-side}). In other words, if $E_{y'}$ lie in a $(s,t)$-mincut then $E_{y'}$ must be in one-side of the inherent partition of $y'$. Note that both sides of inherent partition of $y'$ can be determined from $G_\mu$, independent of the query procedure.

% \begin{enumerate}
%     \item If $E_{y'} \subset E'$ and \textsc{Edge-Contained}$(G_\lambda,s,t,E_y)$ is False, then \textsc{Edge-Contained}$(G_\mu,s,t,E')$ is False.
%     \item If $E_{y'} \not\subset E'$ then \textsc{Edge-Contained}$(G_\lambda,s,t,E_y)$ is False (from Lemma \ref{lem:AUB-contains-E_y} and Lemma \ref{lem:E_y-edges-same-side}).
%     \item If \textsc{Edge-Contained}$(G_\lambda,s,t,E_y)$ is True, then $y \not\in s_t^N$ in G (from Lemma \ref{lem:y-in-nearest-r-s-mincut}).
% \end{enumerate}

% Suppose $\ell = LCA(s,t)$ is the node at which Line $3$-$4$ is being executed.
Suppose $y \not \in s_t^N$ in $G$. In this case, one of the following must happen -- ~$(i)$ $y \not \in s_t^N$ in $G_\ell$ or ~$(ii)$ \textsc{Edge-Contained}$(s,t,E'_\mu)$ evaluates to true for some node $\mu$ in the path from root to $\ell$. Clearly if $(i)$ is true, Algorithm \ref{algo:check-nearest-mincut-efficient} returns false (Line $3$). Thus, assume $(i)$ is false and $(ii)$ is true. 
% Suppose $\mu$ is one such node at which \textsc{Edge-Contained}$(s,t,E'_\mu)$ evaluates to true.
Consider the execution of procedure \textsc{Check-Nearest-Mincut} at node $\mu$. After Line $13$, it must be the case that $E_y \subseteq E_\mu'$. Thus, if $E_\mu'$ lies in some $(s,t)$-mincut then $E_y$ must also lie in some $(s,t)$-mincut. In other words, $\textsc{Edge-Contained}(s,t,E_{y})$ must also be true. Observe that in each subsequent recursive call of \textsc{Check-Nearest-Mincut} procedure, that is from $\nu$ (child of $\mu$) to $\ell$, the \textit{if} condition in Line $11$ will never be true.
This is because at every subsequent call, $E_y$ will always lie in same side of partition of unit containing $y$ in ${\cal D}_{A,z}$ as explained above.
Thus, the set of edges $E_y$ we receive in the final call (level $\ell$) of this procedure is simply the set of edges obtained from repeated query-transformation on $E_{y}$ (at node $\mu$). In other words, if $E_{y}$ (at node $\mu$) lie in a $(s,t)$-mincut in $G_\mu$ then $E_{y}$ (at level $\ell$) lie in a $(s,t)$-mincut in $G_\ell$. Thus, \textsc{Edge-Contained}$(s,t,E_y)$ (at Line $3$) evaluates to true. As a result, our algorithm will return false.

Suppose $y \in s_t^N$ in $G$. In this case, $y \in s_t^N$ in $G_\ell$ (from Lemma \ref{algo:y-lies-in-nearest-r-s-mincut}). Moreover, \textsc{Edge-Contained}$(G_\mu,s,t,E'_\mu)$ evaluates to false for all possible nodes $\mu$ from root to $parent(\ell)$. Thus, \textsc{Edge-Contained} query at level $\ell$ on Line 3 will evaluate to false. Thus, our algorithm will return true.

\end{proof}

\begin{algorithm}%[H]
    \caption{An efficient way to heck if $y$ lies in nearest $s$ to $t$ mincut}
    \label{algo:check-nearest-mincut-efficient}
    For initial call $\mu = r$ (root node in ${\cal T}_G$) and $E_y = $ all edges incident on $y$ in $G$. Note that $\mu$ and $E_y$ are auxiliary parameters in this procedure.
    \hrule
    \begin{algorithmic}[1] % The number tells where the line numbering should start
        \Procedure{Check-Nearest-Mincut}{$\mu,s,t,y,E_{y}$}
        \If{$\mu = LCA(s,t)$}
        \IIf{$y\not \in s_t^N$ in $G_\mu$ \textbf{or} \textsc{Edge-Contained}$(s,t, E_y)$} \textbf{return} \textbf{False}
        \State \textbf{return} \textbf{True}
        \EndIf
        \State $\nu \gets$ Node to which $s,t$ are mapped in skeleton ${\cal H}_{S(\mu)}$
        \IIf{$y$ is mapped to $\nu$} \textbf{return} \textsc{Check-Nearest-Mincut}$(\nu,s,t,y,E_y)$
        \State $y' \gets$ contracted vertex in $G_\nu$ to which $y$ is compressed
        \State ${\bar A} \gets$ set of vertices compressed to $y'$
        \State Pick $z \gets {\bar A}\cap {S(\mu)}$ and construct the strip ${\cal D}_{A,z}$
         \If{$E_y$ does not lie in one-side of partition of $y$ in ${\cal D}_{A,z}$}
        \State $E_{y} \gets$ one-side of partition of $y$ in ${\cal D}_{A,z}$
        \EndIf
        \State $E_{y'} \gets$ Query-Transformation on $E_y$ from $G_\mu \rightarrow G_\nu$
        % \If{$E_{y'}$ does not lie in one-side of inherent partition of $y'$}
        % \State $E_{y'} \gets$ one side of inherent partition of $y'$
        % \EndIf
        \State \textbf{return} \textsc{Check-Nearest-Mincut}$(\nu,s,t,y',E_{y'})$
        \EndProcedure
        
    \end{algorithmic}
\end{algorithm}

An \textsc{in-Mincut}$(s,t,(x,y))$ query can be reported using four instances of \textsc{Check-Nearest-Mincut} queries (from Lemma \ref{lem:edge-insertion-increases-mincut}) and the old value of $(s,t)$-mincut. For reporting a $(s,t)$-mincut upon insertion, please refer to Appendix \ref{appendix:edge-insertion-reporting-mincut}. We can thus state the following theorem.

\begin{theorem}
Given an undirected unweighted multigraph $G=(V,E)$ on $n=|V|$ vertices and $m=|E|$ edges, there exists a data structure of ${\cal O}(m)$ size that can report the value of $(s,t)$-mincut for any $s,t\in V$ upon failure (or insertion) of any edge. The time taken to answer this query is ${\cal O}(\min(m,nc_{s,t}))$.
\label{thm:O(m)-size-data-structure}
\end{theorem}

The detailed analysis of the query time of edge-containment query and space taken by the data structure stated in Theorem \ref{thm:O(m)-size-data-structure} is given in Appendix \ref{appendix:size-time-analysis-compact-ds}.

\section{Distributed Sensitivity Oracle}
\label{sec:distributed-sensitivity-oracle}
In this section, we shall restrict our attention to determining if failure/insertion of an edge changes the value of $(s,t)$-mincut. We shall show how the ${\cal O}(n^2)$ space sensitivity oracle in Section \ref{sec:n^2-space-sensitivity-oracle} can be distributed evenly among the vertices. The key insight we use while designing this distributed structure is as follows. To determine if failure (or insertion) of edge $(x,y)$ decreases (or increases) the value of $(s,t)$-mincut for any $s,t \in V$ we only require ~$(i)$ the hierarchy tree ${\cal T}_G$, ~$(ii)$ the skeleton stored at $LCA$ of $s$ and $t$, and ~$(iii)$ the projection mapping of units containing $x$ and $y$ in this skeleton. This insight follows from Algorithm \ref{algo:quadratic-space-query} and Appendix \ref{appendix: edge-insertion-n^2-space-ds}. Using this insight, we store a data structure ${\cal L}_v$ at each vertex $v \in V$ described as follows. \\

${\cal L}_v$ consists of the hierarchy tree ${\cal T}_G$ where each internal node $\nu$ stores --
\begin{enumerate}
    \item the skeleton tree $T({\cal H}_{S(\nu)})$ for the skeleton defined for the Steiner set $S(\nu)$.
    \item the projection mapping of unit corresponding to $v$ in skeleton ${\cal H}_{S(\nu)}$, i.e. $\pi_{S(\nu)}(v)$.
\end{enumerate}

% ~$(i)$ the skeleton tree $T({\cal H}_{S(\nu)})$ and ~$(ii)$ the projection mapping of unit corresponding to $v$ in skeleton ${\cal H}_{S(\nu)}$, i.e. $\pi_{S(\nu)}(v)$. 
Note that the projection mapping of Steiner units in $T({\cal H}_{S(\nu)})$ can be directly inferred from the children of node $\nu$ in hierarchy tree ${\cal T}_G$.

\subsection{Reporting the pairs of vertices whose mincut is affected by failure/insertion of an edge}

In this section, we shall show that ${\cal L}_x$ and ${\cal L}_y$ alone can be used to report all pairs of vertices whose mincut is changed upon failure (or insertion) of edge $(x,y)$.
% In this section, we shall show that our ${\cal O}(n^2)$ data structure can also efficiently report, for any given failed edge (likewise a new edge), all-pairs of vertices whose mincut decreases (likewise increases).
We start with the following lemma that states an important observation about Steiner mincuts. Its proof follows from the submodularity of cuts (Lemma \ref{lem:submodularity}).
%For a given Steiner set $S\subseteq V$, let ${\cal H}_S$ be the skeleton and ${\cal F}_S$ be the flesh  graph.  

\begin{lemma}
Let $S\subset V$ be any Steiner set, and let $x,y\in V$ be any two vertices belonging to the same unit, say $\mathbf{x}$, in ${\cal F}_S$. For any two vertices $u,v\in S$ such that $u,v\notin \mathbf{x}$, each $(u,v)$-mincut keeps $x$ and $y$ intact.
\label{lem:intact-nu-in-each-u,v-mincut}
\end{lemma}

%It is easy to observe that no Steiner mincut is affected by either the insertion of edge $(x,y)$ or the failure of the edge $(x,y)$, if exists. 
Let $\nu$ be the LCA of $x$ and $y$ in ${\cal T}_G$. It can be observed that for each ancestor $\mu$ of $\nu$ in ${\cal T}_G$, $x$ and $y$ belong to the same Steiner unit in the flesh graph associated with Steiner set $S(\mu)$.
So applying Lemma \ref{lem:intact-nu-in-each-u,v-mincut}, it can be inferred that if $(u,v)$ is a pair of vertices whose mincut value changes due to the failure/insertion of edge $(x,y)$, then $u$ as well as $v$ must appear as leaf nodes in subtree ${\cal T}_G(\nu)$ only.

The reader may refer to the conditions for the decrease (increase) in the value of the mincut between any pair of vertices upon failure (or insertion) of an edge. These conditions in conjunction with Lemma \ref{lem:intact-nu-in-each-u,v-mincut} lead to the following insight.
\begin{lemma}
 Let $S\subset V$ be any Steiner set. There exists a pair of vertices in $S$ belonging to different Steiner units in ${\cal F}_S$ whose mincut value changes upon the failure of the edge $(x,y)$, if exists, or the insertion of a new edge $(x,y)$ if and only if $x$ and $y$ belong to different units in ${\cal F}_S$.
\label{lem:processing-a-node}
\end{lemma}

We introduce a terminology here.
We say that a Steiner set $S$ remains unaffected by the failure of edge $(x,y)$ (likewise insertion of edge $(x,y)$) if there is no pair of vertices from $S$ whose mincut value changes due to the failure or insertion of edge $(x,y)$.
Let $\mu$ be $\nu$ or a descendant of $\nu$ in 
${\cal T}_G$. It follows from the construction of ${\cal T}_G$ that if two vertices belong to the same unit in the flesh graph associated with ${S(\mu)}$, then they continue to belong to the same unit in the flesh graph
of the Steiner set associated with each descendant of $\mu$ in ${\cal T}_G$. The following lemma can be viewed as a corollary of this observation and Lemma \ref{lem:processing-a-node}.
\begin{lemma}
Let $\mu$ be a descendant of $\nu$ in ${\cal T}_G$. If it is found that there does not exist any pair of vertices belonging to different nodes in ${\cal H}_{S(\mu)}$ whose mincut value changes upon failure of an existing edge $(x,y)$  (likewise insertion of a new edge $(x,y)$), then $S(\mu)$ remains unaffected by the failure or insertion of edge $(x,y)$.
\label{lem:contiguity}
\end{lemma}

Lemmas \ref{lem:intact-nu-in-each-u,v-mincut} and \ref{lem:contiguity} suggest the following insight for computing all-pairs of vertices whose mincut value changes due to the failure of edge $(x,y)$, if exists (likewise insertion of a new edge $(x,y)$). We begin with the processing of node $\nu$ in ${\cal T}_G$. While processing a node $\mu$ ($\mu$ is either $\nu$ or a descendant of $\nu$), if it is found that there are non-zero pair of vertices belonging to distinct nodes (in the corresponding skeleton stored at $\mu$) whose mincut value is changed then we report all such pairs and process children of $\mu$ recursively; Otherwise we there is no need to process any child of $\mu$. Thus the nodes that need to be processed form a truncated subtree of ${\cal T}_G(\nu)$. We shall now describe the processing of a node $\mu$ for handling the failure of an edge $(x,y)\in E$ (likewise insertion of a new edge $(x,y)$). 

\subsubsection*{Processing node $\mu$ for the insertion of edge $(x,y)$}
Consider any node $\mu$ which is $\nu$ or a descendant of $\nu$ in ${\cal T}_G$. In the skeleton tree $T({\cal H}_{S(\mu)})$, let $P$ be the unique path that joins $\pi(x)$ and $\pi(y)$ such that it has exactly one node in common with $\pi(x)$ and one node in common with $\pi(y)$. If $\pi(x)$ and $\pi(y)$ share one or more nodes, $P$ is defined as any singleton node common to $\pi(x)$ and $\pi(y)$. The endpoints of path $P$ can be computed in ${\cal O}(1)$ using a couple of LCA queries on the skeleton tree $T({\cal H}_{S(\mu)})$. 

%Based on Lemma \, 
We now state the necessary and sufficient condition for $(u,v)$-mincut to increase upon the insertion of edge 
$(x,y)$ for any $u,v\in S_\mu$.

\begin{lemma}
Let $u$ and $v$ be any two vertices from $S_\mu$ that belong to distinct nodes in ${\cal H}_{S(\mu)}$. The value of $(u,v)$-mincut increases upon insertion of edge $(x,y)$ if and only if the nodes containing $u$ and the node containing $v$ belong to path $P$.
\label{lem:P-and-x,y-insertion}
\end{lemma}
\begin{proof}
Let $P=\langle \mu_1, \ldots, \mu_k \rangle$ with $\mu_1$ in common with $\pi(x)$ and $\mu_k$ in common with $\pi(y)$.

% If $u$ and $v$ belong to the same node in skeleton ${\cal H}_{S(\mu)}$, then there exists a Steiner mincut that keeps $\omega$ on one side and, $x$ and $y$ on the other side. Applying Lemma \ref{lem:GH}, the $(u,v)$-mincut remains preserved if we compress the other side (containing $x$ and $y$) into a single node. So the $(u,v)$-mincut remains unchanged upon insertion of edge $(x,y)$. So henceforth, we assume that $u$ and $v$ belong to distinct nodes in skeleton ${\cal H}_{S(\mu)}$.

If the node containing $u$ (or $v$) does not belong to path $P$, there already exists a $(u,v)$-mincut that keeps $x$ and $y$ on the same side of this cut. So the $(u,v)$-mincut value remains unchanged by the insertion of $(x,y)$ in this case. 

 Let us consider the case when the node containing $u$ as well as the node containing $v$ lie on path $P$. Without loss of generality, assume $u$ and $v$ belong to $\mu_i$ and $\mu_j$ respectively with $1\le i<j\le k$. It follows from the construction of $P$ and Lemma \ref{lem:nearest-mincut-and-skeleton-tree} that $x\in u^N_v$ and $y\in v^N_u$. Therefore, as implied by Lemma \ref{lem:edge-insertion-increases-mincut}, $(u,v)$-mincut will increase upon insertion of edge $(x,y)$ in this case.
\end{proof}
% Case 2: $u,v$ belong to the same node in the skeleton 
% ${\cal H}_S$\\
% If the node containing $u$ and $v$, say $\nu$ does not belong to path $P$, then there already exists a Steiner mincut that keeps entire path $P$ on one side of the cut and $u,v$ on the other side. Using Lemma
% \ref{lem:y-in-nearest-(r,s)-G_A}, we can infer that 
% there exists a $(u,v)$-mincut that keeps $x$ and $y$ on one side of the cut only.

Let $\omega$ be any node in ${\cal H}_{S(\mu)}$. If $\omega$ does not belong to $P$, there exists a Steiner mincut that separates $\omega$ from both $x$ and $y$. So it follows from Lemma \ref{lem:GH} that for each pair of vertices $a,b\in S_\mu$ mapped to $\omega$, there exists a $(a,b)$-mincut that keeps $x$ and $y$ on one side of the cut. So $(a,b)$-mincut value will remain unchanged after the insertion of edge $(x,y)$. So instead of processing all children of $\mu$, we need to process only those children whose Steiner set corresponds to a node on path $P$. 
%%%%%%%%%%%%%%%%%%%%%%%%%%%%%%%%%%%%%%%
\subsubsection*{Processing node $\mu$ for the failure of edge $(x,y)$}
%For a given Steiner set $S\subseteq V$, let 
For the Steiner set $S(\mu)$, let $P$ be the path in skeleton ${\cal H}_{S(\mu)}$ to which edge $(x,y)$ is projected.
 Recall that the endpoints of path $P$ can be computed in ${\cal O}(1)$ time since it takes only a couple of LCA queries on the skeleton tree. In order to compute all-pairs of vertices whose mincut decreases upon the failure of edge $(x,y)$, we employ the link structure $L(P)$  (refer to Appendix \ref{appendix:link-associated-to-a-path} for the detailed description).
%  introduced in \cite{DBLP:journals/siamcomp/DinitzV00,StructureThesis}. 
 %We use the notion of link associated to path $P$, denoted by $L(P)$, to simplify our analysis. 
%  Link $L(P)$ is a compact tree structure that retains all Steiner mincuts which contain the edge $(x,y)$. It is obtained by appropriate compressions carried out on skeleton tree $T({\cal H}_S)$ along path $P$. (refer to Appendix \ref{appendix:link-associated-to-a-path} for the detailed description).
 %  Let $u,v$ be any two Steiner vertices belonging to distinct nodes in ${\cal H}_{S(\mu)}$. 
Using $L(P)$, we now state the  necessary and sufficient condition for $(u,v)$-mincut to decrease upon failure of edge $(x,y)$. Proof of the lemma follows from Lemma \ref{lem:link-stores-all-pairs} and Algorithm \ref{algo:quadratic-space-query}.

\begin{lemma}
For any $u,v\in S(\mu)$ belonging to distinct nodes in skeleton ${\cal H}_{S(\mu)}$, $(u,v)$-mincut decreases upon failure of edge $(x,y)\in E$
if and only if $u$ and $v$ belong to distinct nodes in link $L(P)$.
\label{lem:P-and-x,y-deletion}
\end{lemma}

% Let $(x,y)$ be the edge that is inserted and we have to report all those pairs of vertices in the graph whose mincut increases due to this insertion. We compute LCA($x,y$) in the hierarchy tree
% ${\cal T}$. Let this node be $\nu$. 

\subsubsection{Analysis of the algorithm}
 For each node $\mu$ of ${\cal T}_G$ that is processed, we spend ${\cal O}(1)$ time in its processing in addition to reporting the pairs of vertices from $S(\mu)$ belonging to distinct nodes whose mincut value is changed due to the failure/insertion of $(u,v)$. So in order to show ${\cal O}(k)$ bound,  we just need to account for the ${\cal O}(1)$ time that is spent in processing those nodes $\mu'$ of ${\cal T}_G$ that don't report any pair, that is,  $S(\mu')$ turns out to be unaffected by the failure/insertion of edge $(x,y)$. We can show that this time can be subsumed by the time spent in processing the parent of $\mu'$ in ${\cal T}_G$ as follows.
 
 For the case of insertion of edge, recall that the node $\mu'$ belonged to the path $P$ that was processed at the parent of $\mu'$ in ${\cal T}_G$. At least one pair reported during this processing can be uniquely assigned to $\mu'$. 
 Let us analyse the case of failure of edge.
 Suppose the processing of a node $\mu$ results in reporting non-zero pairs of vertices whose mincut is affected. 
 It follows from Lemma \ref{lem:P-and-x,y-deletion} that the number of children of $\mu$ in ${\cal T}_G$ is asymptotically upper bounded by the number of the pairs reported during the processing of $\mu$. 
 %Thus the cost of processing of those children $\mu$ of $\mu'$ such that $S(\mu)$ is unaffected is subsumed by the cost spent at the parent node. 
 Thus the time complexity of the algorithm is ${\cal O}(k)$. Now notice that the affected pairs of vertices at each node of ${\cal T}_G$ have a compact structure -- path $P$ in the case of insertion and the link $L(P)$ in the case of failure. So instead of reporting the affected pairs explicitly, we may as well output these compact structures only. The combined size of these structures is ${\cal O}(n)$ only. Moreover, only the projection mappings of units containing $x$ and $y$ are required in this algorithm. So we can state the following lemma.
 
 \begin{lemma}
 Given a failed (or inserted) edge $(x,y)$ 
 \label{thm:reporting-all-pairs-edge-insertion} all pairs of vertices whose mincut value changes upon the failure (or the insertion) can be compactly reported in ${\cal O}(min(n,k))$ time using ${\cal L}_x$ and ${\cal L}_y$, where $k$ is the total number of such pairs.
 \end{lemma}

% \begin{theorem}
% There exists an ${\cal O}(n^2)$ size data structure for an undirected graph $G=(V,E)$ that, for any given $x,y\in V$, takes ${\cal O}(min(n,k))$ time to report all pairs of vertices whose mincut value changes upon failure of edge $(x,y)$, if exists (likewise the insertion of edge $(x,y)$), where $k$ is the total number of pairs whose mincut increases.
% \label{thm:reporting-all-pairs-edge-insertion}
% \end{theorem}

We summarize the results of this section in the following theorem.

\begin{theorem}
\label{th:distributed-data-structure}
Given an undirected unweighted multigraph $G = (V,E)$ on $n = |V|$ vertices we can associate each vertex $v\in V$ with an ${\cal O}(n)$ space data structure ${\cal L}_v$ that supports the following operations.

\begin{enumerate}
    \item Given a failed/inserted edge $(x,y)$, we need only ${\cal L}_x$ and ${\cal L}_y$ to determine if the value of $(s,t)$-mincut has changed for any $s,t \in V$ in ${\cal O}(1)$ time.
    \item Given a failed/inserted edge $(x,y)$ we need only ${\cal L}_x$ and ${\cal L}_y$ to report a compact encoding of all pairs of vertices whose mincut is changed in ${\cal O}(\min(n,k))$ time where $k$ is the number of such pairs. 
\end{enumerate}
\end{theorem}

If the endpoints of the failed/inserted edges are confined to a subset $R\subseteq V$, we need to keep ${\cal L}_v$ only for $v\in R$. The following theorem can be seen as a corollary of Theorem \ref{th:distributed-data-structure}.

\begin{theorem}
 An undirected graph $G=(V,E)$ on $n=|V|$ vertices and any subset $R\subseteq V$ can be preprocessed to build an ${\cal O}(n|R|)$ 
 size data structure that takes ${\cal O}(1)$ time to report the value of $(s,t)$-mincut for any given $s,t\in V$ upon 
\begin{enumerate} 
\item 
 failure of any given edge $(x,y)\in (R\times R) \cap E$. ~~Or
\item insertion of any given edge $(x,y)\in R\times R$.
 \end{enumerate}
\label{thm:parametrized-sensitivity-oracle}
\end{theorem}
\section{Conclusion}
\label{sec:conclusion}
We have presented the first data structures for all-pairs mincut sensitivity (failure as well as insertion of an edge). We conclude with the following open problems. 
% {\color{blue} I removed the portion "an open problem is ..." from the following bullets since the were redundant in the light of the previous sentence.}

\begin{enumerate}
    \item Our first data structure achieves optimal query time but uses ${\cal O}(n^2)$ space. The second data structure, on the other hand, takes linear space but has ${\cal O}(\min(m,nc_{s,t}))$ query time. Does there exist a data structure that can achieve a space-time trade-off on these two extremes ?
    \item Can we have a compact data structure for all-pairs mincut sensitivity for $k$ edge failures/insertions where $k \geq 2$? For the related and much studied problem of fault-tolerant {\it exact} all-pairs distance/connectivity oracle for {\it directed} graphs, the best known result \cite{DBLP:conf/soda/DuanP09a} can only handle dual edge failures ($k=2$). 
\end{enumerate}

%{\color{blue} It is natural to ask if we can improve the query time and space simultaneously. In other words, the following problem is still open.}
% Therefore, the following problem still remains open. 

% \textit{For a sparse undirected graph, does there exist an $o(n^2)$ size data structure that can answer all-pairs mincut sensitivity queries in $o(m)$ time?} 
%%%%%%%%%%%%%%
% TODO

% 2. Go through new sections appendix F,I,J,K for consistency in terminology.
% 4. Algorithm 3 and its correctness proof.

%%%%%%%%%%%%
% \section{Conclusion and Future Work}

\pagebreak

\bibliography{refs}

% \pagebreak
% \section{TODO}
% \subfile{src/todo}

\pagebreak
\appendix

\section{Comparison of Query Time with Static Algorithms}
\label{appendix:non-trivial-query-time}
Goldberg and Rao \cite{DBLP:conf/focs/GoldbergR97a} gave a deterministic algorithm for computing $(s,t)$-mincut in an undirected unweighted graph that runs in ${\cal O}(n^{3/2}m^{1/2})$ time. Karger and Levine \cite{DBLP:conf/stoc/KargerL98} gave an algorithm that runs in ${\cal O}(nm^{2/3}c_{s,t}^{1/6})$ time for the same problem. 

We show that the query time of our data structure is ${\Omega}(\sqrt{n})$ times faster than both of them. The query time we achieve is ${\cal O}(\min(m,nc_{s,t}))$.
It is clear that $\frac{n^{3/2}m^{1/2}}{m}\geq \sqrt{n}$ and thus, the query time is $\Omega(\sqrt{n})$ times faster than Goldberg and Rao's algorithm \cite{DBLP:conf/focs/GoldbergR97a}. We can show the same for Karger and Levine's algorithm \cite{DBLP:conf/stoc/KargerL98} by separately considering the cases $m<nc_{s,t}$ and $m\geq nc_{s,t}$. Recently, Liu and Sidford \cite{DBLP:journals/corr/abs-2003-08929} gave a deterministic algorithm for computing $(s,t)$-mincut on unweighted graphs in ${\cal O}(m^{4/3 + o(1)})$ time.

\section{Link {structure associated} to a path in skeleton tree} \label{appendix:link-associated-to-a-path}
% {\color{red} Should we not cite Dinitz Vainshtein in addition to Belous for link structure ? Didn't they coin this term in SICOMP 2000 paper ?}
In this section we describe the link { structure} associated to a path in the skeleton tree { introduced in \cite{DBLP:journals/siamcomp/DinitzV00,StructureThesis}}. Suppose we are given a skeleton tree $T({\cal H}_S)$ and let $P$ be any path in this tree between $\nu_1$ and $\nu_2$. Recall that each node in the skeleton tree is either a \textit{cycle node} or a \textit{non-cycle node}. Cycle nodes correspond to cycles in skeleton ${\cal H}_S$ whereas non-cycle nodes correspond to nodes in skeleton ${\cal H}_S$. Note that $\nu_1$ and $\nu_2$ are non-cycle nodes.

The link $L(P)$ corresponding to path $P$ is constructed as follows. For each node $\nu$ in path $P$ we perform the following compression.
\begin{enumerate}
    \item If $\nu$ is a non-cycle node, we compress the subtree rooted at $\nu$, which is hanging from path $P$, into a single node. 
    \item If $\nu$ is a cycle node, we pick each of its neighbouring node $\nu'\not\in P$ and compress the subtree rooted at $\nu'$, which is hanging from edge $(\nu,\nu')$, into a single node.
\end{enumerate}

\begin{figure}
    \centering
    \includegraphics[width=\textwidth]{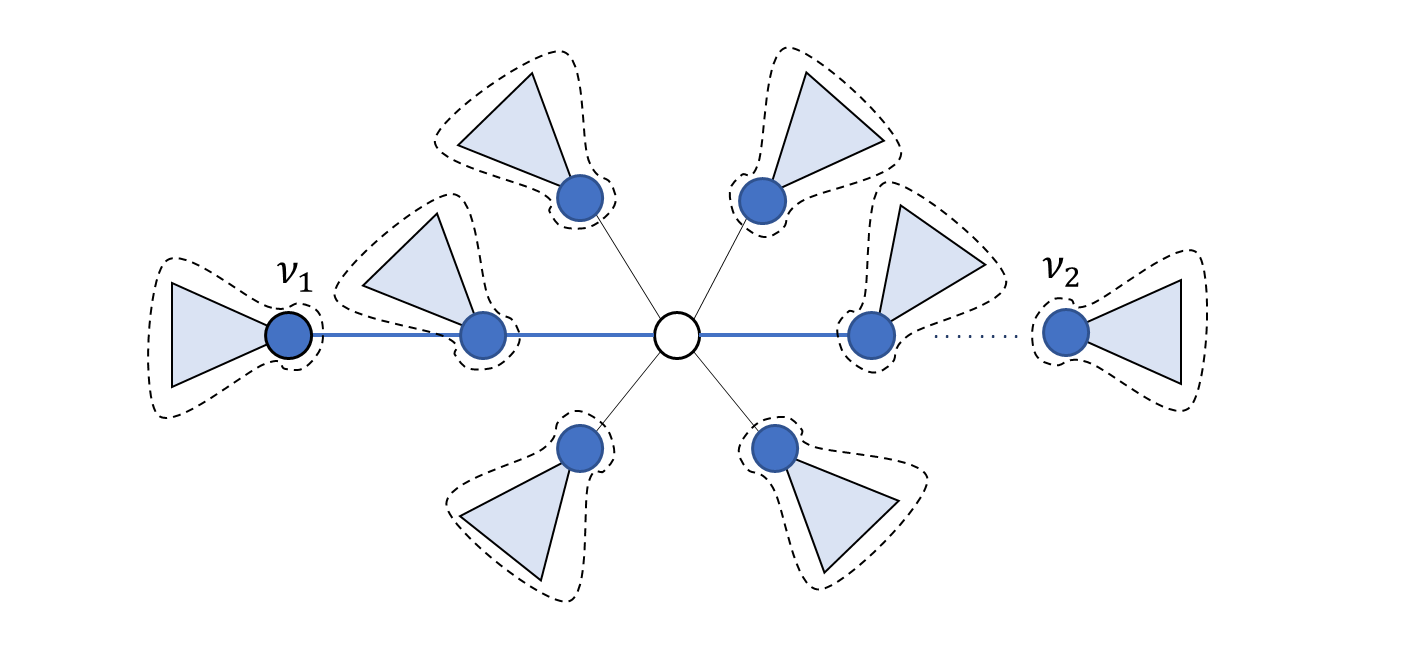}
    \caption{Link $L(P)$ associate to path $P = (\nu_1,\nu_2)$. Hollow node depicts cycle node while solid nodes depict non-cycle nodes. Dotted lines denote the nodes of $L(P)$.}
    \label{fig:link-associated-to-path}
\end{figure}

% {\color{blue} Should we enumerate these 2 steps ? and then state the sentence "Thus, $L(P)$ can be viewed as ...".}  
Thus, $L(P)$ can be viewed as a tree obtained from $T({\cal H}_S)$ after carrying out the compressions described above. Figure \ref{fig:link-associated-to-path} describes the construction of link $L(P)$.

The following property of link is useful for reporting all pairs of nodes whose path intersects $P$ at a tree edge or a cycle in ${\cal H}_S$. The proof follows from the construction of link.

\begin{lemma}
\label{lem:link-stores-all-pairs}
The path between any two nodes $\nu$ and $\nu'$ intersects the path $P$ between $\nu_1$ and $\nu_2$ in skeleton ${\cal H}_S$ at a tree edge or a cycle if and only if $\nu$ and $\nu'$ lie in different nodes in link $L(P)$.
\end{lemma}

\section{Proof of Lemma \ref{lem:s-t-mincut-containing-x-y-topological-fixed}}\label{appendix:s-t-mincut-containing-x-y-topological-fixed}
\begin{proof}
Consider $u \in X$ to be a non-terminal in strip ${\cal D}_{s,t}$. We shall show that ${\cal R}_s(u) \setminus \mathbf{s} \subseteq X$, i.e. reachability cone of $u$ towards source ${\mathbf s}$ in the strip ${\cal D}_{s,t}$ avoiding $\mathbf s$ is a subset of $X$. Consider any non-terminal $v \in {\cal R}_s(u) \setminus \mathbf{s}$. Since $u$ is reachable from $v$ in direction $\mathbf t$, therefore $\tau(v) < \tau(u)$. Therefore, $v \in X$. Therefore, ${\mathbf s} \cup X$ defines a transversal in the strip ${\cal D}_{s,t}$ (from Lemma \ref{lem:mincut-transversal}) and thus defines a $(s,t)$-mincut. The fact that $(x,y)$ lies in this $(s,t)$-mincut follows from the fact that $\tau(y) > \tau(x)$ and thus, $y \not \in X$.
\end{proof}

\section{Edge insertion on \texorpdfstring{${\cal O}(n^2)$}{quadratic} space data structure} \label{appendix: edge-insertion-n^2-space-ds}

We shall now present an algorithm 
to determine using our ${\cal O}(n^2)$ size data structure whether insertion of any edge increases the mincut between any pair of vertices. It follows from Lemma \ref{lem:edge-insertion-increases-mincut} that in order to accomplish this objective in ${\cal O}(1)$ time, it suffices if we can determine whether $x\in s^N_t$ in ${\cal O}(1)$ time for any $s,x,t\in V$. 

Let $\nu$ be the LCA of $s$ and $t$ in the hierarchy tree ${\cal T}_G$. Let ${\mathbf s}$,
${\mathbf t}$, and ${\mathbf x}$ be the units in the flesh graph ${\cal F}_{S(\nu)}$ for $s,t$, and $x$ respectively. Recall that we do not store ${\cal F}_{S(\nu)}$ at $\nu$. Rather, we just store only the units of the flesh, their projection mapping $\pi$, and the skeleton associated with ${S(\nu)}$ at $\nu$. Observe that $x$ belongs to $s^N_t$ if and only if ${\mathbf x}$ gets mapped to the source node in
the $(s,t)$-strip. { The following lemma states the necessary and sufficient conditions for $x$ to belong to $s^N_t$.}

%Refer to the algorithm given in the previous section for the construction of the $(s,t)$-strip. As a result, the following conditions in terms of the skeleton tree at $\nu$ capture the necessary and sufficient condition for $x\in s^N_t$. 

\begin{lemma}
% {\color{brown} The following are the necessary and sufficient conditions for $x$ to belong to $s^N_t$ depending upon whether ${\mathbf{x}}$ is a Steiner unit or a non-Steiner unit.
% \begin{enumerate}
%     \item ${\mathbf x}$ is a Steiner unit:~
%     $x\in s^N_t$ if and only if $\pi(s)$
%     lies on the path between $\pi(t)$ and $\pi(x)$ in the skeleton tree. 
%     \item If ${\mathbf x}$ is a non-Steiner unit:~
%     Let ${\mathbf x}$ be mapped to path between $\omega_1$ and $\omega_2$ in skeleton ${\cal H}_{S(\nu)}$. 
%     $x\in s^N_t$ if and only if $\pi(s)$ lies on the path between ${\omega_1}$ and $\pi(t)$ as well as the path between ${\omega_2}$ and $\pi(t)$ in the skeleton tree.
% \end{enumerate}
% }
{
Let $s$ and $t$ be any two Steiner vertices mapped to different nodes in skeleton ${\cal H}_S$. Let $x$ be any vertex from $V$ and let $(\omega_1,\omega_2)$ be the endpoints of the proper path to which it is mapped in ${\cal H}_S$. 
$x$ will belong to $s^N_t$ if and only if $\pi(s)$ lies on the path between ${\omega_1}$ and $\pi(t)$ as well as the path between ${\omega_2}$ and $\pi(t)$ in the skeleton tree.
}
\label{lem:nearest-mincut-and-skeleton-tree}
\end{lemma}
\begin{proof}

{ In this proof, we shall use the following fact that follows from the construction of the skeleton tree.
The path from $\omega_1$ to $\pi(t)$ in the skeleton tree $T({\cal H}_S)$ passes through $\pi(s)$   if and only if each path from $\omega_1$ to $\pi(t)$ in the skeleton ${\cal H}_S$ passes through $\pi(s)$.}

Let $(A,{\bar A})$ be any arbitrary $(s,t)$-mincut with $s\in A$. Let ${\cal B}$ be the subbunch such that $(A, {\bar A})$ is present in it. 
% Consider the cut ${\cal C}$ in the skeleton defined by $({\cal H}_S({\cal B}), {\bar {\cal H}_S({\cal B})})$ 
% % ($\pi(s) \in {\cal H}_S({\cal B})$) 
% corresponding to subbunch ${\cal B}$. 
{In the corresponding cut $({\cal H}_S({\cal B}), {\bar {\cal H}_S({\cal B})})$ in the skeleton, let us assume without loss of generality that $\pi(s) \in {\cal H}_S({\cal B})$}. 
Clearly, any path between $\omega_1$ or $\omega_2$ to $\pi(t)$ passes through $\pi(s)$ (from hypothesis). Therefore, $\omega_1,\omega_2 \in {\cal H}_S({\cal B})$. 
% {\color{brown} Thus,  $\mathbf{x}$ lies in the terminal containing $s$ in the strip ${\cal D}_{\cal B}$ (positioning of units in strip, Section \ref{subsec:connectivity-carcass}). }
{
It thus follows from the construction of the strip ${\cal D}_{\cal B}$ (Section \ref{subsec:connectivity-carcass}) that
$\mathbf{x}$ lies in the terminal containing $s$ in ${\cal D}_{\cal B}$.}
Therefore, $x \in A$.

Now, we show the other side of proof. Without loss of generality, assume that there exists a path $P_1$ from $\omega_1$ to $\pi(t)$ with $\pi(s)$ not lying on this path. We take a path from $\pi(s)$ and $\pi(t)$, take its union with $P_1$, and call it $P'$. Suppose $e$ is the skeleton edge incident on $\pi(s)$ in $P'$. Suppose ${\cal C}$ is a cut in the skeleton ${\cal H}_S$ defined as follows-- ~ $(i)$ If $e$ is a tree-edge ${\cal C}=\{e\}$ and ~$(ii)$ if $e$ is a cycle-edge then ${\cal C}=\{e,e'\}$ where
$e'$ is the other edge incident on $\pi(s)$ in the same cycle. Suppose ${\cal B}$ is the subbunch corresponding to ${\cal C}$. Clearly, $\omega_1$ and $\pi(t)$ lie on the same side of ${\cal C}$ and $\pi(s)$ lies on the other side. 
% {\color{brown} Therefore, one of the following cases may arise ~$(i)$ $\mathbf x$ will either be a non-terminal unit in ${\cal D}_{\cal B}$, i.e. $\omega_2$ is on side of $\pi(s)$ or ~$(ii)$ in the terminal unit containing $t$ in ${\cal D}_{\cal B}$, i.e. $\omega_2$ is on the side of $\pi(s)$ (positioning of units in strip, Section \ref{subsec:connectivity-carcass}). The terminal containing $s$ defines a $(s,t)$-mincut that separates $s$ from $x$.}
{ If $\omega_2$ lies on the side of $\pi(s)$, it follows from the construction of the strip ${\cal D}_{\cal B}$ described in Section \ref{subsec:connectivity-carcass} that $\mathbf x$ will be a non-terminal unit in ${\cal D}_{\cal B}$. If $\omega_2$ lies on the side of $\pi(t)$, $\mathbf x$ will be in the terminal unit containing $t$ in ${\cal D}_{\cal B}$. In both these case, the terminal containing $s$ defines a $(s,t)$-mincut in which $x$ lies on the side of $t$. Hence $x\notin s^N_t$.}

% \end{itemize}

\end{proof}
It just requires a couple of LCA queries on the skeleton tree to check the conditions mentioned in the lemma stated above. We can thus state the following theorem.

\begin{theorem}
For any undirected and unweighted graph on $n$ vertices, there exists an ${\cal O}(n^2)$ size data structure that can answer 
any sensitivity query for all-pairs mincuts in ${\cal O}(1)$ time.
\end{theorem}

\section{Extended Preliminaries} \label{appendix:extended-preliminaries}

\begin{definition}[Nearest mincut from $s$ to $t$]
An $(s,t)$-mincut $(A,{\bar A})$ where $s\in A$ is called the nearest mincut from $s$ to $t$ if and only if for any $(s,t)$-mincut $(A',{\bar A'})$ where $s \in A'$, $A\subseteq A'$. The set of vertices $A$ is denoted by $s_t^N$.
\label{def:nearest-s-t-mincut}
\end{definition}

The following lemma gives necessary and sufficient condition for an edge $(x,y)$ to increases the value of $(s,t)$-mincut.

\begin{lemma}[\cite{DBLP:journals/mp/PicardQ82}]
The insertion of an edge $(x, y)$ can increase the
value of $(s,t)$-mincut by unity if and only if $x\in s_t^N$ and $y\in t_s^N$ or vice versa.
\label{lem:edge-insertion-increases-mincut}
\end{lemma}

The following lemma exploits the undirectedness of the graph.
\begin{lemma}
Let $x,y,z$ be any three vertices in $G$. If $c_{x,y}>c$ and $c_{y,z}>c$, then $c_{x,z}>c$ as well. 
\label{lem:triangle-inequality}
\end{lemma}

When there is no scope of confusion, we do not distinguish between a mincut and the set of vertices defining the mincut. 
We now state a well-known property of cuts.
% \begin{lemma}[Submodularity of cuts]
% For any two subsets $A,B\subset V$, the following inequality holds.
% \[ c(A) +c(B) \ge c(A\cup B) + c(A\cap B) \]
% \label{lem:submodularity}
% \end{lemma}
\begin{lemma}[Submodularity of cuts]
For any two subsets $A,B\subset V$, ~
$ c(A) +c(B) \ge c(A\cup B) + c(A\cap B)$.
\label{lem:submodularity}
\end{lemma}

%The proof of the following lemma exploits just the property of a $(u,v)$-mincut.

The following lemma follows from the fact that a $(s,t)$-mincut is the minimum cardinality cut that separates $s$ and $t$.

\begin{lemma}
Let $S \subset V$ define an $(s,t)$-mincut with $s\in S$. For any subset $S'\subset V\setminus S$ with $t\notin S'$,
\[ 
c(S,S') \le c(S,V\setminus (S\cup S'))
\]
\label{lem:subset-property-of-min-cut}
\end{lemma}

\subsection*{Compact Representation for all $(s,t)$-mincuts}

Suppose ${\cal D}_{s,t}$ is a strip containing all $(s,t)$-mincuts and $x$ is a non-terminal in this strip. The $(s,t)$-mincut defined by ${\cal R}_s(x)$ is the nearest mincut from $\{s,x\}$ to $t$. Interestingly, each transversal in ${\cal D}_{s,t}$, and hence each $(s,t)$-mincut, is a union of the reachability cones of a subset of nodes of ${\cal D}_{s,t}$ in the direction of $s$. We now state the following Lemma that we shall crucially use.

\begin{lemma}[\cite{DBLP:journals/siamcomp/DinitzV00}]
If $x_1,\ldots, x_k$ are any non-terminal nodes in strip ${\cal D}_{s,t}$,  the union of the reachability cones of $x_i$'s in the direction of ${\mathbf s}$ defines the nearest mincut between $\{s, x_1,\ldots, x_k\}$ and $t$.
\label{lem:reachability-cones}
\end{lemma} 

We also use the following Lemma.

\begin{lemma} 
If $A\subset V$ defines a $(s,t)$-mincut with $s\in A$, then $A$ can be merged with the terminal  node ${\mathbf s}$ in ${\cal D}_{s,t}$ to get the strip ${\cal D}_{A,t}$ that stores all those $(s,t)$-mincuts that enclose $A$.
\label{lem:strip-A}
\end{lemma}

Another simple observation helps us describe the nearest mincuts in the strip.

\begin{lemma}
The mincuts defined by $\mathbf{s}$ and $\mathbf{t}$ are the nearest mincut from $s$ to $t$ and $t$ to $s$ respectively.
\label{lem:nearest-mincut-strip}
\end{lemma}

\subsection*{Compact Representation for all Steiner Mincuts}
It is important to note that nearest $s$ to $t$ and $t$ to $s$ mincuts are easier to identify in the connectivity carcass. The following lemma conveys the fact.

\begin{lemma}[\cite{DBLP:conf/stoc/DinitzV94}]
\label{lem:u-nearest-s-t-mincut}
Let $s,t \in S$ such that $c_{s,t}=c_S$. Determining if a unit $u$ lies in nearest $s$ to $t$ mincut (or vice-versa) can be done using skeleton ${\cal H}_S$ and projection mapping $\pi_S$ using ${\cal O}(1)$ LCA queries on skeleton tree.
\end{lemma}

The size of flesh ${\cal F}_S$ is ${\cal O}(\min(m,\tilde{n}c_S))$ where $\tilde{n}$ is the number of units in ${\cal F}_S$. The size taken by skeleton is linear in the number of Steiner units. Thus, overall space taken by the connectivity carcass is ${\cal O}(\min(m,\tilde{n}c_S))$.

\section{Proof of Lemma \ref{lem:y-in-nearest-(r,s)-G_A} } \label{appendix:y-sNr-G}

\begin{proof}
It follows from Lemma \ref{lem:GH} that 
each $(s,r)$-mincut in $G'$ is also a $(s,r)$-mincut in $G$. Suppose $(B,\bar{B})$ is the ($s,r$)-mincut  nearest to $s$ in $G'$ such that $\bar{A} \not \in B$ and $s \in B$. Viewing this cut in $G$, it follows that $y \not \in B$. Therefore, $y$ does not lie in nearest $s$ to $r$ mincut in $G$. 
\end{proof}

\section{Proof of Lemma \ref{lem:contracted-subcactus-mincut}} \label{appendix:contracted-subcactus-mincut}

Let $c$ be any cycle (or tree edge) passing through (incident on) $\nu$ in the skeleton
${\cal H}_S$. Let ${\cal D}_{s,t}$ be the strip corresponding to the sub-bunch defined by the structural edge(s) incident on $\nu$ by $c$.
Let $\nu$ be on the side of the source $\mathbf{s}$ in this strip.
Let ${\cal H}_S(c)$ be the subcactus formed by removing the structural edge(s) from $c$ incident on $\nu$ and not containing $\nu$. 
Recall that the subcactus ${\cal H}_S(c)$ was contracted into a vertex, say $v_c$, in the graph 
$G_{S'}$.
%Moreover, suppose $v_c$ is the contracted vertex corresponding to contracted subcactus ${\cal H}_S(c)$.
%Also, assume that $\nu$ is on the side of the source $\mathbf{s}$ in this sub-bunch.

\begin{lemma}
Let $u$ and $u'$ be any two non-terminal units in ${\cal D}_{s,t}$ such that none of them is compressed to $v_c$ in $G_{S'}$. If one of them is reachable from the other in the direction of ${\mathbf{s}}$, then both of them will be compressed to the same contracted vertex in $G_{S'}$.
%
%Let $u$ and $u'$ be any two non-terminal units in ${\cal D}_{s,t}$ such that $u'$ is reachable from $u$ in the direction of ${\mathbf{s}}$ and $u$ is not contracted to vertex $v_c$. $u'$ and $u$ will be compressed to the same contracted node in $G_{S'}$.
\label{lem:u-u'-in-G-nu}
\end{lemma}
\begin{proof}
Assume without loss of generality that $u'$ is reachable from $u$ in the direction of ${\mathbf{s}}$.
Let the proper paths associated with each of $u$ and $u'$ in ${\cal H}_S$ be $P(\nu_1,\nu_2)$ and $P(\nu_1',\nu_2')$ respectively. 
It follows from the construction of ${\cal D}_{s,t}$ that
$P(\nu_1,\nu_2)$ as well as $P(\nu_1',\nu_2')$ will pass through one of the structural edge(s) from $c$ on $\nu$. Without loss of generality,  assume that $P(\nu_1,\nu_2)$ passes through $e$. Since $P(\nu_1,\nu_2)$ is a proper path, this implies that this is the only structural edge in this cut (of skeleton) through which this path passes.
Since $u'$ is reachable from $u$ in flesh ${\cal F}_S$, so $P(\nu_1',\nu_2')$ will also have to pass through $e$ (from Lemma \ref{lem:path-extendable}).
It again follows from Lemma \ref{lem:path-extendable}, that $P(\nu_1,\nu_2)$ as well as $P(\nu_1',\nu_2')$ are subpaths of a path, say $P(\nu',\nu'')$, in skeleton
${\cal H}_S$. This combined with the above discussion establishes that $P(\nu',\nu'')$ has the structure shown in Figure \ref{fig:structure-of-p(nu',nu'')}.

Observe that any path in skeleton that passes through a node $\nu$ can intersect at most 2 cycles or tree-edges that are passing though $\nu$. We know that suffix of $P(\nu',\nu'')$ after $e$ lies in ${\cal H}_S(c)$, so the prefix upto $e$ must have endpoint in subcactus ${\cal H}_S(c')$ where $c'\neq c$. This implies that $u$ must be compressed to $v_{c'}$ because it is not compressed to $v_c$. Thus, $c'$ precedes $c$ in total order. It follows from the structure of path $P(\nu_1',\nu_2')$ that it will have an endpoint in ${\cal H}_S(c')$. Thus, $u'$ will be compressed to the same compressed vertex $v_{c'}$ in $G_{S'}$. This completes the proof.

\begin{figure}%[H]
\centering
\includegraphics[width=0.95\textwidth]{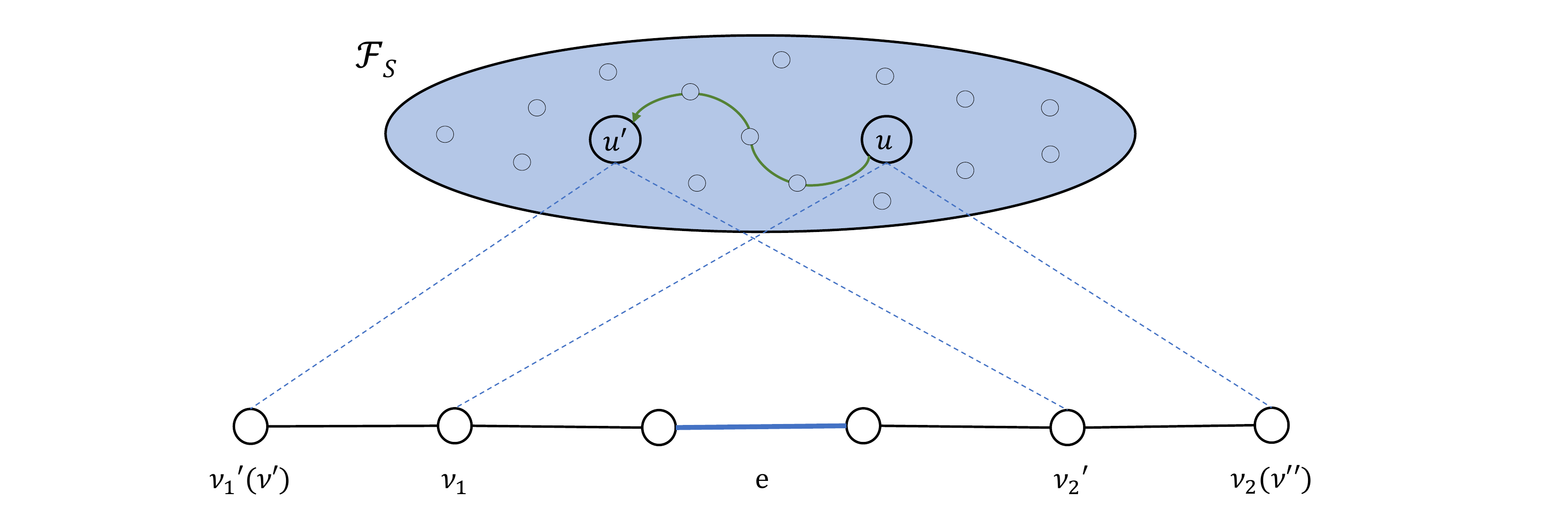}
    \caption{The structure of path $P(\nu',\nu'')$.}
\label{fig:structure-of-p(nu',nu'')}
\end{figure}

\end{proof}
% Lemma \ref{lem:contracted-subcactus-mincut} implies that vertex set compressed to each contracted node defines a Steiner mincut. This completes the proof of Lemma \ref{lem:u-u'-in-G-nu}.

Consider the set of non-terminals in the strip ${\cal D}_{s,t}$ that are not compressed to contracted vertex $v_c$. Let this set be $U$. Observe that the set of units $\bigcup_{u\in U} {\cal R}_s(u)$ form a Steiner mincut (using Lemma \ref{lem:reachability-cones}). Moreover, it follows from Lemma \ref{lem:u-u'-in-G-nu} that each non-terminal unit in the set $\bigcup_{u\in U} {\cal R}_s(u)$ is not compressed to contracted vertex $v_c$. Thus, $U = \bigcup_{u\in U} {\cal R}_s(u) \setminus \{\mathbf{s}\}$. All the set of vertices compressed to $v_c$ forms the complement of set $\bigcup_{u\in U} {\cal R}_s(u)$, and thus defines the same Steiner mincut. Therefore, the set of vertices corresponding to each contracted vertex defines a Steiner mincut.

\begin{figure}
    \centering
    \includegraphics[width=0.9\textwidth]{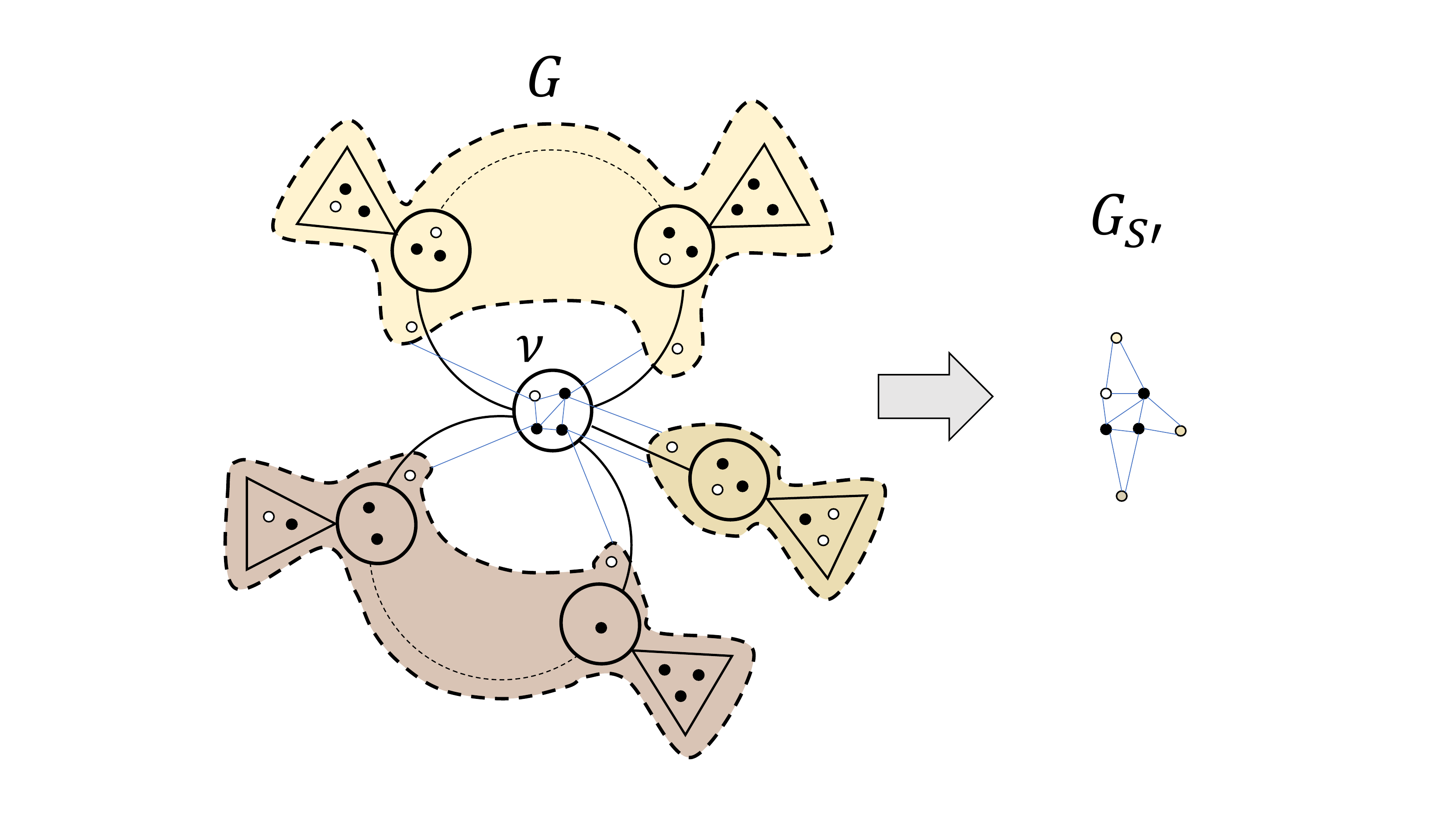}{}
    \caption{$2$-step contraction procedure to construct $G_{S'}$. We only show the vertices and relevant edges of graph along with the skeleton ${\cal H}_S$. Solid vertices belong to Steiner set $S$ and hollow vertices are non-Steiner vertices. All Steiner vertices inside node $\nu$ form the set $S'$.}
    \label{fig:image-contraction}
\end{figure}

It follows from the construction that $G_{S'}$ is a quotient graph of $G$. Moreover, the number of contracted vertices equals the number of cycles and tree edges incident on node $\nu$ in the skeleton. Figure \ref{fig:image-contraction} gives a nice illustration of the contraction procedure.

\section{Proof of Lemma \ref{lem:linear-time-qt} and \ref{lem:mincut-qt}}
\label{appendix:linear-time-qt}
\begin{proof}
Consider the case when $y$ does not belong to any contracted vertex. In this case, all edges in $E_y$ remain intact in $G_{S'}$ and thus $E_{y'}=E_{y}$.

Now, suppose $y$ belong to contracted vertex $y'$. Let $\bar A$ be the set of vertices compressed to contracted vertex $y'$. We select a vertex $t\in {\bar A} \cap S$ and construct the ${\cal D}_{A,t}$ strip using the flesh ${\cal F}_S$ and skeleton ${\cal H}_S$ in time linear in the size of flesh (using Lemma \ref{lem:strip-from-carcass}). Using the construction outlined in Lemma \ref{lem:query-transformation} we can obtain the set of edges $E_{A}$ by computing reachability cone(s) in strip ${\cal D}_{A,t}$. This takes time linear in size of ${\cal D}_{A,t}$. All edges in $E_A$ share same endpoint $y'$ in $G_{S'}$. Thus, we get the set of edges $E_{y'}$ which is simply all edges in $G_{S'}$ corresponding to set $E_A$. Clearly, this process can be accomplished in time linear in the size of flesh ${\cal F}_S$.

Suppose we have a $(r,s)$-mincut in $G_{S'}$, say $B$ such that $s,t \not\in B$ that contains all edges in $E_{y'}$. If $y$ does not belong to any contracted vertex, this cut itself can be reported as $E_y=E_{y'}$. Suppose $y$ belong to contracted vertex $y'$. We can construct another $(r,s)$-mincut $B\cup R$ (recall the definition of $R$ in Proof of Lemma \ref{lem:query-transformation}). This procedure also involves construction of ${\cal D}_{A,t}$ strip and computation of reachability cone(s) in this strip. This process can also be accomplished in time linear in the size of flesh ${\cal F}_S$.
\end{proof}

\section{Reporting \texorpdfstring{$(s,t)$}{(s,t)}-mincut upon insertion of edge \texorpdfstring{$(x,y)$}{(x,y)}} \label{appendix:edge-insertion-reporting-mincut}

We now show how to report a $(s,t)$-mincut upon insertion of edge $(x,y)$ in ${\cal O}(m)$ space data structure. 

If $\textsc{In-Mincut}(s,t,(x,y))$ evaluates to true, we can report the set of vertices defining an old $(s,t)$-mincut (follows from Fact \ref{fact:(x,y)-insertion}). This can be easily reported using Gomory-Hu \cite{GH61} tree. However, if $\textsc{In-Mincut}(s,t,(x,y))$ evaluates to false, we have to report an old $(s,t)$-mincut that does not separate $x$ from $y$.

%%%%%%%%%%%%%%%%%%%%%%| New Addition |%%%%%%%%%%%%%%%%%%%%

If $\textsc{In-Mincut}(s,t,(x,y))$ is false, one of the following cases may arise (wlog) ~$(i)$ $x,y \in s_t^N$ or ~$(ii)$ $y \not\in s_t^N$ and $y \not\in t_s^N$. If $x,y \in s_t^N$, then reporting any old $(s,t)$-mincut will do because no $(s,t)$-mincut separates $x$ from $y$. Thus, assume $(ii)$ is true. 

% {\color{red} Add remark below this paragraph as to why we need to move up the parent level.}

Consider the executions of Algorithm \ref{algo:check-nearest-mincut-efficient} of \textsc{Check-Nearest-Mincut}$(s,t,y)$ and \textsc{Check-Nearest-Mincut}$(t,s,y)$. At $\ell = LCA(s,t)$, the condition on Line $3$ must be true in both executions. Firstly, assume \textsc{Edge-Contained}$(s,t,E_y)$ is false. In this case, $y \not\in s_t^N$ in $G_\ell$ and $y \not\in t_s^N$ in $G_\ell$. Thus, we can construct $(s,t)$-strip in $G_\ell$ and report a $(s,t)$-mincut that does not separate $x$ from $y$. Now, consider the case when \textsc{Edge-Contained}$(s,t,E_y)$ evaluates to true.  Suppose $\mu$ is the last node from root to $\ell$ at which Line $11$ evaluated to true. Suppose $\mathbf{y}$ is the unit containing $y$ in strip ${\cal D}_{A,z}$ (ref Line $10$) at node $\mu$. Let $E_{y,A}$ be the side-$A$ of $\mathbf{y}$ at this node. Since all subsequent calls will only transform this edge-containment query,  \textsc{Edge-Contained}$(s,t,E_{y,A})$ is true. We can produce a $(s,t)$-mincut ${\cal C}_1 = (B, {\bar B})$ that contains all edges in $E_{y,A}$ and $\mathbf{y} \in B$.  Observe that ${\cal C}_2 = (B \setminus\{{\mathbf y}\}, {\bar B}\cup \{{\mathbf y}\})$ is another $(s,t)$-mincut and at least one of ${\cal C}_1$ or ${\cal C}_2$ must not separate $x$ from $y$.

% \begin{remark}
% It is worthwhile to mention that in the later case there may not exist any $(s,t)$-mincut in $G_\ell$ that does not separates $x$ from $y$. Thus, we have to move to an ancestral level where such a $(s,t)$-mincut can be produced.
% \end{remark}

\begin{remark}
If $(s,t)$-mincut value remains unchanged after the insertion of $(x,y)$, our aim is to report a $(s,t)$-mincut in $G$ that does not separate $x$ and $y$. In the algorithm described above, if \textsc{Edge-Contained}$(s,t,E_y)$ turns out to be true, it may happen that each $(s,t)$-mincut at level $\ell$ separates $x$ and $y$. Therefore, we have to move to an ancestor $\mu$ of $\ell$ where such a $(s,t)$-mincut can be produced.
\end{remark}

Thus, we can report a $(s,t)$-mincut that does not separates $x$ from $y$ in ${\cal O}(\min(m,nc_{s,t}))$ time.

\section{Size and Time analysis of compact data structure}
\label{appendix:size-time-analysis-compact-ds}

The data structure doesn't seem to be an ${\cal O}(m)$ size data structure at first sight. Observe that augmentation at any internal node can still take ${\cal O}(m)$ space individually. Interestingly, we show that collective space taken by augmentation at each internal node will still be ${\cal O}(m)$.

We begin with the following lemma which gives a tight bound on the sum of weights of edges in Gomory-Hu tree is $\Theta(m)$. We also give the proof for the same which was suggested in \cite{DBLP:conf/stoc/HariharanKPB07,DBLP:journals/siamcomp/DinitzV00}.

\begin{lemma}[\cite{DBLP:conf/stoc/HariharanKPB07,DBLP:journals/siamcomp/DinitzV00}]
\label{fact:GH-weight}
The sum of weights of all edges in the Gomory-Hu tree is ${\Theta}(m)$.
\end{lemma}
\begin{proof}
Consider any edge $(u,v)\in E$. This edge must be present in every $(u,v)$-mincut. Thus, the sum of weights of all edges in Gomory-Hu tree is at least $m$. Now, root the Gomory-Hu tree at any arbitrary vertex $r$. Let $f$ maps each edge in this tree to its lower end-point. It is easy to observe that $f$ is a one-to-one mapping. Let $e$ be an edge in the Gomory-Hu tree. Observe that $w(e) \leq deg(f(e))$ (where $deg()$ is the degree of vertex in $G$). Thus the sum of weights of all edges in Gomory-Hu tree is at most $2m$. This comes from the simple observation that the sum of the degree of all vertices in $G$ equals $2m$.
\end{proof}

Let us assign each edge $(\mu,\mu')$ in the hierarchy tree ${\cal T}_G$ weight equal to the Steiner mincut value for the Steiner set $S(\mu)$ (if $\mu$ is the parent of $\mu'$). We shall show that sum of the weight of edges in hierarchy tree ${\cal T}_G$ is $\Theta(m)$. %This will help in our analysis later.

To establish this bound refer to algorithm \ref{Construct Tree} that gives an algorithm to construct the hierarchy tree from Gomory-Hu tree. Observe that the variable $ctr$ in this algorithm stores the sum of weights of all edges in ${\cal T}_G$. It is clear that for $k$ edges removed from the Gomory-Hu tree, we add $k+1$ edges of equal weight in ${\cal T}_G$. Thus, the sum of the weight of all edges in ${\cal T}_G$ is at most $4m$ (since $k+1 \leq 2k$). Therefore, we state the following lemma.

\begin{algorithm}%[H]
    \caption{Construct Hierarchy Tree ${\cal T}_G$ from Gomory-Hu Tree $\hat{ \cal T}_{G}$}
    \label{Construct Tree}
    \begin{algorithmic}[1] % The number tells where the line numbering should start
        \State $ctr \gets 0$
        \Procedure{Construct-Tree}{$\hat{ \cal T}_{G}$}
            \If{$\hat{ \cal T}_{G}$ has single node}
            \State Create a node $\mu$
            \State $val(\mu) \gets val(\hat{ \cal T}_{G})$
            \State \textbf{return} $\mu$
            \EndIf
            \State $c_{\min} \gets$ $\min_{e\in \hat{ \cal T}_{G}}w(e)$
            \State Let there be $k$ edges with weight $c_{\min}$
            \State Remove all edges of weight $c_{\min}$ in $\hat{ \cal T}_{G}$ to get $(k+1)$ trees $T_1,..,T_{k+1}$
            \State Create a node $\mu$
            \State $ctr \gets ctr + c_{\min}\times(k+1)$
            \State $children(\mu) \gets \{\textsc{Construct-Tree}(T_i)\;|\;\forall i\in [k+1]\}$
            \State \textbf{return} $\mu$
        \EndProcedure
    \end{algorithmic}
\end{algorithm}

\begin{lemma}
\label{lem:hierarchy-tree-weight}
The sum of weights of all edges in the tree ${\cal T}_G$ is ${\Theta}(m)$.
\end{lemma}
% The size and time analysis of the data structure crucially exploits the following observations,
We shall now give a bound on the size of augmentation at each internal node. {\color{black} Using this bound, we shall give an upper bound on the total size of the data structure.} The following lemma gives a bound on the size of flesh graph ${\cal F}_S$ for any Steiner set $S$.

\begin{lemma}
Let ${\cal V}_S$ and ${\cal W}_S$ denote the set of Steiner and non-Steiner units respectively in the flesh graph ${\cal F}_S$ of $G$ with Steiner set $S\subseteq V$. The size of ${\cal F}_S$ is  ${\cal O}(|{\cal V}_S|c_S + \sum_{u\in {\cal W}_S}deg(u))$.
\label{lem:size-of-flesh}
\end{lemma}
\begin{proof}
Consider the Gomory-Hu tree of the flesh graph ${\cal F}_S$, say ${\cal T}$. It is evident that the value of mincut between any two units is at most $c_S$. This follows from the definition of a unit. Now, root this tree ${\cal T}$ at some Steiner unit. Let $f$ maps each edge in this tree to its lower end-point. Any edge in this tree has weight at most $c_S$. However, for any non-Steiner unit $u$, $w(f^{-1}(u)) \leq deg(u)$ (where $deg()$ is the degree of vertex in flesh ${\cal F}_S$).  Thus, the sum of weight of all edges in ${\cal T}$ is bounded by $|{\cal V}_S|c_S + \sum_{u\in {\cal W}_S}deg(u)$.
% {\color{blue} 
The weight of Gomory-Hu tree ${\cal T}$ is an upper bound on the size of graph ${\cal F}_S$ (see Lemma \ref{fact:GH-weight}). So the lemma holds true.
% }
% Using Lemma \ref{fact:GH-weight}, it follows that size of flesh ${\cal F}_S$ is ${\cal O}(|{\cal V}_S|c_S + \sum_{u\in {\cal W}_S}deg(u))$.
\end{proof}

% {\color{red} @Sir, Kindly go through this blue draft, I have tried to incorporate all your suggestions. I have also improved the projection mapping. $LA$ queries can be used to easily determine the unit to which any contracted vertex is mapped in $O(1)$ time.}

{ Recall that while constructing the graphs for each level we introduced \textit{contracted vertices} as we moved down from an internal node to its child. In particular, as we moved down from node $\mu$ to $\mu'$, the graph $G_{\mu'}$ was constructed from $G_\mu$ using ~ $(i)$ all vertices mapped to node $\mu'$ in skeleton ${\cal H}_{S(\mu)}$, and ~$(ii)$ contracted vertices introduced while compressing various subcactuses. Moreover, the number of contracted vertices equals the number of cycles and tree edges incident on node $\mu'$ in skeleton ${\cal H}_{S(\mu)}$ (using Lemma \ref{lem:contracted-subcactus-mincut} $(2)$). Consider the flesh graph ${\cal F}_{S(\mu)}$ stored at some internal node $\mu$ in the tree and ${\cal V}_{S(\mu)}$ and ${\cal W}_{S(\mu)}$ denote the set of Steiner and non-Steiner units in ${\cal F}_{S(\mu)}$ respectively. We make the following key observation. The observation follows from the fact that ${\cal H}_{S(\mu)}$ is a cactus graph and hence, it has ${\cal O}(|{\cal V}_{S(\mu)}|)$ edges.

\begin{observation}
\label{obs:number-of-contracted-vertices}
The total number of contracted vertices introduced by an internal node $\mu$ across all its children is ${\cal O}(|{\cal V}_{S(\mu)}|)$.
\end{observation}

Suppose $u \in {\cal W}_{S(\mu)}$. It is evident that $u$ consists of only contracted vertices obtained as a result of the contraction procedure at ancestors of $\mu$. It follows from the contraction procedure that the non-Steiner unit $u$ gets compressed to a new contracted vertex in $G_{\mu'}$ where $\mu'$ is a child of $\mu$, and in a sense \textit{disappears}. Therefore, each contracted vertex appears in at most one non-Steiner unit across flesh graphs stored at \textit{all} internal nodes of ${\cal T}_G$. We summarize this insight in the following observation.

\begin{observation}
\label{obs:contracted-vertex-appears-once}
Any contracted vertex introduced by an internal node $\mu$ appears at most once in a non-Steiner unit in flesh graph stored at one of the descendants of $\mu$. 
\end{observation}

Using Observation \ref{obs:number-of-contracted-vertices} and \ref{obs:contracted-vertex-appears-once}, We shall now give an upper bound on the total number of units across all flesh graphs stored in ${\cal T}_G$. It follows from Observation \ref{obs:number-of-contracted-vertices} that total number of contracted vertices introduced by all internal nodes is ${\cal O}(\sum_{\mu \in {\cal T}_G} |{\cal V}_{S(\mu)}|) = {\cal O}(n)$. Further, it follows from Observation \ref{obs:contracted-vertex-appears-once} that the number of non-Steiner units across all flesh graphs is bounded by the total number of contracted vertices introduced at every internal node. Thus, the total number of non-Steiner units across all flesh graphs is ${\cal O}(n)$. The total number of Steiner units, on the other hand, is also 
% ${\cal O}(n)$
${\cal O}(\sum_{\mu \in {\cal T}_G} |{\cal V}_{S(\mu)}|) = {\cal O}(n)$.
Thus, we draw the following inference.
% since each Steiner unit is uniquely mapped to a node in the skeleton. Thus, we draw the following inference.

\begin{inference}
\label{inf:units-O(n)}
The total number of units across all flesh graphs stored in ${\cal T}_G$ is ${\cal O}(n)$.
\end{inference}

Now, we shall bound the sum of the degree of all non-Steiner units across all flesh graphs stored in ${\cal T}_G$. Since each contracted vertex appears in at most one non-Steiner unit, we can sum the degree of all contracted vertices to get an upper bound. It again follows from Lemma \ref{lem:contracted-subcactus-mincut} that the degree of contracted vertex introduced by node $\mu$ is exactly $c_{S(\mu)}$. Thus, the sum of degree of all contracted vertices introduced by node $\mu$ is ${\cal O}(|{\cal V}_{S(\mu)}|c_{S(\mu)})$. The following lemma gives a bound on the sum of degrees of all non-Steiner units across all flesh graphs stored in ${\cal T}_G$.

\begin{lemma}
\label{inf:sum-degree}
The sum of degree of non-Steiner units across flesh graphs stored at all internal nodes is ${\cal O}(m)$.
\end{lemma}

\begin{proof}
The sum of degree of all contracted vertices introduced by node $\mu$ is ${\cal O}(|{\cal V}_{S(\mu)}|c_{S(\mu)})$. Thus, the sum of degree of contracted vertices introduced at each internal node is ${\cal O}(\sum_{\mu \in {\cal T}_G}{|{\cal V}_{S(\mu)}|}c_{S(\mu)})$ = ${\cal O}(m)$ (using Lemma \ref{lem:hierarchy-tree-weight}). Recall that each non-Steiner unit comprises of only contracted vertices and each contracted vertex appears in at most one non-Steiner unit (from Observation \ref{obs:contracted-vertex-appears-once}). Therefore, the sum of degree of non-Steiner units in all flesh graphs, that is $\sum_{\mu \in {\cal T}_G} \sum_{u \in {\cal W}_{S(\mu)}}deg(u)$ 
% ($deg()$ means the degree of unit in flesh graph ${\cal F}_{S(\mu)}$) 
is ${\cal O}(m)$.
\end{proof}

The size of mincut-sensitivity data structure can be bounded by combining Lemma \ref{lem:size-of-flesh} and \ref{inf:sum-degree} as follows.

\begin{lemma}
The size taken by mincut-sensitivity data structure is ${\cal O}(m)$.
\end{lemma}
 \begin{proof}
 The sum of sizes of flesh graphs stored at all internal nodes of ${\cal T}_G$ is given by the following expression.
 
 \begin{equation*}
    \begin{split}
        \nonumber
        \sum_{\mu \in {\cal T}_G}|{\cal F}_{S(\mu)}| &\leq c_1 \times \sum_{\mu \in {\cal T}_G} (|{\cal V}_{S(\mu)}|c_{S(\mu)} + \sum_{u\in {\cal W}_{S(\mu)}}deg(u)) \hspace{3mm}\text{(using Lemma \ref{lem:size-of-flesh})}\\
        &\leq c_2 \times m \hspace{57mm}\text{(using Lemma \ref{inf:sum-degree})}
    \end{split}
\end{equation*}
 
 \end{proof}

\subsubsection*{A note on mapping from vertices to flesh units}

It is important to observe that the mapping from graph vertices to flesh units cannot be stored individually for each internal node in ${\cal T}_G$. This is because we may end up using ${\cal O}(n^2)$ space only for storing this mapping. In order to optimize the space taken by this mapping, we critically examine the relevant internal nodes in ${\cal T}_G$ where a contracted vertex, say $u$, lies in the flesh graph. Suppose $u$ is introduced at internal node $\mu$ while constructing the graph $G_{\mu'}$ ($\mu'$ is a child of $\mu$). We have already observed that $u$ can appear as a non-Steiner unit in the flesh graph stored at at most one descendant of $\mu$. Moreover, if $u$ is mapped to a Steiner unit in flesh ${\cal F}_{S(\mu')}$, it is evident that $u$ appears in the child node $\mu''$ to which this Steiner unit is mapped. Thus, the relevant flesh graphs in which $u$ appears form a path descending from $\mu$ to a descendant, say $\nu$, such that ~$(i)$ $u$ appears in a non-Steiner unit in ${\cal F}_{S(\nu)}$ or ~$(ii)$ $\nu$ is a leaf node. Thus, the mapping $\varphi$ of a contracted vertex can be stored in ${\cal O}(1)$ space. It consists of ~ $(i)$ the internal node $\mu'$ where it was first introduced and ~ $(ii)$ the internal node $\nu$ where it finally appears, and ~$(iii)$ the non-Steiner unit it appears in, if any. Using this information, we can determine the unit to which $u$ is mapped in any of the relevant flesh graphs with the help of level-ancestor (LA) queries in constant time \cite{DBLP:journals/jal/BenderFPSS05}. Determining the mapping $\varphi$ of Steiner vertices at any internal node can be done easily using LA queries from appropriate leaf-nodes. Thus, the total space taken by projection mapping $\varphi$ is ${\cal O}(n)$ (using Lemma \ref{inf:units-O(n)}).

\subsubsection*{Analysis of edge-containment query time}

A trivial bound on the query time follows from the size analysis itself. Since the total size of our data structure is ${\cal O}(m)$, it follows that the sum of size of all flesh graphs stored from the root node to $LCA(s,t)$ will also be ${\cal O}(m)$. Dinitz and Vainshtein \cite{DBLP:conf/stoc/DinitzV94} showed that size of flesh graph ${\cal F}_S$ is ${\cal O}(\tilde{n}c_S)$ for a Steiner set $S$, where $\tilde{n}$ is the number of units in the flesh graph. Total number of units across all flesh graphs in ${\cal T}_G$ is only ${\cal O}(n)$ (Inference \ref{inf:units-O(n)}). The value of Steiner mincut increases as we traverse from the root towards a leaf. Thus, $c_{s,t}$ is the maximum Steiner mincut value in the path from root node to $LCA(s,t)$ . Thus, the sum of sizes of all flesh graphs in this path is bounded by ${\cal O}(nc_{s,t})$. Thus, the query time we achieve is ${\cal O}(\min(m,nc_{s,t}))$.

}

\section{Reporting \texorpdfstring{$(s,t)$}{(s,t)}-strip for steiner mincuts} \label{appendix:xxx}

Let $s$ and $t$ be any two arbitrary vertices. We shall now provide an algorithm to report $(s,t)$-strip using our data structure presented in Section 3. 

We first compute the LCA of $s$ and $t$
in the hierarchy tree ${\cal T}_G$. Let this node be $\nu$. Let $S(\nu)$ be the Steiner set associated with $\nu$, and ${\cal F}_{S(\nu)}$ be the corresponding flesh graph. Recall that we augment $\nu$ with the skeleton ${\cal H}_{S(\nu)}$ and the projection mapping $\pi$ of all the units present in ${\cal F}_{S(\nu)}$.

Each $(s,t)$-mincut is a Steiner mincut for $S(\nu)$ that separates $s$ and $t$ as well, and the $(s,t)$-strip will store precisely all these cuts. Observe that each Steiner mincut that separates $s$ and $t$ is defined by a tree edge or a (suitable) pair of edges from the cycle that  appears on the path between the node containing $s$ and the node containing $t$ in ${\cal H}_{S(\nu)}$. Using this observation, the $(s,t)$-strip will be a quotient graph of the flesh graph ${\cal F}_{S(\nu)}$ that preserves only these Steiner mincuts.
% {\color{brown} We shall compute it using the skeleton tree for ${\cal H}_{S(\nu)}$ and the projection mapping of various units. Without loss of generality we assume that the skeleton tree is rooted at the node $\pi(s)$. }
% {\color{red} The preceding brown text is not required since we are using link structure. The preceding black text can be extended to give motivation for using the link structure. }
% {\color{blue} justification for the previous claim?}
We use the link $L(P)$ (see Appendix \ref{appendix:link-associated-to-a-path}) associated to path $P$ between $\pi(s)$ and $\pi(t)$ in skeleton ${\cal H}_{S(\nu)}$ to build the $(s,t)$-strip. We process the units of the flesh graph ${\cal F}_{S(\nu)}$ as follows.

\begin{enumerate}
    \item Processing the terminal units:\\
    {
    % Let $\mu$ be a node in the link $L(P)$. For any terminal unit $u$ mapped to $\mu$, we compress it into a single vertex in the original graph. {\color{red} This statement needs refinement.}
    % Each terminal unit $u$, such that $\pi(u)$ is compressed to $\mu$ in $L(P)$, is compressed to a single vertex in original graph.
    For each node $\mu$ in the link $L(P)$ we compress all the terminal units $u$ such that $\pi(u) \in \mu$ to a single vertex. Therefore, each node in link $L(P)$ will correspond to a compressed vertex.
    
    }
    % {\color{red}
    % Let $u$ be a unit in the flesh and $\mu = \pi(u)$ be the node to which it is mapped in skeleton ${\cal H}_{S(\nu)}$.
    % % Let $\mu$ be any Steiner unit in the flesh graph ${\cal F}_{S(\nu)}$. 
    % Let $\omega$ be the LCA of $\mu$ and $\pi(t)$ in the skeleton tree. If $\omega$ is a tree node in the skeleton tree, we merge $\mu$ with the unit corresponding to $\omega$ in the flesh graph. If $\omega$ is a cycle node, we merge $u$ with the child of $\omega$ that is ancestor of $\mu$. It requires a single LCA query and a level ancestor query \cite{DBLP:journals/tcs/BenderF04} on the skeleton tree. Both of them can be accomplished in ${\cal O}(1)$ time to accomplish this task. 
    % }
    \item Processing the stretched units:\\
    Let $u$ be any stretched unit in the flesh graph ${\cal F}_{S(\nu)}$. Let $\pi(u)$ be the proper path between $\omega_1$ and $\omega_2$ in the skeleton to which $u$ is mapped. 
    {
    $u$ appears as a non-terminal in $(s,t)$-strip if and only $\omega_1$ and $\omega_2$ lie in different nodes in link $L(P)$. If unit $\omega_1$ and $\omega_2$ lie in different nodes in $L(P)$ we retain $u$ as it is in the original graph. Otherwise, we merge it with the compressed vertex corresponding to the node of $L(P)$ which contains both $\omega_1$ and $\omega_2$.
    }
    % {\color{red}
    % $\mu$ will appear as nonterminal in $(s,t)$-strip if and only if $\pi(\mu)$ intersects the $(s,t)$-path in the skeleton ${\cal H}_{S(\nu)}$. This requires $O(1)$ LCA queries only on the skeleton tree to determine it. If $\mu$ turns out to intersect the $(s,t)$-path, we retain it as it is in the quotient graph. Otherwise, we merge $\mu$ with
    % the same node in the quotient graph to which both the two endpoints of $\pi(\mu)$ are merged.
    % }
\end{enumerate}

Once we have the quotient graph corresponding to $(s,t)$-strip, we need to determine the direction of each edge in the quotient graph. The resulting graph will be a DAG such that each $(s,t)$-mincut appears as transversal in this DAG. We now describe the procedure of assigning the direction to any edge $(a,b)$ in the quotient graph.
Let $(u,v)$ be the corresponding edge in the flesh graph ${\cal F}_{S(\nu)}$; if there are multiple such edges pick any one of them arbitrarily. 
Let $u$ be compressed to $a$ and $v$ be compressed to $b$ while forming the quotient graph. 

Let $P$ be the proper path to which the edge $(u,v)$ is mapped in the skeleton. $P$ is obtained by extending path $\pi(u)$ to $\pi(v)$ in a specific direction. Direct the path $P$ in arbitrary direction. Without loss of generality, assume that $\pi(u)$ is a prefix and $\pi(v)$ is a suffix of the directed $P$. Note that $P$ must be intersecting the path joining the unit containing $s$ and the unit containing $t$ in the skeleton. Let one such edge be $(x,y)$
with $x$ lying on the side of the unit containing $s$ and $y$ lying on the side of unit containing $t$. While traversing the directed path $P$, if edge $(x,y)$ is traversed along $x\rightarrow y$, we assign the edge $(a,b)$ direction $a\rightarrow b$; otherwise we assign the direction $b \rightarrow a$.

% {
\begin{theorem}
For an undirected unweighted graph $G$, there exists an ${\cal O}(n^2)$ space data structure which can report the $(s,t)$-strip for any $s,t \in V$ in ${\cal O}(m)$ time.
\end{theorem}
\end{document}